\documentstyle[12pt,a41,epsfig]{article}

\newcommand\YINT{\int_{-1}^{+1}}
\newcommand{\DIRJ}{{\cal Q}^J_{\alpha}(p_2,p_1)}
\newcommand{\DIRJF}{{\cal Q}^{J,5}_{\alpha}(p_2,p_1)}

\newcommand{\bea}{\begin{eqnarray}}
\newcommand{\bq}{\begin{equation}}
\newcommand{\eea}{\end{eqnarray}}
\newcommand{\eq}{\end{equation}}

\newcommand\ka{\kappa_1}
\newcommand\kb{\kappa_2}
\newcommand\kap{\kappa_1'}
\newcommand\kbp{\kappa_2'}
\newcommand\xx{\tilde{x}}

\newcommand\AAA{\alpha_1}
\newcommand\AB{\alpha_2}

\newcommand\Pvec{\mbox{\boldmath $P$}}

\newcommand\bx{\overline{x}}
\newcommand\by{\overline{y}}
\newcommand\FFA{\mbox{$\widetilde{F}^a$}}
\begin{document}
\noindent
\sloppy
\thispagestyle{empty}
\begin{flushleft}
DESY 99--020 \hfill
{\tt hep-ph/9903520}\\
NTZ  16/99\\
March 1999
\end{flushleft}
%
\vspace*{\fill}
\begin{center}
{\LARGE\bf  The Virtual Compton Amplitude in the
}\\
\vspace{2mm}
{\LARGE\bf
Generalized Bjorken Region:}
\\
\vspace{2mm}
{\LARGE\bf Twist--2 Contributions}
\\
\vspace{2cm}
\large
Johannes Bl\"umlein$^a$, Bodo Geyer$^b$, and
Dieter Robaschik$^{a,c}$
\\
\vspace{2em}
\normalsize
{\it $^a$~DESY -- Zeuthen, Platanenallee 6, D--15735 Zeuthen, Germany}
\\
\vspace{2em}
{\it $^b$~Naturwissenschaftlich--Theoretisches Zentrum,
Universit\"at Leipzig,}\\
{\it Augustusplatz 10, D--04109 Leipzig, Germany}
 \\
\vspace{2em}
{\it $^c$~Institut f\"ur Theoretische Physik, 
 Karl--Franzens--Universit\"at Graz,}\\
{\it Univerist\"atsplatz 5, A--8010 Graz, Austria} \\
\today
\end{center}
\vspace*{\fill}
%
\begin{abstract}
\noindent
A systematic derivation is presented of the twist--2 anomalous dimensions
of the general quark and gluon light--ray operators in the generalized
Bjorken region in leading order both for unpolarized and polarized 
scattering. Various representations of the anomalous dimensions are 
derived in the non--local and local light cone expansion and their 
properties are discussed in detail. Evolution equations for these
operators are derived using different representations. General two-- and 
single--variable evolution equations are presented for the expectation 
values of these operators for non--forward scattering. The Compton 
amplitude is calculated in terms of these distribution amplitudes. In the
limit of forward scattering a new derivation of the integral relations 
between the twist--2 contributions to the structure functions is given. 
Special limiting cases which are derived from the general relations are 
discussed, as the forward case, near--forward scattering, and 
vacuum--meson transition. Solutions of the two--variable evolution 
equations for non--forward scattering are presented.
\end{abstract}
\vspace*{\fill}
\newpage
\section{Introduction}
\renewcommand{\theequation}{\thesection.\arabic{equation}}
\setcounter{equation}{0}
\label{sec-1}

\vspace{1mm}
\noindent
The Compton amplitude for the scattering of a virtual photon off a hadron
\begin{equation}
\gamma^* + p_1 \rightarrow \gamma^{'*} + p_2
\end{equation}
provides one of the basic tools to understand the short--distance
behavior of the nucleon and to test Quantum Chromodynamics (QCD) at
large space--like virtualities. The Compton amplitude for the general case of
non--forward scattering is given by
\begin{equation}
\label{COMP}
T_{\mu\nu}(p_+,p_-,q)
= i \int d^4x \,e^{iqx}\,
\langle p_2, S_2\,|T (J_{\mu}(x/2) J_{\nu}(-x/2))|\,p_1,
S_1\rangle~.
\end{equation}
Here,
\begin{eqnarray}
p_+ &=& p_2 + p_1,~~~~~~~~~~p_- = p_2 - p_1 = q_1 - q_2, \\
q   &=& \hbox{\large $\frac{1}{2}$}
        \left(q_1 + q_2\right),~~~~~~~~p_1 + q_1 = p_2 + q_2~,
\end{eqnarray}
where $q_1~(q_2)$ and $p_1~(p_2)$ denote the four--momenta of the
incoming~(outgoing) photon and hadron, respectively, and $S_{1}, S_{2}$
are the spins of the initial-- and final--state hadron.
The {\it generalized Bjorken region} is defined by the conditions
\begin{eqnarray}
\label{gBr1}
\nu =  qp_+ \longrightarrow \infty, \qquad
Q^2 = - q^2 \longrightarrow \infty~,
\end{eqnarray}
keeping the variables
\begin{eqnarray}
\label{gBr2}
\xi  =  \frac{Q^2 }{qp_+}, \qquad
\eta = \frac{qp_-}{qp_+} = \frac{q_1^2 - q_2^2}{2\nu}
\end{eqnarray}
fixed.
Relations between these
kinematic variables are given in
Appendix~\ref{sec-A}
For $q_2^2 = 0$ one obtains $\eta = - \xi$.
In distinction to deep inelastic forward scattering,
the scaling variable $\xi$ is not restricted to the
interval $0 \leq \xi \leq 1$. For forward scattering
$q = q_{1}= q_2, p_+ = 2p, \eta = 0$,
$\xi \rightarrow x_{\rm Bj} = Q^2/(2 pq)$, holds.
In principle the results presented below can be analytically continued 
to time--like processes with two space--like initial state photons as
a {\sf necessary} condition.

The (renormalized) time--ordered product  in Eq.~(\ref{COMP}) can be
represented in terms of the operator product expansion. Unlike the local
operator product expansion~\cite{OPEL}, which is widely used in the case
of forward scattering, the non--local operator product
expansion~[2--5], to which we refer in the present paper,
leads to  compact expressions for coefficient functions and operators in
the non--forward case.
As was shown in Ref.~\cite{SLAC} the expressions of the local operator
product expansions can be obtained by a Taylor
expansion of
the non--local operator product expansion~(cf. also \cite{AS,ZAV}). In
lowest order in the coupling constant the expansion contains only quark
operators with two external legs. These operators can be decomposed into
operators of different twist. In the present paper
we will consider the contribution of the
twist--2 operators only and calculate the anomalous dimensions which
emerge in the corresponding renormalization group equations.

Our goal is to demonstrate that quite different processes, such as deep
inelastic forward scattering, the different channels in deeply virtual
Compton scattering, i.e. non--forward scattering at a general kinematics,
cf. 
[7--9]
for early studies, and vacuum--meson
transitions contain as an essential input the same light--ray operators
of twist--2. Therefore the scaling violations of these processes are
described by the same anomalous dimensions. The non--perturbative
partition functions are obtained as the Fourier transforms of the
corresponding matrix elements of these operators. With the help of the
Fourier representations we derive a new representation for the Compton
scattering amplitude in the generalized Bjorken region which
extends the representation known for forward scattering. Moreover, the
partition functions as the non--perturbative input distributions are
related to the Fourier transforms of their matrix elements. Therefore,
the evolution equations result directly from the renormalization group
equation of the light--ray operators involved. Consequently, all the
evolution kernels can be determined from the anomalous dimensions of the
corresponding operators.
As special cases one obtains the evolution equations for forward
scattering and for the case of vacuum--meson transitions. Furthermore
the equivalence to other representations in the recent literature is
shown, in which different kinematic variables have been used.

In the generalized Bjorken region
the distribution amplitudes for the twist--2
contributions depend on two scaling variables unlike the case of
forward scattering. One of these parameters is related to the sum, the
other to the difference of the initial and final state nucleon momenta.
In \cite{RADL,RAD} a solution of the evolution equation for the non--singlet
case was obtained. 
In the literature mostly
the one--parametric equations have been discussed so far
by imposing further
conditions on the momenta.
The solution for the singlet case was  given in
Ref.~\cite{BGR2} 
for the two--variable equations. 

The paper is organized as follows. In Section~2 we derive the non--local
light cone expansion of the Compton amplitude for non--forward scattering
in lowest order. The twist decomposition of the non--local operators
is performed in Section~3. In Section~4 the non--forward matrix elements
of the non--local operators are calculated and related to the
corresponding distribution amplitudes. The double and single--variable
distribution amplitudes contributing to the Compton amplitude are
discussed in Section~5, where also a derivation of special relations
holding in the case of forward
scattering off
unpolarized and polarized nucleons is presented.
In Section~6 a detailed derivation
of the anomalous dimensions of the quark and gluon operators contributing
at the leading twist level to the non--forward Compton amplitude are
given in
$O(\alpha_s)$.
The evolution kernels are presented in different
representations both for unpolarized and polarized scattering.
The corresponding evolution equations are discussed in Section~7, both
for the case of two and one--variable evolution equations. Here also
the kernels for the single--variable equations are given.
Special cases of evolution equations into which the general equations
transform in a series of kinematic limits are discussed in Section~8,
including the cases of forward scattering and vacuum--meson transitions.
The solutions of the general two--variable evolution equations are
provided in Section~9. Section~10 contains the conclusions. In the
Appendices details of the calculation are presented. A brief summary
of part of our results has already appeared~\cite{BGR1}. In this paper
the  derivation is presented in detail.
\section{The non--local light--cone expansion in Born approximation}
\renewcommand{\theequation}{\thesection.\arabic{equation}}
\setcounter{equation}{0}
\label{sec-2}

\vspace{1mm}
\noindent
The Compton amplitude will be evaluated using the non--local light cone
expansion (LCE) in terms of bi--local light--ray operators. This
expansion is a
summed--up version of the local LCE. The latter
can be obtained performing a Taylor expansion of the
former representation.
The rigorous proof of the non--local LCE is rather complicated.
Here, following the prescription of Refs.~\cite{AS,ZAV,LEIP,SLAC},
it will be introduced heuristically with some
comments on the essential steps.

Firstly, one considers the
renormalized ($R$) time ordered ($T$) product of two electromagnetic
currents
\begin{equation}
RT\left(J_\mu(x/2)J_\nu(-x/2)S \right),
\end{equation}
 which in perturbative QCD is represented on the Hilbert space
of free fields. The electromagnetic current reads
\begin{eqnarray}
J_\mu(x) = \overline \psi(x) \gamma_\mu \lambda^{\rm em} \psi(x)~,
\label{tcur}
\nonumber
\end{eqnarray}
where $\psi(x)$ are the quark fields and
\begin{eqnarray}
\lambda^{\rm em} =
\hbox{\large$\frac{1}{2}$} \Big(\lambda^{3}_f +
\hbox{\large$\frac{1}{\sqrt 3}$}
\lambda^{8}_f\Big)
\quad{\rm for} \quad  SU(3)_{\rm flavor}~;
\nonumber
\end{eqnarray}
 $S$ denotes the renormalized S--matrix.

In lowest order in the coupling constant we have $S = 1$ and,
expressing the $T$--product in terms of normal products, we obtain
\begin{eqnarray}
\label{JJ1}
RT (J_\mu(x/2)J_\nu(-x/2)S)  
&=&
i 
:\,\overline\psi(x/2) \{
   {\hat S}_{\mu \nu \rho \sigma} 
   - i\varepsilon_{\mu \nu \rho \sigma} \gamma_5\} \gamma^\sigma
(i\partial_x^{\rho} D^c(x)) c^a \lambda^a_f \psi(-x/2)\,:
\nonumber
\\
& &
- \left[ (x/2,~\mu) \leftrightarrow (-x/2,~\nu) \right]
\\
& & + ~{\rm higher~order~terms~,}
\nonumber
\end{eqnarray}
\vspace*{-.3cm}
where
\vspace*{-.3cm}
\begin{eqnarray}
{\hat S}_{\mu \nu \rho \sigma} \equiv - S_{\mu \rho \nu \sigma}
= g_{\mu\nu} g_{\rho \sigma} - g_{\mu\rho} g_{\nu \sigma}
    - g_{\mu\sigma} g_{\rho \nu}
\nonumber
\end{eqnarray}
and $\varepsilon_{\mu\nu\rho\sigma}$ denotes the Levi--Civita symbol.
The $SU(3)_{\rm flavor}$ vector
\begin{eqnarray}
c^a = \left(
\hbox{\large$\frac{2}{9}$} \delta^{a0} +
\hbox{\large$\frac{1}{6}$} \delta^{a3} +
\hbox{\large$\frac{\sqrt{3}}{6}$} \delta^{a8} \right) e^2
\nonumber
\end{eqnarray}
with $e$ the electric charge,
determines the flavor content and
$D^c(x) = (-i/4\pi^2)(x^2 - i\varepsilon)^{-1}$ is the free scalar
propagator. Because this expression is restricted to leading order
no phase factors occur.

Secondly, since the non--local LCE of composite operators appears
as a (multiple) integral representation we hint at it by
 introducing two auxiliary $\kappa$--integrations
\cite{LEIP}~\footnote{
Here, and in the following, the normal product symbols
for the relevant operator products will be omitted.}
\begin{eqnarray}
\label{LCB}
RT \left(J_\mu(x/2)J_\nu(-x/2)S\right)    &=&
\frac{1}{2\pi^2} \frac{c^a x^\rho }{(x^2 - i\varepsilon)^2}
 \int_{-\infty}^{+\infty}  d\kappa_1
 \int_{-\infty}^{+\infty}  d\kappa_2 \\
&\times& \hspace*{-3mm}
\left\{
\hbox{\large $\frac{i}{2}$}
\left[
\overline\psi(\kappa_1 x)\lambda^a_f
\gamma^\sigma \psi(\kappa_2x)
   - \overline\psi(\kappa_2 x)\lambda^a_f \gamma^\sigma
\psi(\kappa_1 x)\right]
   (- {\hat S}_{\mu \nu \rho \sigma}) \Delta_-(\kappa_1,\kappa_2)
   \right. \nonumber \\
& & \left. \hspace*{-5mm}
+ \hbox{\large $\frac{i}{2}$}
\left[\overline\psi(\kappa_1 x)\gamma_5 \lambda^a_f
\gamma^\sigma \psi(\kappa_2 x)
+ \overline\psi(\kappa_2 x)\gamma_5\lambda^a_f
\gamma^\sigma \psi(\kappa_1 x)\right]
i\varepsilon_{\mu \nu \rho \sigma}
\Delta_+(\kappa_1,\kappa_2)\right\} \nonumber\\ & &
\hspace*{-5mm} + ~\ldots~,
\nonumber
\end{eqnarray}
where
\begin{eqnarray}
\Delta_{\pm}(\kappa_1, \kappa_2) =
  \left[\delta(\kappa_1 -\hbox{$\frac{1}{2}$})
  \delta(\kappa_2 + \hbox{$\frac{1}{2}$})
\pm
  \delta(\kappa_2 -\hbox{$\frac{1}{2}$})
  \delta(\kappa_1 + \hbox{$\frac{1}{2}$})\right]~.
\nonumber
\end{eqnarray}

The next step consists in the projection of the above expressions
onto the light--cone by replacing  $x \rightarrow \tilde x$ with
${\tilde x}^2 = 0$, leaving $x^2$ unchanged. The light--like vector
\begin{eqnarray}
\xx = x + \frac{\zeta}{\zeta^2}
\left(
\sqrt{{(x\zeta)}^2 - {x^2 \; \zeta^2}} - {(x\zeta)}
\right),
\nonumber
\end{eqnarray}
is related to $x$ and a fixed non--null subsidiary four--vector $\zeta$
the dependence of which drops out in leading order expressions. It has to
be noted that because the light--cone expansion is considered for the
operators and {\em not}, as often used,
 for their {\em matrix elements}, a light--like
reference vector has necessarily to be introduced in configuration
space.\footnote{$\xx$ and its pendant $\xx^\ast$ with
$\xx \xx^\ast= 1$ correspond to the light--like vectors $n$ and $p$ of
dimension ${\rm mass}^{-1}$ and mass, respectively, usually introduced in
deep inelastic scattering according to $P_\mu = p_\mu + (M^2/2) n_\mu$
\cite{JJ}. Furthermore, $\xx$ and $\xx^\ast$ define in a sense a
`comoving' reference system on the light cone.}

In general, also non--leading contributions related to higher order terms
of the $S$--matrix and, in certain gauges, also of the phases
factors $U(x/2, -x/2)$ have to
be taken into account. Then the coefficient functions like
$(c^a / 2\pi^2) (x^2 - i \varepsilon)^2 \Delta_\pm(\ka, \kb)$  depend
non--trivially on $\kappa_i$. In this way one arrives at a pre--form of
the non--local light--cone expansion being given by a sum over the
singular coefficient functions $C_a(x^2, \kappa_i) $ times
 the light--ray operators $O^a(\kappa_i\xx)$,
\begin{eqnarray}
\sum_a C_a(x^2, \kappa_i) \times  O^a(\kappa_i\xx)~.
\nonumber
\end{eqnarray}
%
It should be remarked that even if the perturbative expansion of the
renormalized time ordered product of the two currents has been correctly
performed, there is no reason to assume that the appearing light--ray
operators are the renormalized ones: The expansions resulting from
Eq.~(\ref{LCB}) are not yet the true light--cone expansions.

A complete proof which shows that the free field operators appearing in
Eq.~(\ref{LCB}) have to be substituted by renormalized light--ray
operators is much more involved~\cite{AS,ZAV}. Firstly, an
{\it operator identity} for the product of two composite operators,
e.g., the two currents, will be proven which holds on the whole Hilbert
space. Then, introducing a new light--cone adapted renormalization
procedure $R$ it can be shown that the perturbative functional of the
renormalized product of two currents may be split into an asymptotically
relevant part -- the light--cone expansion up to a definite order in the
light--cone singularity  -- and a well--defined remainder being less
singular:
\begin{eqnarray}
\label{GLCE}
\hspace{-.6cm}
{R} T\left(\!J_\mu(x/2)J_\nu(-x/2)S\!\right)\!
\approx\!
\int^{+1}_{-1} \!\! d^2 \underline\kappa\,
C_\Gamma(x^2, \underline\kappa; \mu^2)
{R }T\!\left( O^\Gamma(\ka \xx, \kb \xx) S\right)\!
+{\rm higher~order~terms,}
\end{eqnarray}
where the free light--ray operators with a specified
$\Gamma$--structure are given by
\begin{eqnarray}
\label{GLCEOP}
O^\Gamma(\ka \xx, \kb \xx)
=
\int 
\frac{dp_1}{(2\pi)^4}
\frac{dp_2}{(2\pi)^4}
\, e^{i\ka \xx p_1 + i \kb \xx p_2}
:\overline\psi(p_1) \Gamma \psi (p_2):~.
\end{eqnarray}
Here, the coefficient functions and the corresponding light--ray operators
appear quite naturally as renormalized ones. In the following, for
brevity, we usually
do not explicitly indicate that they are renormalized quantities.

Let us point to the fact that, in principle, within this formalism the
non--local as well as the usual local light--cone expansion may be
obtained (see also~\cite{SLAC}). The only difference consists in the
following: In the {\it non--local LCE}, as an intermediate step
to obtain the final representation Eq.~(\ref{GLCE}), 
a {\em Fourier transformation} of the renormalized, perturbatively determined
coefficient functions $F_\Gamma(x^2, \xx p_i, \mu^2)$ which multiply the
various normal products of free operators in momentum space, like
$:\overline\psi(p_1) \Gamma \psi (p_2):$ in Eq.~(\ref{GLCEOP}),
has been performed. Primarily, these functions are given as
%
\begin{eqnarray}
F_\Gamma(x^2, \xx p_i, \mu^2) = \int^{+1}_{-1} d^2\underline\kappa \;
e^{i\ka \xx p_1 + i \kb \xx p_2} \,
C_\Gamma(x^2, \underline\kappa; \mu^2).
\end{eqnarray}
In addition, by the $\alpha$--representation it can be proven that
they are entire analytic functions with respect to
the variables $\xx p_i$ \cite{AS,ZAV} and therefore the range of
$\kappa_i$ is
restricted to
\begin{equation}
-1 \leq \kappa_i \leq +1~.
\end{equation}
On the other hand, in the {\it local LCE} one applies a
{\em Taylor expansion}
of $F_\Gamma$ with respect to $\xx p_i$ which leads to a (twofold)
infinite series over the local operators.

These local and non--local
operators are connected by the following relations:
\begin{eqnarray}
\label{taylor}
O^\Gamma(\ka \xx, \kb \xx)
&=&
\sum_{n_1 n_2}
\frac{\ka^{n_1}}{n_1!}
\frac{\kb^{n_2}}{n_2!} \;
O^\Gamma_{n_1 n_2}(\xx),
\\
\label{dtaylor}
O^\Gamma_{n_1 n_2}(\xx)
&=&
\frac{\partial^{n_1}}{\partial \ka^{n_1}}
\frac{\partial^{n_2}}{\partial \kb^{n_2}}
\left. O^\Gamma(\ka \xx, \kb \xx)
\right|_{\ka=\kb=0}.
\end{eqnarray}
With the convention
\begin{eqnarray}
(\xx \vec{\partial})^n \equiv
\xx^{\mu_1}...\xx^{\mu_n}
\vec{\partial}_{\mu_1}...
\vec{\partial}_{\mu_n}
\nonumber
\end{eqnarray}
and choosing the axial gauge, $\xx^\mu A_\mu(x) = 0$,
we obtain
\begin{equation}
O^\Gamma_{n_1 n_2}(\xx) =
{\overline\psi}(0)
(\stackrel{\leftarrow}{\partial}\xx)^{n_1}
\Gamma
(\xx\stackrel{\rightarrow}{\partial})^{n_2}
\psi(0)~.
\nonumber
\end{equation}

Summarizing the foregoing procedure and introducing, for later
convenience, the variables
\begin{equation}
\kappa_{\pm} = \hbox{\large
$\frac{1}{2}$}(\kappa_2 \pm \kappa_1)
\qquad {\rm with} \qquad
\kappa_{1,2} = \kappa_+ \mp \kappa_-,
\end{equation}
the general expression will be, taking into account the explicit form of
${\hat S}_{\mu\nu\rho\sigma}$,
\begin{eqnarray}
\label{tpro}
\lefteqn{
RT(J_{\mu}(x/2) J_{\nu}(-x/2) S) \approx
\frac{1}{2}
\int D\kappa
}
\\
\hspace{-1.5cm}
& & \times
\left[C_a(x^2,\kappa_+,\kappa_-,\mu^2)
\left(g_{\mu \nu} O^a (\kappa_+ \xx, \kappa_- \xx, \mu^2)
 -\xx_{\mu} O^a_\nu (\kappa_+ \xx, \kappa_- \xx, \mu^2)
 -\xx_{\nu} O^a_\mu (\kappa_+ \xx, \kappa_- \xx, \mu^2)
\right) \right.
\nonumber\\
\hspace{-1.5cm}
& &
\left.
+ \,i\,C_{a,5}(x^2, \kappa_+, \kappa_-, \mu^2)
{\varepsilon_{\mu\nu}}^{\rho\sigma}
\xx_{\rho} O_{5, \sigma}^a (\kappa_+ \xx, \kappa_- \xx, \mu^2)
\right ] ~+~{\rm higher~order~terms},\nonumber
\end{eqnarray}
where the measure
\bea
\label{Dkappa}
D\kappa
&=&
d\ka d\kb \theta(1-\ka) \theta(1+\ka) \theta(1-\kb) \theta(1+\kb) \\
&=&  2
d\kappa_+ d\kappa_-
\theta(1+\kappa_+ + \kappa_-) \theta(1+\kappa_+ - \kappa_-)
\theta(1-\kappa_+ + \kappa_-) \theta(1-\kappa_+ - \kappa_-)
\nonumber
\eea
has been introduced.
Here  $C_{a (5)}$ are the renormalized coefficient functions
which in Born approximation read
\begin{eqnarray}
\label{coe1}
C_a(x^2, \kappa_+,\kappa_-)&=&
\frac{1}{2\pi^2} \frac{c_a}{(x^2 - i\varepsilon)^2}
  \delta(\kappa_+)
  \left[\delta(\kappa_- -\hbox{$\frac{1}{2}$})
  -\delta(\kappa_- +\hbox{$\frac{1}{2}$})\right],
\\
\label{coe2}
C_{a,5}(x^2, \kappa_+,\kappa_-) &=&
\frac{1}{2\pi^2} \frac{c_a}{(x^2 - i\varepsilon)^2}
  \delta(\kappa_+)
  \left[\delta(\kappa_- -\hbox{$\frac{1}{2}$})
  +\delta(\kappa_- +\hbox{$\frac{1}{2}$})\right],
\end{eqnarray}
and $O^a_{(5)\sigma}$ are the renormalized
(anti)symmetric bi--local LC-operators
\footnote{In the following $(\ka, \kb)$ and $(\kappa_+, \kappa_-)$
are chosen interchangeably, i.e. $O(\kappa_+\xx, \kappa_-\xx)$ is written
instead of the more lengthly expression $O((\kappa_+ - \kappa_-)\xx,
(\kappa_+ + \kappa_-)\xx)$. Furthermore, very often $\xx$ and the
renormalization scale $\mu^2$ will be suppressed in the arguments of the
operators and related quantities for brevity.}
\begin{eqnarray}
\label{qop}
\hspace{-1cm}
 O^a_\sigma(\kappa_1,\kappa_2)
\!\!&=&\!\!
\hbox{\large $\frac{i}{2}$}
RT \left [
\overline\psi(\kappa_1\xx)\lambda^a_f\gamma_\sigma
U(\ka,\kb) \psi(\kappa_2\xx)
-
\overline\psi(\kappa_2 \xx)\lambda^a_f\gamma_\sigma
U(\kb,\ka) \psi(\kappa_1\xx)\right]S,
\\
\label{qop5}
\hspace{-1cm}
 O^a_{5,\sigma}(\kappa_1,\kappa_2)
\!\!&=&\!\!
\hbox{\large $\frac{i}{2}$}
RT\left[
\overline\psi(\kappa_1 \xx)\lambda^a_f\gamma_5\gamma_\sigma
U(\ka,\kb) \psi(\kappa_2\xx)
+
\overline\psi(\kappa_2 \xx)\lambda^a_f\gamma_5\gamma_\sigma
U(\kb,\ka) \psi(\kappa_1\xx)\right]S,
\end{eqnarray}
respectively, with the phase factors
\begin{eqnarray}
\label{phase}
U(\ka, \kb)
\equiv
U(\ka \xx, \kb \xx)
 =
{\cal P} \exp\left\{-ig
\int^{\ka}_{\kb} d\tau \,\xx^\mu A_\mu (\tau \xx)
\right\},
\end{eqnarray}
where ${\cal P}$ denotes the path ordering, $g$ is the strong coupling
constant and $ A_\mu = A_\mu^a t^a$ is the gluon field with $t^a$ being
the generators of $SU(3)_{\rm color}$ in the fundamental representation
spanned by the quark fields. In the explicit expressions we have
indicated by $RT[O]S$ that these operators are renormalized.

The general expression Eq.~(\ref{tpro}) contains a flavor singlet as well
as the
flavor non--singlet (NS) parts. In the case of flavor non--singlet
operators, i.e., for $\lambda^a_f \neq 1$, the consideration of the quark
operators Eqs.~(\ref{qop}) and (\ref{qop5}) is sufficient. However, the
singlet operators ($\lambda^0_f \equiv 1$), in the following denoted by
$O^q$, mix with certain gluon operators which will be denoted by $O^G$.
In addition, if we would consider the Compton amplitude beyond leading
order also the gluon operators would appear in the LCE.

The gluon operators related to the  quark operators Eqs.~(\ref{qop})
and (\ref{qop5}) read
%
\begin{eqnarray}
\label{GG0}
\hspace{-.8cm}
O^G_{\mu\nu}(\ka , \kb )
\!\!&=&\!\!
\hbox{\large $\frac{1}{2}$}
RT\left[
F^{a \lambda}_\mu(\ka \xx)U^{ab}(\ka ,\kb )
F^b_{\nu\lambda}(\kb\xx)
+
F^{a\lambda}_\mu(\kb \xx)U^{ab}(\kb ,\ka )
F^b_{\nu\lambda}(\kb \xx)
\right]S,
\\
\label{GG05}
\hspace{-.8cm}
O^G_{5\mu\nu}(\ka , \kb )
\!\!&=&\!\!
\hbox{\large $\frac{1}{2}$}
RT\left[
F^{a\lambda}_\mu(\ka \xx)U^{ab}(\ka ,\kb )
\widetilde F^b_{\nu\lambda}(\kb \xx)
-F^{a\lambda}_\mu(\kb \xx)U^{ab}(\kb ,\ka )
\widetilde F^b_{\nu\lambda}(\ka \xx)
\right]S,
\end{eqnarray}
where,
\vspace*{-.3cm}
\begin{eqnarray}
F_{\mu\nu} \equiv F_{\mu\nu}^a t^a~~~{\rm and}~~~
\widetilde{F}_{\mu\nu} =
\hbox{\large$\frac{1}{2}$}
\varepsilon_{\mu\nu\rho\sigma} F^{\rho\sigma}
\nonumber
\end{eqnarray}
is the gluon field strength and the dual field strength, respectively.
$U^{ab}(\ka,\kb)$ is the phase factor (\ref{phase})
in the adjoint representation.
To obtain exact correspondence to the above quark operators
the gluon operators have to be multiplied by
$\hbox{$\frac{1}{2}$}
(g_\sigma^{~\mu}\xx^\nu + g_\sigma^{~\nu}\xx^\mu)$.
\section{Twist decomposition of non--local operators}
\renewcommand{\theequation}{\thesection.\arabic{equation}}
\setcounter{equation}{0}
\label{sec-3}

\vspace{1mm}
\noindent
The non--local operators Eqs.~(\ref{qop}, \ref{qop5}, \ref{GG0},
\ref{GG05}) contain contributions of different twist. Here we use the
notion of twist in its original definition \cite{GROSS} as
{\em geometric twist}, i.e. canonical dimension  minus spin of the
corresponding operator. Let us now extract the leading
twist parts, which are relevant for the discussion of virtual Compton
scattering. This concerns the contributions of twist two and three,
the latter being especially relevant for the antisymmetric part of the
scattering amplitude which concerns the polarized scattering.

We consider first the quark operators $O^a_\sigma$ and
$O_{5, \sigma}^{a}$, which contain contributions of twist--2, 3 and 4:
%
\begin{eqnarray}
\label{twist}
O^a_\sigma
 &=&
O_\sigma^{a,\,\rm twist 2}
  + O_\sigma^{a,\,\rm twist 3}
  + O_\sigma^{a,\,\rm twist 4},
\\
\label{twist5}
O_{5,\sigma}^a
 &=&
O_{5,\sigma}^{a,\,\rm twist 2}
 + O_{5,\sigma}^{a,\,\rm twist 3}
 + O_{5,\sigma}^{a,\,\rm twist 4}.
\end{eqnarray}
%
In principle also higher twist contributions appear, but they vanish on
the light--cone since they are proportional to
nonvanishing integer powers of $\xx^2$.

To obtain the bi--local operators (\ref{twist}) of definite twist we
consider the {\em centered  operators}
\begin{equation}
\overline\psi(-\kappa x)\lambda^a_f\gamma_\sigma
U(-\kappa x, \kappa x) \psi(\kappa x)
\end{equation}
with {\em arbitrary} values of $x\neq \xx$, i.e.,~we consider the
operators as
not restricted to the light--cone. They are represented through
the Taylor expansion Eq.~(\ref{taylor}) by a series of {\em local}
tensor operators of order $n = n_1 + n_2$ which, in the axial gauge, may
be written as
\begin{eqnarray}
\sum_n \frac{\kappa^n}{n!} x^{\mu_1} \ldots  x^{\mu_n} \overline\psi(0)
\stackrel{\leftrightarrow}{\partial}_{\mu_1} \ldots
\stackrel{\leftrightarrow}{\partial}_{\mu_n} \lambda^a_f
\gamma_\sigma \psi(0)~.
\nonumber
\end{eqnarray}
Here we used the abbreviation $\stackrel{\leftrightarrow}{\partial}
\equiv
\stackrel{\rightarrow}{\partial} -\stackrel{\leftarrow}{\partial} $.
By purely group theoretical methods these local operators are decomposed
into traceless tensor operators of definite dimension and spin carrying
an irreducible representation of the Lorentz group. 
The symmetry behavior
of these tensors is characterized by a few special classes of Young
tableaux'. As is well known, the twist--2 terms correspond to the total
symmetric representation. Finally, the local operators of definite twist
are resummed to give  non--local harmonic tensor operators of definite twist.

Technically, these  non--local harmonic tensor operators are obtained
from the
initial ones by first projecting onto traceless operators and then
applying appropriate differential operators related to the corresponding
symmetry class. Finally, the resulting expressions are projected onto the
light--ray $\xx$. In principle this method can also be applied for
arbitrary values of $\kappa_i$.

This approach has been worked out in Refs.~\cite{LAZAR,GLR} for
various light--ray operators. Here, we refer only to those representations
which are relevant for the subsequent considerations. First the
projections onto the traceless operators are defined. For the non--local
scalar and vector quark operators Eqs.~(\ref{qop}) and (\ref{qop5}),
respectively, they are given by
\begin{eqnarray}
\label{proj_tw2}
O^{q,\;\rm traceless}(-\kappa x,\kappa x)
&=&
\hbox{\Large $\frac{i}{2}$}
\left[\overline{\psi}(-\kappa x)(x\gamma)\psi(\kappa x)
-\overline{\psi}(\kappa x)(x\gamma)\psi(-\kappa x)\right]
\\
&+&
\sum_{k=1}^{\infty}\int_0^1 dt
\left(\frac{1-t}{t}\right)^{k-1}
\left(\frac{-x^2}{4}\right)^k
\frac{\Box^k}{k!(k-1)!}
\nonumber\\ & & \times
\hbox{\Large $\frac{i}{2}$}
\left[\overline{\psi}(-\kappa tx)(x\gamma)\psi(\kappa tx)
-
\overline{\psi}(\kappa tx)(x\gamma)\psi(-\kappa tx)\right],
\nonumber\\
\label{proj_tw3}
O^{q,\;\rm traceless}_\sigma(-\kappa x,\kappa x)
&=&
\hbox{\Large $\frac{i}{2}$}
\left[\overline{\psi}(-\kappa x)\gamma_\sigma \psi(\kappa x)
-\overline{\psi}(\kappa x)\gamma_\sigma \psi(-\kappa x) \right]
\\
&+&
\sum_{k=1}^\infty\int_0^1\frac{dt}{t}
\left(\frac{1-t}{t}\right)^{k-1}
\left(\frac{-x^2}{4}\right)^k
\frac{\Box^k}{k!(k-1)!} \nonumber\\ & &
\times
\hbox{\Large $\frac{i}{2}$}
\left[\overline{\psi}(-\kappa tx)\gamma_\sigma \psi(\kappa tx)
-
\overline{\psi}(\kappa tx)\gamma_\sigma \psi(-\kappa tx)
\right]
\nonumber\\
&+&
\hbox{\Large $\frac{1}{2}$}\,
\Big[
( x_\sigma x_\rho \Box + x^2\partial_\sigma\partial_\rho)
-
2 x_\sigma\partial_\rho (x\partial)
\Big] \nonumber\\
& &
\times
\sum_{k=0}^\infty\int_0^1 d\tau\int_0^1 dt
\left(\frac{1-t}{t}\right)^k
\left(\frac{-x^2}{4}\right)^k
\frac{\Box^k}{k!k!}
\nonumber
\\
& &
\times
\hbox{\Large $\frac{i}{2}$}
\left[
\overline{\psi}(-\kappa\tau tx)\gamma^\rho
\psi(\kappa\tau tx)
-
\overline{\psi}(\kappa\tau tx)\gamma^\rho
\psi(-\kappa\tau tx)
\right].
\nonumber
\end{eqnarray}
%
Here the phase factors and the generators $\lambda^a_f$ of the flavor
group have been suppressed and the operators generically are indicated
by $q$. The sum over terms containing the d'Alembertian
are the subtractions of the traces. The representation for
$O^{q,\;\rm traceless}$, Eq.~(\ref{proj_tw2}), has been obtained
already earlier in  Ref.~\cite{BB}. Taking the limit
$x \rightarrow \xx$ in Eq.~(\ref{proj_tw2}) the additional terms vanish
because of $\xx^2 = 0$. Therefore the operator
$\xx^\sigma O^q_\sigma (-\kappa \xx, \kappa \xx)$ is of twist two:
\bq
\label{tw2qS}
O^{q,\;\rm twist 2}(-\kappa \xx,\kappa \xx)
=
O^{q,\;\rm traceless}(-\kappa \xx,\kappa \xx)
\equiv
O^q(-\kappa \xx,\kappa \xx),
\eq

Because of their tracelessness the above operators fulfill the following
relations
%
\begin{eqnarray}
\Box O^{q,\;\rm traceless}(-\kappa x,\kappa x) = 0,
\quad
\partial^\sigma O^{q,\;\rm traceless}_\sigma(-\kappa x,\kappa x)
=0,
\quad
\Box O^{q,\;\rm traceless}_\sigma(-\kappa x,\kappa x) =0.
\end{eqnarray}
which proves that they are harmonic scalar and vector operators,
respectively.
The differential operators, to be applied to $O^{q,\;\rm traceless}$ and
$O^{q,\;\rm traceless}_\rho$ in order to generate the correct
symmetry behavior, are $\partial_\sigma$ for twist--2 and
$(g_{\sigma\rho} (x\partial) - x_\sigma \partial_\rho)$ for twist--3, 
respectively.
Note that taking these derivatives before carrying out the light--cone
limit leads to non--trivial contributions which result from the trace terms.

The explicit computations straightforwardly lead to the following
expressions for the twist--2, twist--3 and twist--4 light--ray
vector operators \cite{LAZAR,GLR}:
%
\begin{eqnarray}
\label{optw2}
O_{\sigma}^{q,\, \rm twist 2} (- \kappa \xx, \kappa \xx)
& = & \int_0^{1} d {\tau} \;
\partial_\sigma
\left.
O^q_{\rm traceless}(-\kappa\tau x, \kappa\tau x)
\right|_{x \rightarrow \xx}
\nonumber\\
& = & \int_0^{1} d {\tau}
 \left[ \partial_\sigma
+ \hbox{$\frac{1}{2}$}
(\ln \tau) x_\sigma \Box\right]
\left. O^q(-\kappa\tau x, \kappa\tau x)\right|_{x=\xx},
\\
\label{optw3}
 O_{\sigma}^{q,\, \rm twist 3} (- \kappa \xx, \kappa \xx)
& =&  \int_0^1 d {\tau}\;
(g_{\sigma\rho} x\partial- x_\sigma \partial_\rho)
\left. O^{q\,\rho}_{ \rm traceless}(-\kappa\tau x, \kappa\tau x)
\right|_{x=\xx},
\nonumber\\
& =&  \int_0^1 d {\tau}\,
\left[
\left(g_{\sigma\rho} x\partial- x_\sigma \partial_\rho\right)
\right.
\\
&  &\quad \; \left.
- \;x_\sigma ((1 + 2 \ln\tau) \partial_\rho
     + (\ln\tau) x_\rho \Box) \right]
\left. O^{q,\,\rho}(-\kappa\tau x,\kappa\tau x)\right|_{x=\xx},
\nonumber
\\
\label{optw4}
 O_{\sigma}^{q,\, \rm twist 4} (-\kappa \xx, \kappa \xx)
 &= &
x_\sigma \int_0^1 d {\tau}
\left[(1 + \ln\tau) \partial_\rho
+ \hbox{$\frac{1}{2}$} (\ln \tau) x_\rho \Box\right]
\left. O^{q\,\rho}(-\kappa\tau x, \kappa\tau x)\right|_{x=\xx},
\end{eqnarray}
%
where the last expression is obtained using Eq.~(\ref{twist}).
Note that the operators appearing at the r.h.s.~of
Eqs.~(\ref{optw2} -- \ref{optw4}) are to be taken with the phase factors
before carrying out the derivatives.

Important properties of these operators are
\begin{eqnarray}
\xx^\sigma O_{\sigma}^{q,\, \rm twist 2} = O^{q,\, \rm twist 2},
\qquad \xx^\sigma O_{\sigma}^{q,\,\rm twist 3} = 0,
\qquad \xx^\sigma O_{\sigma}^{q,\,\rm twist 4} = 0.
\end{eqnarray}
Analogous representations hold for $O_{5,\sigma}^{q,\, \rm twist 2}$,
$O_{5,\sigma}^{q,\, \rm twist 3}$ and $ O_{5,\sigma}^{q,\, \rm twist 4}$,
which have similar properties.
Note that the operators (\ref{optw2}\,--\,\ref{optw4}) include the trace
terms which are proportional to $\xx_\sigma$. In fact they are again
summed--up local operators fulfilling Eq.~(\ref{taylor}).

Let us now consider the gluon operators Eqs.~(\ref{GG0}) and
(\ref{GG05}). Again they contain a twist--2, twist--3 and twist--4 part,
related to the traceless symmetric, the antisymmetric and the trace part
of the tensor $O^G_{(5)\mu\nu}$, respectively. Since we are concerned with
the scalar and vector gluon operators,
\begin{eqnarray}
\xx^\mu\xx^\nu O^G_{(5)\mu\nu}~~~~~{\rm and}~~~~~
\hbox{\large $\frac{1}{2}$}
(g_\sigma^{\;\mu} \xx^\nu + g_\sigma^{\;\nu} \xx^\mu) O^G_{(5)\mu\nu},
\nonumber
\end{eqnarray}
only the symmetric twist--2 part will be considered here. In the same
manner as in the case of quark operators we first define the projection
onto the centered traceless operator. It is given by
%
\begin{eqnarray}
\label{proj_tw2G}
O^{G,\;\rm traceless}(-\kappa x,\kappa x)
&=&
\hbox{\large $\frac{1}{2}$}\, x_\mu x_\nu
\left[
F^{a \lambda}_\mu(-\kappa x)
F^a_{\nu\lambda}(\kappa x)
+
F^{a\lambda}_\mu(\kappa x)
F^a_{\nu\lambda}(-\kappa x)
\right]
\\
&+&
\sum_{k=1}^{\infty}\int_0^1 dt \,t
\left(\frac{1-t}{t}\right)^{k-1}
\left(\frac{-x^2}{4}\right)^k
\frac{\Box^k}{k!(k-1)!}
\nonumber
\\
& &
\times
\hbox{\large $\frac{1}{2}$}\,x_\mu x_\nu
\left[
F^{a \lambda}_\mu(-\kappa tx)
F^a_{\nu\lambda}(\kappa tx)
+
F^{a\lambda}_\mu(\kappa tx)
F^a_{\nu\lambda}(-\kappa tx)
\right],
\nonumber
\end{eqnarray}
where again the phase factors have been suppressed. It fulfills the
relation
\begin{eqnarray}
\Box\,O^{G,\;\rm traceless}(-\kappa x,\kappa x) = 0.
\end{eqnarray}
In comparison with Eq.~(\ref{proj_tw2}) only an additional factor of $t$
appears under the integral.
This results from the fact that the operator (\ref{GG0}) relative to
(\ref{qop})  contains one more factor of $x$ not being accompanied by
the parameter $\kappa$. The scalar and vector twist--2 gluon light--ray
operators are now given by
%
\begin{eqnarray}
\label{tw2GS}
O^{G,\;\rm twist 2}(-\kappa \xx,\kappa \xx)
&=&
O^{G,\;\rm traceless}(-\kappa \xx,\kappa \xx)
\equiv
O^G(-\kappa \xx,\kappa \xx),
\\
\label{tw2GV}
O_\mu^{G,\;\rm twist 2}(-\kappa \xx,\kappa \xx)
&=&
\int^1_0 d\tau \, \tau \,\partial_\mu
\left.
O^{G,\;\rm traceless}(-\kappa\tau x,\kappa\tau x)
\right|_{x = \xx}
\nonumber
\\
&=&
\int^1_0 d\tau \, \tau
\left[\partial_\mu
+ \hbox{$\frac{1}{2}$} (\ln \tau) x_\mu\Box\right]
\left. O^G(-\kappa\tau x,\kappa\tau x)\right|_{x = \xx}.
\end{eqnarray}
Here we gave for simplicity the
expressions for operators with the special
arguments $\kappa_1 = - \kappa,\, \kappa_2 = \kappa$. However, any of the
Eqs.~(\ref{proj_tw2} -- \ref{tw2GV}) remains true if general arguments are
chosen. As an example the singlet vector operators of twist--2 read
\begin{eqnarray}
\label{optw2A}
O_\mu^{A,\;\rm twist 2}(\ka \xx,\kb \xx)
&=&
\int^1_0 d\tau \, \tau^{d_A-1} \left[\partial_\mu
+ \hbox{$\frac{1}{2}$} (\ln \tau) x_\mu\Box\right] \left.
O^A(\ka\tau x,\kb\tau x)\right|_{x = \xx},
\end{eqnarray}
where $A = (q,G)$ and $d_q = 1, d_G = 2$. In addition, let us emphasize
that in the various integrals defining operators of definite twist
the variables $\kappa_i$ have to be scaled but {\em not} the coordinates
$x$ of the corresponding (original) operators.

The general form of harmonic operators of definite twist
is necessary for an investigation of their matrix elements
whereas the special centered form is sufficient to determine their anomalous
dimensions, as will be shown in  Sections~4 and 6, respectively. Inserting
the result obtained for the
twist--2 terms, 
Eq.~(\ref{optw2}), into Eq.~(\ref{tpro})
we obtain an asymptotic representation of the Compton scattering
amplitude
Eq.~(\ref{COMP}) which is valid in the generalized Bjorken region, as
will
be shown in Section~5 in detail. Concerning the renormalization of the
twist--2 singlet operators and the evolution equations of the
corresponding
distribution amplitudes also the representations Eq.~(\ref{tw2GS}) and
(\ref{tw2GV}) necessarily come into play.
\section{Matrix elements of non--local operators}
\renewcommand{\theequation}{\thesection.\arabic{equation}}
\setcounter{equation}{0}
\label{sec-4}

\vspace{1mm}
\noindent
The Compton scattering amplitude is obtained by taking the matrix
elements of the operator, Eq.~(\ref{tpro}), between the hadron states
$|p_i, S_i \rangle$. Thereby the matrix elements of the light--ray
operators
Eqs.~(\ref{qop}) and (\ref{qop5}) appear as non--perturbative
inputs. These quantities are represented by partition
functions in the forward case and distribution amplitudes in the case
of non--forward scattering.
Here, and for the description
of the evolution equations below, the general form of the operators
$O^\Gamma(\kappa_1 \xx,\kappa_2 \xx)$ has to be taken into account.
It will be shown that their Fourier transforms with respect to
$\kappa_-\xx p_+$ and $\kappa_-\xx p_-$ define the above mentioned
distribution amplitudes. For the following considerations it is
also useful, as it was in the twist decomposition, to study
the non--local operators
temporarily to arbitrary values of $x$ and to perform the light cone
limit only afterwards.

In the following we consider the contributions of twist--2 only. In the
case of purely electromagnetic interactions the matrix elements at lowest
order in the coupling constant, using the equations of motion, can be
represented by two basic hadronic matrix
elements, including also the spin dependence:
\begin{eqnarray}
\label{dirac}
{\cal Q}^{D}(x;p_2,p_1) &\equiv& x^{\mu}{\cal Q}_{\mu}^D(p_2,p_1)
= x^{\mu}\overline{u}(p_2,S_2)  \gamma_{\mu} u(p_1, S_1)
\\
\label{pauli}
{\cal Q}^{P}(x;p_2,p_1) &\equiv& x^{\mu}{\cal Q}_{\mu}^P(p_2,p_1)
= x^{\mu}\overline{u}(p_2, S_2)
\hbox{\large$\frac{1}{M}$}
\sigma_{\mu\nu} p_-^{\nu} u(p_1, S_1)~,
\end{eqnarray}
with
\begin{eqnarray}
\sigma_{\mu\nu} = \hbox{\large$\frac{i}{2}$}
\left[\gamma_{\mu}, \gamma_{\nu} \right]~;
\nonumber
\end{eqnarray}
$M$ is the nucleon mass,
and $u(p,S)$ denotes a free hadronic Dirac spinor. Here the
indices $D$ and $P$ refer to the  Dirac and  Pauli structure,
 respectively.

Let us first consider the matrix elements of the
 scalar twist--2 quark operator. Because of translation invariance,
\begin{eqnarray}
\label{translation}
\langle p_2 \,| O^\Gamma(\ka x, \kb x) |\,p_1\rangle
&\equiv&
\langle p_2 \,| O^\Gamma((\kappa_+ - \kappa_-) x,
 (\kappa_+ + \kappa_-) x) |\,p_1\rangle
\nonumber \\
&=&
e^{i\kappa_+ xp_-}
\langle p_2 \,|O^\Gamma(-\kappa_- x,\kappa_- x)|\,p_1\rangle~,
\end{eqnarray}
it is completely
sufficient to consider the matrix elements of the  centered
operators: 
\begin{eqnarray}
\label{kdec1}
\hspace{-.5cm}
\langle p_2|O^{q,\;\rm twist2}(-\kappa_- x,\kappa_- x)|p_1\rangle
 = i
\tilde f^q_J(\kappa_- xp_+,\kappa_- x p_-,\kappa_-^2 x^2, p_1p_2,\mu^2)
{\cal Q}^J(x; p_2, p_1),
\end{eqnarray}
where the summation is taken over $J=(D,P$).
If considered independently these
functions will be denoted by $f_D \equiv g$ and $f_P \equiv h$.
The generalization to arbitrary values of $\kappa_i$ is trivial due to
translation invariance, Eq.(\ref{translation}), through which a factor
$\exp\{i\kappa_+ \xx p_-\}$ occurs on both sides of Eq.~(\ref{kdec1}).
The projection onto the light--cone immediately gives
\begin{eqnarray}
\label{kdec1lc}
\hspace{-.5cm}
\langle p_2|O^{q,\;\rm twist2}(-\kappa_- \xx,\kappa_- \xx)|p_1\rangle
 =  i
\tilde f^q_J(\kappa_- \xx p_+,\kappa_- \xx p_-,
p_1p_2,\mu^2)
{\cal Q}^J(\xx; p_2, p_1)~.
\end{eqnarray}

The twist--2 vector operators,
because of the additional derivative in Eq.~(\ref{optw2}),
 do not have such a simple representation. Using that
relation between the vector operator and the scalar operator for
general arguments we obtain
\begin{eqnarray}
\label{kdecv}
\lefteqn{\hspace{-1cm}
\langle p_2|O^{q,\;\rm twist 2}_\mu
(\kappa_1 \xx, \kappa_2 \xx)|p_1\rangle
=
\int_0^1 \! d \lambda \, \partial_\mu^x \,
\langle p_2|O^{q,\;\rm twist 2}(\kappa_1 \lambda x,
\kappa_2 \lambda x)|p_1\rangle\Big|_{x=\xx}
}
\\
&=& i
\int_0^1 \! d \lambda \,
\partial_\mu^x \left(
\tilde f^q_J(\kappa_- \lambda x p_+,\kappa_- \lambda x p_-,
\kappa^2_- \lambda^2 x^2, p_1p_2, \mu^2)
{\cal Q}^J(x;p_2,p_1) e^{i\kappa_+\lambda xp_-}
\right)\Big|_{x=\xx}.
\nonumber
\end{eqnarray}
Let us now perform the Fourier transformations of $\tilde f_J$
with respect to $\kappa_-\lambda xp_\pm \equiv \kappa xp_\pm $,
\begin{eqnarray}
\label{zrep}
\tilde f(\kappa x p_+, \kappa x p_-, \kappa^2 x^2)
 = \int Dz
e^{-i\kappa x(p_iz_i)} f(z_+,z_-, \kappa^2 x^2)~,
\end{eqnarray}
where
\begin{eqnarray}
\label{eqpz}
(p_iz_i) \equiv p(z) = p_+z_+ +p_- z_-
\quad {\rm with} \quad z_{\pm} =
\hbox{\large$\frac{1}{2}$} \left(z_2 \pm z_1 \right)~,
\end{eqnarray}
defines a scalar product in the space of two--vectors labeled either by
$(1,2)$ or $(+,-)$. Also these functions,
if projected onto the light--cone, are entire analytic functions with
respect
to $\kappa_- \xx p_i$ and therefore the support of their Fourier
transforms is restricted to the range
\bea
-1\leq z_i \leq 1~.
\nonumber
\eea
Thereby, we assumed that the kinematic decomposition of the matrix
elements does not introduce additional kinematic singularities.
Therefore, the integration measure in Eq.~(\ref{zrep}), analogous to
Eq.~(\ref{Dkappa}), is given by
\bea
\label{Dz}
Dz
&=& \frac{1}{2}
d z_1 d z_2 \theta(1-z_1) \theta(1+z_1) \theta(1-z_2) \theta(1+z_2) \\
&=&
d z_+ d z_-
\theta(1 + z_+ + z_-) \theta(1 + z_+ - z_-)
\theta(1 - z_+ + z_-) \theta(1 - z_+ - z_-)
\nonumber
\eea

The $\lambda$--integration in Eq.~(\ref{kdecv}) can be performed leading
to
\begin{eqnarray}
\label{MAIN}
\lefteqn{\hspace{-1.5cm}
\langle p_2|O^{q,\;\rm twist 2}_\mu
(\kappa_1 \xx, \kappa_2\xx)|p_1\rangle
=
i \int Dz
e^{-i\kappa_- (\xx p_i)z_i}
}
\\
\!\!&\times&\!\! \Big\{\!
F^q_J(z_+,z_-,\kappa_+ \xx p_-, \kappa^2_- x^2)
\left({\cal Q}^J_{\mu}(p_2,p_1)
-i\kappa_- p_\mu(z){\cal Q}^J(x; p_2,p_1)\right)
\nonumber\\ & &\!\!
+
\left(i\kappa_+ p_-^{\mu} \partial_{i\kappa_+ x p_-}
+ 2 x_\mu \kappa^2_-\partial_{\kappa^2_- x^2}\right)
F^q_J(z_+,z_-,\kappa_+ x p_-, \kappa^2_- x^2)
{\cal Q}^J(x;p_2,p_1)\Big\}\Big|_{x=\xx}.
\nonumber
\end{eqnarray}
Here, the following functions
$F^q_J(z_+,z_-,\kappa_+ xp_-, \kappa^2_- x^2)$ are introduced:
\begin{eqnarray}
\label{tt1}
F(z_+,z_-, \kappa_+ xp_-,\kappa^2_- x^2)
&=&
\int_0^1 \frac{d\lambda}{\lambda^2}
f\left(\frac{z_+}{\lambda},\frac{z_-}{\lambda},
\kappa^2_- \lambda^2 x^2\right)
\Theta (\lambda - |z_+|)
\Theta (\lambda - |z_-|) e^{i \kappa_+ \lambda x p_-}~,\\
\label{tt11}
\stackrel{o}{F}(z_+,z_-, \kappa_+ xp_-,\kappa^2_- x^2)
&\equiv&
\left(\partial_{i\kappa_+ x p_-}
F\right)(z_+,z_-, \kappa_+ xp_-,\kappa^2_- x^2)\\
&=&
\int_0^1 \frac{d\lambda}{\lambda}
 f\left(\frac{z_+}{\lambda},\frac{z_-}{\lambda} ,
\kappa^2_- \lambda^2 x^2\right)
\Theta (\lambda - |z_+|)
\Theta (\lambda - |z_-|)  e^{i \kappa_+ \lambda x p_-}~,
\nonumber\\
\label{tt111}
F'(z_+,z_-,\kappa_+ xp_-, \kappa^2_- x^2)
&\equiv&
\left(\partial _{\kappa^2_- x^2 }
F\right)(z_+,z_-,\kappa_+ xp_-, \kappa^2_- x^2)\\
&=&\int_0^1 d\lambda\,
\partial _{\kappa^2_- \lambda^2 x^2 }
f\left(\frac{z_+}{\lambda},\frac{z_-}{\lambda} ,
\kappa^2_- \lambda^2 x^2\right)
\Theta (\lambda - |z_+|)
\Theta (\lambda - |z_-|)  e^{i \kappa_+ \lambda x p_-}~.
\nonumber
\end{eqnarray}
As will be shown in Section~\ref{sec-5} one consequence of
the representation (\ref{tt1}) are the integral relations in
polarized deeply inelastic scattering in the forward case~\cite{WW} and
\cite{BLK},
           as well as the Callan--Gross relation~\cite{CG}.
For $p_- = 0$ one obtains:
\begin{eqnarray}
\label{x1}
F(z)= \int_0^1 \frac{d\lambda}{\lambda}\,
f\bigg(\frac{z}{\lambda}\bigg)
\Theta(\lambda - z) =  \int^{1}_z \,\frac{dz}{z}f(z).
\nonumber
\end{eqnarray}
%
Thereby the derivation
$f'(z_+,z_-, \kappa^2_- x^2) \equiv
\left(\partial_{\kappa^2_- x^2 }f\right)(z_+,z_-, \kappa^2_- x^2)$
was introduced in Eq.~(\ref{tt111}).
Furthermore, a third Fourier transformation can
be performed either with respect to $\kappa^2_- x^2$~\cite{RAD} or,
as will be used here, with
respect to $\kappa_- x$, which allows to study the $(1/Q^2)$--dependence
of the distribution amplitudes
\begin{eqnarray}
\label{xrep}
\tilde f(\kappa_-x p_+, \kappa_- x p_-, \kappa^2_- x^2)
 = \int Dz \int d^4 q \,
e^{-i\kappa_- (xp_i)z_i} \, e^{-i \kappa_- qx}
f(z_+,z_-, q^2)~.
\end{eqnarray}

In the limiting case of {\em forward scattering} we have:
\begin{eqnarray}
\left.
{\cal Q}_D(x;p,p) \equiv \bar u(p) x^{\mu} \gamma_{\mu}  u(p) = 2 x p
\right.
\quad {\rm and} \quad
\left.
{\cal Q}_P(x;p,p) \equiv \bar u(p_2)
\hbox{\large$\frac{1}{M}$}
x^{\mu} \sigma_{\mu \nu}
p_-^{\nu}  u(p_1) \right|_{p_2 \rightarrow p_1} = 0,
\nonumber
\end{eqnarray}
and one obtains $(g \equiv f_D)$
\begin{eqnarray}
\label{x2}
\langle p\,|O^{q,\;\rm twist2}(-\kappa_- \xx,\kappa_- \xx)|\,p\rangle
&=& i
2(\xx p) \; \tilde g^q(2\kappa_- \xx p)\\
&= & i
2(\xx p) \int^1_{-1} dz_+ \,e^{-2i\kappa_- (\xx p) z_+}
\int^{+1-|z_+|}_{-1+|z_+|} dz_- \, g^q (z_+, z_-)~.
\nonumber
\end{eqnarray}
Analogous decompositions and representations are valid for the matrix
elements of the pseudo--scalar and pseudo--vector twist--2 quark
operators. The corresponding partition functions are denoted by
$f^q_{J,5} = (g^q_5, \, h^q_5)$
and  $F^q_{J,5} = (G^q_5, \, H^q_5)$, respectively, and the Dirac
and Pauli structures are to be replaced by
${\cal Q}^D_5, {\cal Q}^P_5$ instead of Eqs.~(\ref{dirac}, \ref{pauli}),
substituting
$\gamma_{\mu} \rightarrow \gamma_5 \gamma_{\mu}$ and 
$\sigma_{\mu\nu} \rightarrow \gamma_5 \sigma_{\mu\nu}$,
respectively.

Let us now extend our consideration to the twist--2 scalar and vector gluon
operators, Eqs.~(\ref{tw2GS}) and (\ref{tw2GV}). Completely analogous to
Eq.~(\ref{kdec1}) the matrix elements of Eq.~(\ref{tw2GS}) are written as
\begin{eqnarray}
\label{kdecg1}
\hspace{-.5cm}
\langle p_2|O^{G,\;\rm twist2}(-\kappa_- x,\kappa_- x)|p_1\rangle
=  i
(ix p_+)\;\tilde f^{G(1)}_J(\kappa_- x p_+,\kappa_- x p_-,
     \kappa^2_- x^2, p_1p_2, \mu^2)
{\cal Q}^J(x; p_2,p_1) ~.
\end{eqnarray}
The additional factor $i x p_+$ has been introduced in order to
compensate for the different $\kappa$--scale dimensions, $d_q = 1$ and
$d_G = 2$, of the quark and gluon operators, respectively, cf. also
Section 6.1.

The matrix elements of the twist--2 gluonic vector operators,
Eq.~(\ref{tw2GV}), are obtained in the same manner as for the corresponding
quark operators:
\begin{eqnarray}
\label{MAING}
\lefteqn{\hspace{-1cm}
\langle p_2|O^{G,\;\rm twist 2}_\mu(\kappa_1\xx,\kappa_2\xx)|p_1\rangle
=   i
\int Dz
e^{-i\kappa_- (x p_i)z_i}
}
\\
\!\!&\times&\!\! \Big\{ (i x p_+)\Big[\!
F^{G(1)}_J(z_+,z_-,\kappa_+ x p_-, \kappa^2_- x^2)
\left({\cal Q}^J_{\mu}(p_2,p_1)
-i\kappa_- p_\mu(z){\cal Q}^J(x; p_2,p_1)\right)
\nonumber\\ & &\!\!
+
\left(i \kappa_+ p_-^{\mu} \partial_{i\kappa_+ x p_-}
+ 2 x_\mu \kappa^2_-\partial_{\kappa^2_- x^2}\right)
F^{G(1)}_J(z_+,z_-,\kappa_+ x p_-, \kappa^2_- x^2)
{\cal Q}^J(x;p_2,p_1)\Big]
\nonumber\\& &\!\!
+\, ip_+^\mu
F^{G(1)}_J(z_+,z_-,\kappa_+ x p_-, \kappa^2_- x^2)
{\cal Q}^J(x;p_2,p_1)
\Big\}\Big|_{x=\xx}.
\nonumber
\end{eqnarray}

We remark that in our preceding paper, Ref.~\cite{BGR1}, for
simplicity we introduced {\em formal} definitions for the matrix elements
of the scalar quark and gluon operators which are sufficient for the
derivation of evolution equations. There  we introduced the following two
representations referring to different kinematic variables,
\begin{eqnarray}
\label{partap}
{\langle p_2|O^{q,\,\rm twist 2}(\ka, \kb)
|p_1\rangle}
&=&
 e^{i\kappa_+ \xx p_-}{(i\xx p_+)}
\int Dz
e^{-i\kappa_-(\xx p_i)z_i} \hat f_{(1)}^q(z_+,z_-),
\\
\label{partapg}
{\langle p_2|O^{G,\,\rm twist 2}(\ka, \kb)
|p_1\rangle}
&=&
 e^{i\kappa_+ \xx p_-}{(i\xx p_+)^2}
\int Dz
e^{-i\kappa_-(\xx p_i)z_i} \hat f_{(1)}^G(z_+,z_-),
\end{eqnarray}
which are now to be replaced by the explicitly determined expressions 
Eqs.~(\ref{kdec1}) and (\ref{kdecg1}) projected onto the light--cone. 
In the limit of forward scattering these functions describe the hadronic 
parton densities.

However, contrary to the case of the quark operators, for the matrix
element of the gluon operator there exists no natural kinematic
decomposition. A second parametrization used earlier is
\begin{eqnarray}
\label{qx1}
\kappa_- \;
\langle p_2|O^{q,\,\rm twist 2}(\ka, \kb)|p_1\rangle
&=&
e^{i\kappa_+ \xx p_-}
\int Dz
e^{-i\kappa_- (\xx p_i)z_i} \hat f_{(2)}^{q}(z_+,z_-),
\\
\label{x11}
\kappa_-^2 \;
\langle p_2|O^{G,\,\rm twist 2}(\ka, \kb)|p_1\rangle
&=&
e^{i\kappa_+ \xx p_-}
\int Dz
e^{-i\kappa_- (\xx p_i)z_i} \hat f_{(2)}^{G}(z_+,z_-)
\end{eqnarray}
corresponding to
\begin{eqnarray}
\label{kdecg2}
\hspace{-.3cm}
\kappa_-\;
\langle p_2|O^{q,\;\rm twist2}(-\kappa_- \xx,\kappa_- \xx)|p_1\rangle
&=&
(\xx p_+)^{-1}
\tilde f^{q(2)}_{J}(\kappa_- \xx p_+,\kappa_- \xx p_-,
      p_1p_2, \mu^2)
{\cal Q}^J(\xx; p_2,p_1) ,
\nonumber\\
\hspace{-.3cm}
\kappa_-^2\;
\langle p_2|O^{G,\;\rm twist2}(-\kappa_- \xx,\kappa_- \xx)|p_1\rangle
&=&
(\xx p_+)^{-1}
\tilde f^{G(2)}_{J}(\kappa_- \xx p_+,\kappa_- \xx p_-,
      p_1p_2, \mu^2)
{\cal Q}^J(\xx; p_2,p_1).\nonumber\\
\end{eqnarray}
Because the vector operators are not used in the following we do not
write down the respective representations for their matrix elements.

Whereas the first choice in the case of forward scattering leads to the
correct matrix elements being proportional to $(2 \xx p)^2$, the second
choice leads to a much simpler form of the anomalous dimensions. Note,
that in both cases, up to a trivial reduction of powers of $\kappa_-$ on
both sides, everywhere the combination $\kappa_- \xx$ appears as a natural
variable.

Let us point to the fact that for both the twist--2 quark and gluon
operators the (pseudo)vector distribution
amplitude $F_J,\, F_{J,5}$
are directly related through Eqs.~(\ref{tt1}) and (\ref{tt11})
to the (pseudo)scalar distribution
amplitude $f_J,\,f_{J,5}$.
Therefore we need not to investigate the renormalization properties of
the (axial)vector operators independently: It is sufficient to work out
the renormalization of the (pseudo)scalar operators and to derive the
renormalization group equations for them only.
\section{Compton scattering amplitude in leading order}
\renewcommand{\theequation}{\thesection.\arabic{equation}}
\setcounter{equation}{0}
\label{sec-5}

\vspace{1mm}
\noindent
With the prerequisites provided in the preceding Sections we now derive
the asymptotic representation of the Compton scattering amplitude. 
Again, we consider the contributions of twist--2 operators only and
we restrict the analysis to leading order. Starting from the matrix 
elements of the non--local operators which depend on the two independent 
variables $z_\pm$ we arrive, after performing the Fourier transformation 
of Eq.~(\ref{COMP}), at double--variable distributions.\footnote{
The general Lorentz--structure of the Compton amplitude was investigated
in Refs.~\cite{DRDR,COM1},
                          cf. also \cite{COM2}. Real--photon processes
were considered in Ref.~\cite{BATU} before. In general the Compton
amplitude consists of 18 basis elements. For $q_1^2 = q_2^2$ four
spin--independent and eight spin--dependent distribution amplitudes
contribute~\cite{DRDR}. Their number reduces to two and four, 
respectively, for $q_1^2 = q_2^2 = 0$.} In the literature
also single--variable non--forward evolution equations are studied 
by imposing an additional kinematic constraint. The latter description
is suited for special kinematic situations.
\subsection{Double-variable distributions}
\subsubsection{Non--forward scattering}
To obtain the double--variable distributions we use the decomposition
Eq.~(\ref{MAIN}) of the Fourier transform of the Compton amplitude
\begin{equation}
\label{COMP1}
T_{\mu\nu}(p_+,p_-,q) = i \int d^4x e^{iqx}
\langle
p_2, S_2\,|RT\left(J_{\mu}(x/2)J_{\nu}(-x/2)S\right)|\,p_1, S_1\rangle.
\nonumber
\end{equation}
Afterwards we have to carry out the limit $x \rightarrow \xx$. In the 
r.h.s.~of Eq.~(\ref{COMP1}) the representation of 
Eqs.~(\ref{tpro}\,--\,\ref{qop5}) is used. 
Furthermore the matrix elements
are expressed in terms of hadron states, 
Eqs.~(\ref{kdec1},\,\ref{kdecv}).
The arguments $\kappa_{\pm}$ of the coefficient functions 
$C_{a(5)}(x^2,\kappa_+,\kappa_-,\mu^2)$  in
Eqs.~(\ref{coe1},\,\ref{coe2})
are fixed at $\kappa_+ = 0$ and $\kappa_- = \pm 1/2$. For the 
representation of the Compton amplitude we could
use the matrix elements at 
these points. However, for the investigation of the evolution of the
matrix elements their representation at general values of $\kappa_+$ and
$\kappa_-$ is needed.

The Fourier transforms of the individual terms are obtained
by, cf.~\cite{GELFAND},
\begin{eqnarray}
\int\frac{d^4x}{2\pi^2}e^{iqx}
\frac{x^\nu}{(x^2 -i\varepsilon)^2}
 &=&
\frac{q^\nu}{q^2 +i \varepsilon}~,
\nonumber \\
\int\frac{d^4x}{2\pi^2}e^{iqx}
\frac{x^\nu x^\mu}{(x^2-i\varepsilon)^2}
 &=&
- i\;\frac{g^{\nu \mu}}{q^2 +i \varepsilon}
+2i\;\frac{q^\nu q^\mu}{(q^2 +i \varepsilon)^2}~,
\nonumber \\
\int\frac{d^4x}{2\pi^2}e^{iqx}
\frac{x^\nu x^\mu x^\tau}{(x^2 -i\varepsilon)^2}
 &=&
 2\;\frac{g^{\nu \mu} q^\tau  + g^{\mu \tau} q^\nu
        + g^{\tau\nu} q^\mu }{(q^2 +i \varepsilon)^2}
-8\;\frac{q^\nu q^\mu q^\tau}{(q^2 +i \varepsilon)^3}~.
\nonumber
\end{eqnarray}
In the generalized Bjorken region one obtains, noting that 
$\xi = Q^2/qp_+, \eta = {qp_-}/{qp_+}$,
\begin{eqnarray}
\label{COMPAS}
T_{\mu \nu}^{\rm twist 2}(p_+,p_-,q)
&=&
\int Dz
\bigg\{ \,
\bigg(\frac{1}{\xi + z_+ +\eta z_- -i\varepsilon} -
      \frac{1}{\xi - z_+ -\eta z_- -i\varepsilon}\bigg)
 F^{(1)}_{\mu \nu}
          \\
& & \quad +\;
\bigg(\frac{1}{(\xi + z_+ +\eta z_- -i\varepsilon)^2} +
      \frac{1}{(\xi - z_+ -\eta z_- -i\varepsilon)^2}\bigg)
 F^{(2)}_{\mu \nu}
\nonumber \\
& & \quad +\;
\bigg(\frac{1}{(\xi + z_+ +\eta z_- -i \varepsilon)^3} -
      \frac{1}{(\xi - z_+ -\eta z_- -i\varepsilon)^3}\bigg)
F^{(3)}_{\mu \nu}
\nonumber \\
& & \quad +\;
\bigg(\frac{1}{\xi + z_+ +\eta z_- -i \varepsilon} +
      \frac{1}{\xi - z_+ -\eta z_- -i\varepsilon}\bigg)
F^{(1)}_{5,\mu \nu}
\nonumber \\
& & \quad +\;
\bigg(\frac{1}{(\xi + z_+ +\eta z_- -i\varepsilon)^2} -
      \frac{1}{(\xi - z_+ -\eta z_- -i\varepsilon)^2}\bigg)
F^{(2)}_{5,\mu \nu}
\bigg\}.
\nonumber
\end{eqnarray}
In this representation of $T_{\mu\nu}$ we refer to the essential terms
only.  One may generalize Eq.~(\ref{COMPAS}) accounting for current
conservation explicitly. For the case of forward scattering this 
representation was given in Ref.~\cite{BT}.

The functions  $F^{(k)}_{(5)\mu\nu}(p_+, p_-, q; z_+, z_-)$ being
(anti)symmetric w.r.t.~the Lorentz indices are given by
\begin{eqnarray}
F_{\mu \nu}^{(1)}
&=&                                 
\frac{1}{qp_+}
\bigg[q^\alpha g_{\mu\nu}
     - (g^\alpha_{\;\nu} q_\mu + g^\alpha_{\;\mu} q_\nu)\bigg]
\DIRJ  F_J(z_+, z_-),
\\
F_{\mu \nu}^{(2)}
&=&
-\bigg(\frac{1}{qp_+}\bigg)^2
 \bigg[g_{\mu\nu} qp(z)
     - (q_\mu p_\nu(z) + q_\nu p_\mu(z))\bigg]
 q^{\alpha} \DIRJ  F_J(z_+, z_-),
\nonumber \\
F_{\mu \nu}^{(3)}
&=&
8\,q_\mu q_\nu  \bigg(\frac{1}{qp_+}\bigg)^3 
q^{\alpha} \DIRJ  F'_J(z_+, z_-),
\\
F_{5,\mu \nu}^{(1)}
&=&
 i {\varepsilon_{\mu \nu}}^{\alpha \beta}\,
 q_{\alpha} \frac{1}{qp_+}
{\cal Q}_{\beta}^{J,5}
  F_{J,5}(z_+, z_-),
\\
F_{5,\mu \nu}^{(2)}
&=&
- i {\varepsilon_{\mu \nu}}^{\gamma \beta}\, q_{\gamma} p_{\beta}(z)
\bigg(\frac{1}{qp_+}\bigg)^2  q^{\alpha} \DIRJF  F_{J,5}(z_+,z_-),
\end{eqnarray}
where $p(z)$ is defined in Eq.~(\ref{eqpz}). Let us remind that the
functions $F_{J,(5)}$
are taken in the limit $x^2 \rightarrow 0$  and that
$F'_J(z_+,z_-)$ results from the Fourier transformation of
$\partial_{\kappa^2_- x^2} F$. 
The distribution amplitudes $F_{J,(5)}$ and
$F_J'$ are complex quantities in general. 
The representation of the LCE, 
Eq.~(\ref{LCB}), implies due to its construction that the current 
conservation for the incoming and outgoing photon
\begin{eqnarray}
\label{eqCURC}
q_{2\mu} T^{\mu\nu} = 0,~~~~~T^{\mu\nu} q_{1\nu} = 0
\end{eqnarray}
is not obeyed in explicit form yet. However, one may impose these
conditions.
\subsubsection{The forward scattering limit}
In the limit of forward scattering, $p_- = p_2 - p_1 \rightarrow 0,
\eta = 0$, the Compton amplitude does not depend on the distribution 
amplitudes $F_{P,(5)}(z_+,z_-)$ anymore.  Using the
normalizations
\begin{eqnarray}
\label{x33}
 {\bar u}(p) u(p)  = 2M , \qquad
 {\bar u}(p)\gamma_\mu u(p)  = 2p_\mu
\nonumber
\end{eqnarray}
for the free hadronic Dirac spinors we obtain for the {\em symmetric}
part of the Compton amplitude 
\begin{eqnarray}
\label{CV1}
T_{\mu \nu}^{\rm sym}
&=&
\left(g^{\mu\nu} - \frac{p^\mu q^\nu + p^\nu q^\mu}{pq}\right)
 \int Dz\\
& & \times
\bigg\{\bigg(\frac{1}{\xi + z_+ -i\varepsilon}
-\frac{1}{\xi - z_+  -i\varepsilon}\bigg)
- z_+
\bigg(\frac{1}{(\xi + z_+ -i\varepsilon)^2}
+\frac{1}{(\xi - z_+  -i\varepsilon)^2}\bigg)
 \bigg\} G^q(z_+,z_-) \nonumber\\
&+& 2 q_{\mu} q_{\nu} \int Dz \left(
\frac{1}{(\xi + z_+ - i \varepsilon)^3}
-
\frac{1}{(\xi - z_+ - i \varepsilon)^3}\right) {G'}^q(z_+,z_-). \nonumber
\end{eqnarray}
The $z_-$--integral can be performed
\begin{eqnarray}
\label{CV2}
\int^{+1-|z_+|}_{-1+|z_+|} dz_- G^q(z_+,z_-)
&=&
\int^{+1-|z_+|}_{-1+|z_+|} dz_- \int_0^1\frac {d\lambda} {\lambda^2} \;
 g^q\bigg(\frac{z_+}  {\lambda}, \frac{z_-}{\lambda}\bigg)
 \Theta(\lambda -|z_+|) \Theta(\lambda -|z_-|)
 \nonumber\\
&=&
\int_0^1\frac {d\lambda} {\lambda}
\int d\left(\frac{z_-}{\lambda}\right) g^q
  \bigg(\frac{z_+}{\lambda}, \frac{z_-}{\lambda}\bigg)
 \Theta(\lambda -|z_+|) 
 \nonumber\\
&\equiv&  \int_{ z_+ }^{{\rm sign}(z_+)}       \frac{dz'}{z'}
\widehat{g}^q(z').
\end{eqnarray}
After partial integration of Eq.~(\ref{CV1}),
\begin{eqnarray}
\label{HI1}
\int_{-1}^{+1} d z_+ \frac{z_+}{(\xi \pm z_+ - i \varepsilon)^2}
\int_{z_+}^{{\rm sign}(z_+)} \frac{dz}{z} \widehat{g}^q(z)
= \pm \int_{-1}^{+1} d z_+ \frac{1}{\xi \pm z_+ - i \varepsilon}
\left [ \int_{z_+}^{{\rm sign}(z_+)} \frac{dz}{z} \widehat{g}^q(z)
- \widehat{g}^q(z_+)\right],
\nonumber
\end{eqnarray}
we finally obtain
\begin{eqnarray}
\label{CV3}
T_{\mu \nu}^{\rm sym} &=&
\left(g^{\mu\nu} - \frac{p^\mu q^\nu + p^\nu q^\mu}{pq}\right)
 \int_{-1}^1 dz_+
\left(\frac{1}{\xi + z_+ -i\varepsilon}
     -\frac{1}{\xi - z_+ -i\varepsilon}\right)
      \widehat{g}^q(z_+) \nonumber\\
&+& 2 q_{\mu} q_{\nu} \int Dz \left(
\frac{1}{(\xi + z_+ - i \varepsilon)^3}
-
\frac{1}{(\xi - z_+ - i \varepsilon)^3}\right) {G'}^q(z_+,z_-).
\end{eqnarray}
In the same way we can treat the {\em antisymmetric} part of the Compton 
amplitude. Using the normalization condition
\begin{eqnarray}
 {\bar u}(p)\gamma_5\gamma_\beta u(p)  = -2 S_\beta,~~~~~S^2 = - M^2~,
\nonumber
\end{eqnarray}
we obtain analogously
\begin{eqnarray}
\label{x4}
T_{\mu \nu}^{\rm antisym}
&=& ~~i
   {\varepsilon_{\mu\nu}}^{\gamma\beta} 
   \frac{q_{\gamma}p_{\beta}}{(pq)^2}
qS    \int_{-1}^{+1} dz_+ \left[\frac{1}{\xi + z_+ - i\varepsilon} 
+ \frac{1}{\xi - z_+ -i \varepsilon}\right]
                \left[ \int_{z_+}^{{\rm sign}(z_+)} \frac{dz}{z}
\widehat{g}^q_5(z) - \widehat{g}^q_5(z_+)\right]
\nonumber\\ & &
-i {\varepsilon_{\mu\nu}}^{\gamma\beta} \frac{q_{\gamma}S_{\beta}}{(pq)  }
   \int_{-1}^{+1} dz_+
\left[\frac{1}{\xi + z_+ - i\varepsilon} + \frac{1}{\xi - z_+
-i \varepsilon}\right] \int_{z_+}^{{\rm sign}(z_+)} \frac{dz}{z}
\widehat{g}^q_5(z)
\end{eqnarray}
\subsubsection{Relations between structure functions}
The hadronic tensor is related to the Compton amplitude for forward
scattering by
\begin{equation}
\label{eqHAD}
W_{\mu \nu} = \frac{1}{2\pi} {\rm Im}\, T_{\mu \nu}~.
\end{equation}
In the case of purely electromagnetic interactions the Lorentz--structure
of the hadronic tensor is
\begin{eqnarray}
\label{hadrt}
W_{\mu \nu} &=& - g_{\mu\nu} F_1(x,Q^2) 
+ \frac{{p}_{\mu}{p}_{\nu}}{pq} F_2(x,Q^2)
+ \frac{q_{\mu} q_{\nu}}{pq} F_4(x,Q^2)
+ \frac{\left(p_{\mu}q_{\nu} + q_{\nu}p_{\mu}\right)}{pq} F_5(x,Q^2)
\nonumber \\ & & 
+ i \varepsilon_{\mu \nu \lambda \sigma} \frac{q^{\lambda} S^{\sigma}}
{pq} g_1(x,Q^2)
+ i \varepsilon_{\mu \nu \lambda \sigma} \frac{q^{\lambda} \left(pq
S^{\sigma} - Sq p^{\sigma}\right)}{(pq)^2} g_2(x,Q^2)~,
\end{eqnarray}
where $x \equiv x_{\rm Bj} = \xi$.
Whereas the polarized structure functions in Eq.~(\ref{hadrt}) form
already the minimal set, current conservation
\begin{equation}
q^{\mu} W_{\mu \nu} = W_{\mu \nu} q^{\nu} = 0
\end{equation}
relates the structure functions $F_4$ and $F_5$ to $F_1$ and $F_2$
by
\begin{eqnarray}
\label{eqSF1}
F_4(x,Q^2) &=& \frac{1}{4 x^2} \left[2x F_1(x,Q^2) +  F_2(x,Q^2)\right],
\\
\label{eqSF2}
F_5(x,Q^2) &=& \frac{1}{2x} F_2(x,Q^2)~.
\end{eqnarray}
The forward Compton amplitude Eqs.~(\ref{CV1}) and (\ref{x4}) are 
related to the hadronic tensor Eq.~(\ref{hadrt}) using~\cite{GELFAND}
\begin{eqnarray}
\label{sochotz}
\lim_{\varepsilon \rightarrow 0^+}
\frac{1}{(x \pm i\varepsilon)^n} = {\sf P} \frac{1}{x^n} \pm (-1)^n
\frac{i\pi}{(n-1)!}  \delta^{(n-1)}(x)~.
\end{eqnarray}
We now assume that $\widehat{g}^q_{(5)}(z_+)$ are real functions. The
tensor structure of Eq.~(\ref{CV3}) implies
\begin{eqnarray}
F_1(x,Q^2) = F_5(x,Q^2)
\end{eqnarray}
at leading order, from which, cf.~Eq.~(\ref{eqSF2}), the Callan--Gross
relation~\cite{CG}
\begin{eqnarray}
\label{eqCG}
F_2(x,Q^2) = 2x F_1(x,Q^2)
\end{eqnarray}
follows. Furthermore, one finds
\begin{eqnarray}
\label{eqF4}
F_4(x,Q^2) = \frac{1}{x} F_1(x,Q^2),
\end{eqnarray}
which relates the distributions ${G'}^q(z_+,z_-)$ and $G^q(z_+,z_-)$
for forward scattering at leading order.
The structure function $F_1$ obtains the representation
\begin{eqnarray}
\label{eqF1A}
F_1(x,Q^2) = \frac{1}{2} \left[\widehat{g}^q(x) - \widehat{g}^q(-x)
\right]~.
\end{eqnarray}
We suppressed, as in the preceding Section, the scale dependence of the
functions $\widehat{g}^q_{(5)}$. In lowest order of perturbation theory 
the $Q^2$--dependence of the structure functions results from the 
identification $\mu^2 = Q^2$. One may identify this representation with 
results being obtained in the quark--parton model at leading order by 
defining
\begin{eqnarray}
\label{eqPAR1}
\widehat{g}^q(x) &=&
\sum_{q=1}^{N_f} e_q^2 q(x) \\
\label{eqPAR2}
- \widehat{g}^q(-x) &=&
\sum_{q=1}^{N_f} e_q^2 \overline{q}(x)~.
\end{eqnarray}
Here $q(x)$ and $\overline{q}(x)$ denote the unpolarized quark and
antiquark densities, respectively, and $e_q$ is the electric charge
of the quark.

We now turn to the polarized case.  The Lorentz structure of the two
tensors in                           Eq.~(\ref{x4}) refers to the
structure functions $g_1+g_2$ and $g_2$, respectively. After performing
the $z_+$--integration one obtains for the twist--2 terms
the Wandzura--Wilczek relation~\cite{WW}
\begin{eqnarray}
\label{WW}
g_2(x,Q^2) = - g_1(x,Q^2) + \int_x^1 \frac{dz}{z} g_1(z,Q^2),
\end{eqnarray}
where
\begin{eqnarray}
\label{eqg1}
g_1(x,Q^2) = \frac{1}{2} \sum_{q=1}^{N_f} e_q^2 \left[\Delta q(x)
                                 +  \Delta \overline{q}(x)\right]
\end{eqnarray}
and we identified
\begin{eqnarray}
\widehat{g}^q_5(x) &=&
\sum_{q=1}^{N_f} e_q^2 \Delta q(x) \\
\widehat{g}^q_5(-x) &=&
\sum_{q=1}^{N_f} e_q^2 \Delta \overline{q}(x)~.
\end{eqnarray}
In the case of general electro--weak couplings also a second integral
relation,
Ref.~\cite{BLK}, is obtained. For both these relations the
parameter--integrals relating the vector and scalar operators which are
discussed
in Section~\ref{sec-4} form the background in the present approach.
The new derivation of these relations outlined above could circumvent
their usual derivation by means of the moments in the local operator
product expansion and leads to the integral form directly. We note
that recently also three integral relations for the twist--3 contributions
to the polarized structure functions $\left. g_j\right|_{j=1}^5$
for the case of forward scattering were found in lowest order in the
coupling constant,
       cf. Ref.~\cite{BT}. Unlike the case for twist--2~\cite{BLK}
all the relations are integral relations requiring an all-order
resummation of the $(M^2/Q^2)$--terms.
\subsection{Single--variable distributions}
Up to now we have considered representations which contain two 
distribution parameters $z_+,~ z_-$. In some connections another 
representation may be helpful. As an example let us introduce in 
representation Eq.~(\ref{COMPAS}) a new variable $t$ by 
$z_+= t- \eta z_-$. Then the integrations factorize and we can write
\begin{eqnarray}
\label{COMPAT}
\lefteqn{\hspace{-1cm}
T_{\mu \nu}^{\rm twist 2}(p_+,p_-,Q)  ~=~
\int_{-\infty}^{+\infty} dt \bigg\{
\bigg(\frac{1}{\xi + t -i \varepsilon} -
      \frac{1}{\xi - t -i \varepsilon}\bigg)
\hat F^{(1)}_{\mu \nu}(p_+,p_-,q;t,\eta)
} \\
&+& 
\bigg(\frac{1}{(\xi + t -i\varepsilon)^2} +
      \frac{1}{(\xi - t -i\varepsilon)^2}\bigg)
\hat F^{(2)}_{\mu \nu}
+ 
\bigg(\frac{1}{(\xi + t -i \varepsilon)^3} -
      \frac{1}{(\xi - t -i \varepsilon)^3}\bigg)
\hat F^{(3)}_{\mu \nu}
\nonumber \\
&+& 
\bigg(\frac{1}{\xi + t -i \varepsilon} +
      \frac{1}{\xi - t -i \varepsilon}\bigg)
\hat F^{(1)}_{5,\mu \nu}
+
\bigg(\frac{1}{(\xi + t -i\varepsilon)^2} -
      \frac{1}{(\xi - t -i\varepsilon)^2}\bigg)
\hat F^{(2)}_{5,\mu \nu} \bigg\}.
\nonumber
\end{eqnarray}
Also within the functions
$F^{(k)}_{(5),\mu \nu}\left(p_+,p_-,q,F_J(z_+,z_-; p_1p_2, \mu^2);
\mu^2\right)$ the variables $z_+,z_-$ have to
be transformed thereby changing these functions to
$F_{J (5)}(z_+= t_- -\eta z_-, z_-)$. Then the remaining
$ z_-$--integration can be performed to give
\begin{eqnarray}
\label{x3}
\hat F_{J (5)}(t,\eta) &=&
\int_{-1}^{+1} d z_-
\Theta(1 + t - \eta z_- + z_-)
\Theta(1 + t - \eta z_- - z_-) \nonumber\\ & & \times
\Theta(1 - t + \eta z_- + z_-)
\Theta(1 - t + \eta z_- - z_-)
F_{J(5)}(z_+ =t-\eta z_- ,z_-).
\end{eqnarray}
These new partition functions $\hat F_J$ and $\hat F_{J,5}$ 
depend on the variables $t$ and $\eta$ and therefore we have
$\hat{F}^{(k)}_{(5),\mu \nu}(p_+,p_-, q; t, \eta; \mu^2)
=  F^{(k)}_{(5)\mu \nu}\left(p_+,p_-,q,\hat F_J(t,\eta;\mu^2);
\mu^2\right)$. Obviously, in the limit of forward
scattering we obtain $\hat{F}^{(k)}_{(5),\mu \nu}(2p,0, q; t, 0; \mu^2)$.
Note that the partition functions $F,F'$ and $F_5$ are directly related 
to the partition functions $f,f'$ and $f_5$ which are defined by the
matrix elements of scalar and pseudo--scalar twist--2 quark operators,
Eq.~(\ref{kdec1}). This demonstrates again that for the description of 
the virtual Compton scattering, at least in leading order, the properties 
of the scalar operators are sufficient.

The renormalization of the twist--2 operators will be considered in the
following Section. The evolution equations of their expectation values
are dealt with in Section~\ref{sec-7}.
\section{Operators of twist--2 and their anomalous dimensions}
\renewcommand{\theequation}{\thesection.\arabic{equation}}
\setcounter{equation}{0}
\label{sec-6}

\vspace{1mm}
\noindent
In this section we discuss the renormalization properties of the twist--2
light--ray operators. As has been shown above we can restrict these
studies to the case of scalar operators for which any free tensorial index of
the various operators is saturated by multiplying with as many as
necessary vectors of $\xx^\mu$. Moreover, there holds the much wider
statement that the operators $O_{\sigma}^{a, \rm twist 2}$ and
$O^{a, \rm twist 2}$ are completely equivalent concerning their
renormalization properties. This can be seen in the following way. We
consider the renormalization group equation of the vector operator,
\begin{eqnarray}
\label{RGvec}
\hspace{-.5cm}
\mu^2\frac{d}{d\mu^2}
O_{\sigma}^{a,\,\rm twist2}(\kappa_1\xx,\kappa_2\xx; \mu^2)
&=&
\int^{\ka}_{\kb} d\kappa_1' d\kappa_2'
\gamma(\kappa_1,\kappa_2;\kappa_1',\kappa_2';\mu^2)
O_{\sigma}^{a,\,\rm twist 2}(\kappa_1'\xx,\kappa_2'\xx; \mu^2)~,
\end{eqnarray}
cf.~Eq.~(\ref{measureK}).
Obviously, the scalar operator
\begin{eqnarray}
O^{a,\,\rm twist 2}(\kappa_1,\kappa_2)
= \xx^\sigma O_{\sigma}^{a,\,\rm twist 2}(\kappa_1,\kappa_2)
\end{eqnarray}
fulfills exactly the same renormalization group equation with the
same anomalous dimension kernel
$\gamma(\kappa_1,\kappa_2;\kappa_1',\kappa_2')$, since the multiplication
of both sides by $\xx^\sigma$ commutes both with the differentiation on
the left and with the integration on the right hand side. The same
conclusion can be drawn for the gluon operators.

Because the renormalization properties of the various twist--2 operators
are independent of their special Lorentz structure we restrict ourselves
to the scalar operators. Contrary to the previous considerations we
will account for the flavor content of the operators because in the
flavor singlet case twist--2 quark and gluon operators mix
under renormalization.
To simplify the subsequent considerations, and the necessary computations
of Feynman diagrams, we apply the {\em axial gauge} where
 the phase factors are $U(\ka, \kb) = 1$.

We consider the following {\em scalar} flavor non--singlet (NS) and
singlet light--ray operators\footnote{Possible trace terms vanish and the
general dependence $RT[O]S$ on the renormalization procedure is
understood but has been omitted here.}:
\begin{eqnarray}
\label{ONS}
O^{\rm NS}(\ka,\kb)
&=&
\xx^{\mu}\,
\hbox{\large$\frac{i}{2}$}
 \left [
\overline{\psi}(\ka\xx)\lambda_f
\gamma_{\mu} \psi(\kb \xx)
-
\overline{\psi}(\kb\xx)\lambda_f
\gamma_{\mu} \psi(\ka \xx)
\right ]
\\
\nonumber
\\
\label{O5NS}
O^{\rm NS}_5(\ka,\kb)
&=&
\xx^{\mu} \,
\hbox{\large $\frac{i}{2}$}
\left [
\overline{\psi}(\ka\xx) \lambda_f \gamma_5
\gamma_{\mu} \psi(\kb \xx)
+
\overline{\psi}(\kb\xx) \lambda_f \gamma_5
\gamma_{\mu} \psi(\ka \xx)
\right ]
\\
\nonumber
\\
\label{Oq}
O^q_{\rm     }(\ka,\kb)
&=&
\xx^{\mu}\,
\hbox{\large $\frac{i}{2}$}
\left [
\overline{\psi}(\ka\xx)
\gamma_{\mu} \psi(\kb \xx)
-
\overline{\psi}(\kb\xx)
\gamma_{\mu} \psi(\ka \xx)
\right]
\\
\label{OG}
O^G_{\rm     }(\ka,\kb)
&=&
\xx^{\mu}\,\xx^{\nu}\,
\hbox{\large $\frac{1}{2}$}
\left[
F_{\mu}^{a\,\rho}(\ka\xx) F^a_{\nu\rho}(\kb\xx)
+
F_{\mu}^{a\,\rho}(\kb\xx) F^a_{\nu\rho}(\ka\xx)
\right]
\\
\nonumber
\\
\label{O5q}
O^q_{\rm 5   }(\ka,\kb)
&=&
\xx^{\mu} \,
\hbox{\large $\frac{i}{2}$}
\left[
\overline{\psi}(\ka\xx) \gamma_5
\gamma_{\mu} \psi(\kb \xx)
+
\overline{\psi}(\kb\xx) \gamma_5
\gamma_{\mu} \psi(\ka \xx)
\right]
\\
\label{O5G}
O^G_{\rm 5 }(\ka,\kb)
&=&
\xx^{\mu}\,\xx^{\nu}\,
\hbox{\large $\frac{1}{2}$}
\left[
F^{a\,\rho}_{\mu}(\ka\xx) \tilde F^a_{\nu\rho}(\kb\xx)
-
F^{a\,\rho}_{\mu}(\kb\xx) \tilde F^a_{\nu\rho}(\ka\xx)
\right]~.
\end{eqnarray}
Here $\lambda_f$ denotes the generators of the flavor group $SU(N_f)$,
where $N_f$ is the number of active quark flavors. The scalar operators
Eqs.~(\ref{ONS}, \ref{Oq}) and (\ref{OG}) contribute in the case of
unpolarized Compton scattering, whereas the pseudo--scalar operators
Eqs.~(\ref{O5NS}, \ref{O5q}) and (\ref{O5G}) are relevant for Compton
scattering off polarized hadrons.
\subsection{General properties of non--local anomalous dimensions}
\label{sec6.1}
We consider now some general properties of the anomalous dimensions of
these operators which hold {\em at any order of perturbation theory} and
are relevant in the following. In order to cover the general case we
refer to the scalar singlet operators denoted by $O^A$ with
$A = (q,G)$.~\footnote{The respective relations hold
synonymously for the three possible cases of non--singlet evolution
equations~\cite{BV} which have to be suitably
projected out. They are all equivalent in leading order, and two of them
are even equivalent in next-to-leading order.}

The renormalization group equation under consideration reads:
\begin{eqnarray}
\label{RGsing}
\mu^2 \frac{d}{d \mu^2}
O^A(\ka \xx,\kb \xx; \mu^2)
&=&
\int^{\ka}_{\kb} d\kap d\kbp
\gamma^{AB}(\ka, \kb, \kap,\kbp; \mu^2)
O^B(\kap \xx,\kbp \xx; \mu^2)~.
\end{eqnarray}
$\gamma^{AB}$ denotes the {\em non--local} anomalous dimension matrix
which, through the strong coupling constant
$\alpha_s(\mu^2) = g_s^2(\mu^2)/(4\pi)$, depends on the renormalization
scale $\mu$.\footnote{In the remaining part of this and the following
section the explicit
dependence on $\xx$ and $\mu^2$ will be suppressed in the operators,
anomalous dimensions and evolution kernels.} Analogous relations
hold for the pseudo--scalar case. The non--singlet cases are covered by
the respective projection in flavor\,--space of the quark--quark 
submatrix.

Since the variables $\kappa_i$ simply parametrize points on the
light--ray these anomalous dimensions are invariant under
reparametrizations~\cite{BGR}, i.e. under translations and scale
transformations,\footnote{These properties of the anomalous dimensions
may be traced back to the independence of the renormalization properties
of the operators from shifting or scaling their arguments.}
\begin{eqnarray}
\label{shift}
\gamma^{AB}(\ka, \kb; \kap, \kbp)
&=&
\gamma^{AB}(\ka - \kappa, \kb - \kappa;
            \kap - \kappa, \kbp - \kappa)
\\
\label{scale}
&=&
{\lambda}^{d_{AB}}
\gamma^{AB}(\lambda\ka, \lambda\kb;
            \lambda\kap,\lambda\kbp)~,
\end{eqnarray}
where
\begin{eqnarray}
\label{dAB}
d_{AB} &=& 2 + d_A - d_B,
\end{eqnarray}
with $d_A$ being the difference of the canonical and the $\kappa$--scale
dimension of the operators $O^A$. The latter is given by the number of
factors $\xx$ emerging in the operator $O_A$,  i.e.,
\begin{eqnarray}
\label{eqdqG}
d_q = 1~~~{\rm and}~~~d_G = 2~.
\end{eqnarray}
Therefore, the number of relevant $\kappa$--parameters may be reduced by
two. We consider two (related) choices in order to normalize the anomalous
dimension matrix, shifting by either $\ka$ or $\kappa_+$ and scaling by
either $(\kb - \ka)^{-1}$ or $(\kappa_-)^{-1}$, respectively. One obtains
\begin{eqnarray}
\label{repara}
(\kb - \ka)^{d_{AB}} \gamma^{AB}(\ka, \kb; \kap, \kbp)
&=&
\gamma^{AB}(0, 1; \AAA, 1 - \AB)
~\equiv~
\widehat{K}^{AB}(\AAA, \AB),
\\
\label{reparw}
4 (\kappa_-)^{d_{AB}} \gamma^{AB}(\ka, \kb; \kap, \kbp)
&=&
4 \gamma^{AB}(-1, +1; w_1, w_2)
~\equiv~
{\widetilde K}^{AB}(w_1 - w_2, w_1 + w_2)
\end{eqnarray}
\vspace*{-.5cm}
where
\vspace*{-.4cm}
\begin{eqnarray}
\label{alpha}
\AAA = \frac{\kap - \ka}{\kb - \ka},\hspace{2.2cm}
&~~&
- \AB = \frac{\kbp - \kb}{\kb - \ka},\\
\label{w}
w_1 = \AAA -\AB
= \frac{{\kappa'}_+ - \kappa_+}{\kappa_-},
&~~&
\phantom{-}
w_2 = 1-\AAA -\AB = \frac{{\kappa'}_-}{\kappa_-},
\end{eqnarray}
respectively. The variables $\kappa'$ are related to the parameters
$\alpha$ and $w$ by
\begin{eqnarray}
 {\kap \choose \kbp}
~=~
{\ka(1-\AAA)+\kb\AAA \choose \kb(1-\AB)+\ka\AB}
~=~
\frac{1}{2}
 {\ka (1-w_1+w_2) + \kb (1+w_1-w_2) \choose
  \ka (1-w_1-w_2) + \kb (1+w_1+w_2) }~.
\end{eqnarray}

\vspace*{7cm}
\begin{picture}(120,20)(100,100)
\put(20,-320){\epsfig{file=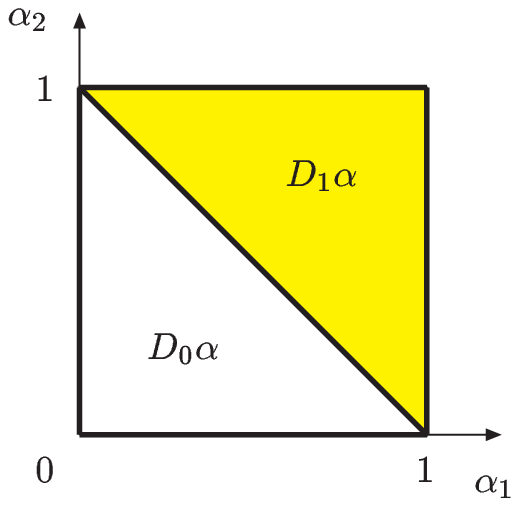,width=16cm}}
\put(250,-280){\epsfig{file=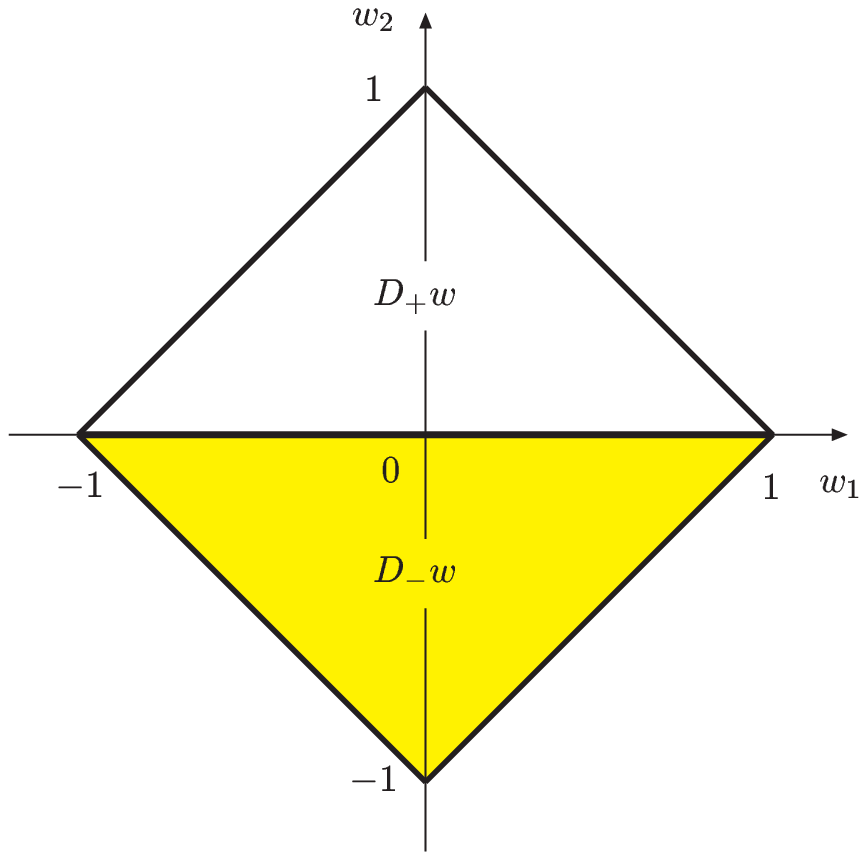,width=16cm}}
\end{picture}

\noindent
\begin{center}
{\sf Figure~1:~Integration ranges of the parameters
$(\alpha_1, \alpha_2)$ and $(w_1, w_2)$,
Eqs.~(\ref{measureA}, \ref{measureW}).}
\end{center}
\normalsize
Therefore, it seems to be quite natural to introduce instead of the
variables $(\kap, \kbp)$ as new variables either $(\alpha_1, \alpha_2)$
or $(w_1, w_2)$. The integration measures are related by
\begin{eqnarray}
D\kappa' ~=~ (\kb - \ka)^2 D\alpha ~=~ 4 (\kappa_-)^2 Dw~,
\end{eqnarray}
where
\begin{eqnarray}
\label{measureK}
D \kappa' 
&\equiv&
d\kap d\kbp \, \Theta(\ka - \kap)\Theta(\kap - \kb)
\Theta(\ka - \kbp)\Theta(\kbp - \kb),
\\
\label{measureA}
D \alpha = D_0\alpha + D_1\alpha
&\equiv&
d\AAA d\AB\; \Big[\Theta(\AAA) \Theta(\AB) \Theta(1-\AAA-\AB)
\\
& & \qquad + \quad \,
\Theta(1-\AAA) \Theta(1-\AB) \Theta(\AAA+\AB-1)\Big],
\nonumber
\\
\label{measureW}
Dw = D_+ w + D_- w
&\equiv&
\hbox{\large $\frac{1}{2}$}\;
dw_1 dw_2 \,\Big[\Theta(1+w_1-w_2) \Theta(1-w_1-w_2) \Theta(w_2)\\
& &  \qquad  + \qquad
\Theta(1+w_1+w_2) \Theta(1-w_1+w_2) \Theta(-w_2)\Big].
\nonumber
\end{eqnarray}
The $w$--representation was derived in Ref.~\cite{LEIP} for all orders
in the coupling constant.
The measures $D\alpha$ and $Dw$ have been divided into two parts related
by $\alpha_i \leftrightarrow 1 - \alpha_i$ and
$w_2 \leftrightarrow - w_2$, respectively. Usually only the first part
(unshadowed in Fig.~1) is
considered in the literature. Because of symmetry requirements for the
anomalous dimensions being discussed below this more general
representation proves, however, to be appropriate.

Let us now consider the {\em symmetry} of the quark and gluon operators,
and likewise of their anomalous dimensions, under changing
$(\ka, \kb) \leftrightarrow (\kb, \ka)$ and
$(\kap, \kbp) \leftrightarrow (\kbp, \kap)$. Looking at
Eqs.~(\ref{Oq}\,--\,\ref{O5G}) it is easily seen that
\begin{eqnarray}
\label{sym_op}
O^A  (\ka, \kb) &=& \phantom{-} (-1)^{d_A} O^A(\kb, \ka),
\\
\label{sym_op5}
O^A_5(\ka, \kb) &=& -(-1)^{d_A} O^A_5(\kb, \ka),
\end{eqnarray}
holds. We first consider the scalar case. From Eq.~(\ref{sym_op}) we
obtain
\begin{eqnarray}
\label{sym_gam}
\gamma^{AB}(\ka, \kb;\, \kap,\kbp)
~=~(-1)^{d_{AB}}
\gamma^{AB}(\kb, \ka;\, \kbp,\kap)
~=~(-1)^{d_B}\gamma^{AB}(\ka, \kb;\, \kbp,\kap)~.
\end{eqnarray}
Using reparametrization invariance, Eqs.~(\ref{shift})
and (\ref{scale}), we obtain
\begin{eqnarray}
\label{sym_KA}
{\widehat K}^{AB}(\AAA, \AB)
&=&
{\widehat K}^{AB}(\AB, \AAA)
~~~~=~(-1)^{d_B}{\widehat K}^{AB}(1-\AAA, 1-\AB),
\\
\label{sym_KW}
{\widetilde K}^{AB}(w_1, w_2)
&=&
{\widetilde K}^{AB}(- w_1, w_2)
~=~(-1)^{d_B}{\widetilde K}^{AB}(w_1, -w_2).
\end{eqnarray}
The renormalization group equation~(\ref{RGsing}) transforms into
\begin{eqnarray}
\label{RGa}
\mu^2 \frac{d}{d \mu^2}
O^A(\ka ,\kb )
&=&
\int D\alpha \,
(\kb -\ka)^{d_B - d_A}
\widehat{K}^{AB}(\AAA, \AB)
O^B(\kap, \kbp),
\\
\label{RGw}
\mu^2 \frac{d}{d \mu^2}
O^A(\ka, \kb)
&=&
\int Dw \,
(\kappa_-)^{d_B-d_A}
{\widetilde K}^{AB}(w_1, w_2)
O^B(\kap,\kbp)
\nonumber
\\
&=&
\int\limits_0^1 dw_2 \int\limits_{-1+w_2}^{1-w_2}dw_1  \,
(\kappa_-)^{d_B-d_A}
{\widetilde K}^{AB}_{\rm sym}(w_1, w_2)
O^B(\kap,\kbp),
\end{eqnarray}
using the kernels ${\widehat K}^{AB}(\AAA, \AB)$ and
${\widetilde K}^{AB}(w_1, w_2)$,
Eqs.~(\ref{repara},\, \ref{reparw}).\footnote{To get rid of the unwanted
$\kappa$--dependent factors in these equations, it seems to be reasonable
to multiply the scalar quark and gluon operators $O^A$ by
$(\kb - \ka)^{d_A}$ or $\kappa_-^{d_A}$ if the $\alpha$-- or
$w$--representation has been chosen. However, this does not really
resolve the problem since an additional factor
$(1 - \alpha_1 - \alpha_2)^{-d_B}$ or $w_2^{-d_B}$ is introduced which
multiplies the kernels $\widehat K$ and $\widetilde K$, respectively.} In
the second line of Eq.~(\ref{RGw}) we have taken into account the above
decomposition (\ref{measureW}) of the integration measure and the second
of the equalities (\ref{sym_KW}) to restrict the integration onto the
usual range by symmetrizing the anomalous dimension kernels in an
appropriate way:
\begin{eqnarray}
\label{symsymW}
{\widetilde K}^{AB}_{\rm sym}(w_1, w_2)
&=&
\hbox{\large $\frac{1}{2}$}
\left[
{\widetilde K}^{AB}_0(w_1, w_2)
+ (-1)^{d_B} {\widetilde K}^{AB}_0(w_1,- w_2)
\right]~.
\end{eqnarray}
Here ${\widetilde K}^{AB}_0(w_1, w_2)$ is defined in the range of the
measure $D_+ w$ only. Even more, with the help of the first of the equalities
(\ref{sym_KW}) the $w_1$--integration can be restricted to the
range $0\leq w_1\leq 1-w_2$. Using these kernels we can write the
renormalization group equation for the centered scalar operators in terms
of the variables $w$ as follows:
\begin{eqnarray}
\label{RGwc}
\hspace{-.3cm}
\mu^2 \frac{d}{d \mu^2}
O^A(-\kappa_-, \kappa_-)
=
2 \!\int\limits^1_0 \!\!dw_2 \!\!\!\int\limits^{1-w_2}_0 \!\!\!\!dw_1\,
\kappa^{d_B - d_A}_-
{\widetilde K}^{AB}_{\rm sym}(w_1, w_2)
O^B\Big(\!(w_1 - w_2)\kappa_-, (w_1 + w_2)\kappa_-\!\Big).
\end{eqnarray}
In terms of the $\alpha$--variables these expressions are more lengthy.

In the pseudo--scalar case the sign in the last expression of
Eqs.~(\ref{sym_gam}\,--\,\ref{sym_KW}), as well as in Eq.~(\ref{symsymW})
has to be changed into $-(-1)^{d_B}$ in accordance with
Eq.~(\ref{sym_op5}), which also implies the behavior of the sign functions
$\sigma_{n\,n'}^{(\pm)}$  defined in Eq.~(\ref{eqSIGM}) below. The
corresponding kernels will be denoted by
$\Delta \gamma, \Delta {\widehat K}$ and $\Delta {\widetilde K}$,
respectively.
\subsection{Relations between non--local and local anomalous dimensions}
Obviously, the general statements proven here for the non--local operators
and their anomalous dimension kernels can be formulated for the
{\em local} operators and their anomalous dimension matrices, too. With
the definition Eq.~(\ref{dtaylor}) of the local operators
$O^\Gamma_{n_1 n_2}$ the renormalization group equations read
\begin{eqnarray}
\label{RGloc}
\mu^2 \frac{d}{d \mu^2} O^A_{n_1 n_2}
=
\sum_{n_1',\,n_2'}
\gamma^{AB}_{n_1, n_2;\, n_1', n_2'} O^B_{n_1' n_2'}~,
\end{eqnarray}
with
\begin{eqnarray}
\label{ADloc}
\gamma^{AB}_{n_1, n_2;\, n_1', n_2'}
=
\frac{\partial^{n_1}}{\partial \ka^{n_1}}
\frac{\partial^{n_2}}{\partial \kb^{n_2}}
\int^{\ka}_{\kb} d\kap \int^{\ka}_{\kb} d\kbp \,
\frac{(\kap)^{n_1'}}{n_1'!}
\frac{(\kbp)^{n_2'}}{n_2'!}
\left.
\gamma^{AB}(\kappa_1,\kappa_2;\kappa_1',\kappa_2')
\right|_{\ka=\kb= 0}~.
\end{eqnarray}
First of all we remark that the symmetry relation (\ref{sym_gam}) is
simply transformed into
\begin{eqnarray}
\gamma^{AB}_{n_1, n_2;\, n_1', n_2'}
~=~
(-1)^{d_{AB}}
\gamma^{AB}_{n_2, n_1;\, n_2', n_1'}
~=~
(-1)^{d_B}
\gamma^{AB}_{n_1, n_2;\, n_2', n_1'}~.
\end{eqnarray}
Furthermore, taking into account the reparametrization invariance,
Eqs.~(\ref{repara}) or (\ref{reparw}), together with the corresponding
representation of $\kappa_i'$, it is easily seen that the values of $n_i$
and $n_i'$ are restricted by the relations
\begin{eqnarray}
N = n_1 + n_2 = n_1' + n_2' + d_B - d_A~,
\end{eqnarray}
since otherwise the anomalous dimension vanishes. From this it is obvious
that only those local operators mix under renormalization which are
specified by $N$. Therefore, we may choose
\begin{eqnarray}
n_1 = N - n,\quad  n_2 = n; \quad
n_1' = N - n',\quad n_2' = n' - d_B + d_A~,
\end{eqnarray}
and the local anomalous dimension can be written as
\begin{eqnarray}
\gamma^{AB}_{n_1, n_2;\, n_1', n_2'}
=
\gamma^{AB}_{N-n,\, n;\, N-n',\, n' - d_B + d_A}
\equiv
{}^N\gamma^{AB}_{n\, n'}~.
\end{eqnarray}
Now we go over to the variables $\kappa_\pm$ in place of $(\ka, \kb)$
which are favored because of the relations (\ref{reparw}) and
(\ref{w}). For simplicity we perform the next steps for the
{\em non--singlet} local anomalous dimensions only. Analogous results are
obtained in the singlet case. Let us define
\begin{eqnarray}
{\widetilde O}_{n_1 n_2}
&=&
\partial_+^{n_1}\,\partial_-^{n_2}
\left. O(\ka \xx, \kb \xx)\right|_{\kappa_\pm = 0},
\\
{\widetilde \gamma}_{n_1, n_2;\, n_1', n_2'}
&=&
\partial_+^{n_1}\,\partial_-^{n_2}
\int Dw \,
\frac{(\kappa_+')^{n_1'}}{n_1'!}
\frac{(\kappa_-')^{n_2'}}{n_2'!}
\left.
{\widetilde K}(w_1, w_2)
\right|_{\kappa_\pm = 0}~,
\end{eqnarray}
with the abbreviation
$\partial_\pm \equiv \partial/{\partial \kappa_\pm}$.
Taking into account the representation of $\kappa_\pm'$ by Eq.~(\ref{w})
and decomposing  $(\kappa_+')^{n'_1} = (\kappa_+ + w_1 \kappa_-)^{n_1'}$
we obtain
\begin{eqnarray}
\label{x5}
{}^N\gamma_{n n'}
\equiv
\gamma_{n n'}
&=&
\frac{\partial_+^{N-n}\,\partial_-^{n}}{(N-n')!\,n'!}
\sum_{l = 0}^{N-n'}
\hbox{\large ${N-n' \choose l}$}
\kappa_+^{N-n' -l}\kappa_-^{l + n'}
\int Dw\,
w_1^l w_2^{n'}
{\widetilde K}(w_1, w_2)
\nonumber
\\
&=&
\label{triang}
\hbox{\large${n \choose n'}$}
\int Dw \, w_1^{n-n'} w_2^{n'}
{\widetilde K}(w_1, w_2)~.
\end{eqnarray}
From this equation some remarkable facts can be read off which hold at
any order of perturbation theory. First of all, the local matrix of
anomalous dimensions is {\em triangular}, i.e. $n' \leq n$ and, secondly,
it is {\em universal}, i.e., independent of $N$. Thirdly, because in
the case of forward scattering, $p_- = 0$,
the variable $\kappa_+$ disappears from the expressions,
cf. Eq.~(\ref{translation}), the relation $N = n = n'$ is implied
with the consequence that the {\em diagonal} elements of
that matrix are the {\em forward} anomalous dimensions. In the
unpolarized case one obtains
\begin{eqnarray}
\label{localAD}
\gamma_n
=
\hbox{\large$\frac{1}{2}$}
\Big(1 - (-1)^{n}\Big)
\int_0^1 dw_2 \, w_2^{n}\; \left\{2
\int^{1- w_2}_{0} dw_1
{\widetilde K}_0(w_1, w_2)\right\}~.
\end{eqnarray}
The anomalous dimensions vanish for even values of $n$.\footnote{
The reader should be reminded that usually in the definition
of the forward anomalous dimensions, Eq.~(\ref{localAD}),
the value of $n$
is shifted to $n-1$. To choose the Mellin transform with the power
$(n-1)$ rather than the power $n$ is motivated by the interpretation
of this quantity in terms of an angular momentum variable. However,
choosing $n$, as we have done here, leads to more compact representations.
}

Since the local anomalous dimensions in the forward case are moments of
the splitting functions, the latter are given by
\begin{eqnarray}
\label{APw}
P(z) &=& 2 \int^{1- z}_{0} dw_1 {\widetilde K}_0(w_1, z)
\end{eqnarray}
performing the inverse Mellin transformation. Here $z$ denotes the
partonic momentum fraction. Analogous relations hold in the pseudo--scalar
case, with the difference that now the even values of $n$ are to be taken.
These considerations show the value of the $w$--representation, which
naturally introduces a {\em partial} diagonalization in the variable
$w_2$ through the Mellin transformation. It becomes also evident
that the $w_1$--dependence is related to the off--diagonality of the
anomalous dimension matrix.

Let us now consider the local anomalous dimensions for the {\em singlet}
case. The local scalar light--ray operators are given by
\begin{eqnarray}
\label{x6}
{}^N{\widetilde O}^q_n
&=&
\partial^{N-n}_+\partial^n_- O^q(\ka \xx, \kb \xx)
\Big|_{\kappa_\pm = 0}
\qquad
{\rm with}
\quad
0 \leq n \leq N,
\\
\label{x7}
{}^N{\widetilde O}^G_{n-1}
&=&
\partial^{N-n}_+\partial^{n-1}_- O^G(\ka \xx, \kb \xx)
\Big|_{\kappa_\pm = 0}
\quad
{\rm with}
\quad
1 \leq n \leq N,
\end{eqnarray}
where $N+1$ equals the total spin
 of the local operators mixing under
renormalization. The difference between quark and gluon operators results
from the fact that their $\kappa$--scale dimensions equals $d_q = 1,
d_G = 2$. Obviously, ${}^NO^q_0$ has no counterpart to mix with. This
situation can be resolved if
\begin{eqnarray}
({\underline O}^q,{\underline O}^G) \equiv
(\partial_- O^q, O^G)
\end{eqnarray}
is used instead of $(O^q, O^G)$. The corresponding renormalization group
equations read
\begin{eqnarray}
\label{x8}
\mu^2 \frac{d}{d\mu^2}
{}^N{\underline O}^A_{n-1}
&=&
\sum^n_{n'=1}
{}^N{\underline \gamma}^{AB}_{n-1,\,n'-1}
{}^N{\underline O}^B_{n'-1}
\quad
{\rm with}
\quad
1 \leq n, n' \leq N,
\end{eqnarray}
where the local anomalous dimension matrix
${}^N{\underline \gamma}^{AB}_{n-1,\,n'-1}$ is given by
\begin{eqnarray}
\label{triangs}
{}^N{\underline \gamma}^{AB}_{n-1,\,n'-1}
&=&
\hbox{ \large ${n-1 \choose n'-1}$}
\int Dw \, w_1^{n-n'} w_2^{n'-1}
{\underline{\widetilde K}}^{AB}(w_1, w_2)~.
\end{eqnarray}
in accordance with Eq.~(\ref{triang}).
Of course the same conclusions concerning the triangularity,
universality and the eigenvalues of these anomalous dimension matrices
can be drawn as in the non--singlet case above. However, the connection
with the kernels ${\widetilde K}^{AB}(w_1, w_2)$ by no means is trivial.
However, computing the elements of the usual local anomalous dimension
matrix $\gamma^{AB}_{n\;n'}$ individually, taking into account that only
operators of the same value of $N$ mix under renormalization, one obtains
the (infinite) triangular matrices \cite{Diss}
\begin{eqnarray}
\label{localqq}
\gamma^{qq}_{nn'}
&=&
\hbox{\large ${n \choose n'}$}
\sigma^{(-)}_{n\,n'} \int_0^1 dw_2 w_2^{n'} \left\{2
\int_0^{1-w_2}dw_1
\,w_1^{n-n'}
{\widetilde K}^{qq}_0(w_1, w_2)\right\},
\\
\label{localqG}
\gamma^{qG}_{nn'}
&=&
n \hbox{\large ${n-1 \choose n'-1}$}
\sigma^{(-)}_{n\,n'} \int_0^1 dw_2 w_2^{n'-1} \left\{2
\int_0^{1-w_2}dw_1
\,w_1^{n-n'}
{\widetilde K}^{qG}_0(w_1, w_2)\right\},
\\
\label{localGq}
\gamma^{Gq}_{nn'}
&=&
\hbox{\large $\frac{1}{n} {n \choose n'}$}
\sigma^{(-)}_{n\,n'} \int_0^1 dw_2 w_2^{n'} \left\{2
\int_0^{1-w_2}dw_1
\, w_1^{n-n'}
{\widetilde K}^{Gq}_0(w_1, w_2)\right\},
\\
\label{localGG}
\gamma^{GG}_{nn'}
&=&
\hbox{\large ${n-1 \choose n'-1}$}
\sigma^{(-)}_{n\,n'} \int_0^1 dw_2 w_2^{n'-1} \left\{2
\int_0^{1-w_2}dw_1
\, w_1^{n-n'}
{\widetilde K}^{GG}_0(w_1, w_2)\right\},
\end{eqnarray}
with
\begin{eqnarray}
\label{eqSIGM}
\sigma^{(\pm)}_{n\,n'}
~=~
\hbox{\large$\frac{1}{4}$}
\Big(1 + (-1)^{n-n'}\Big)
\Big(1 \pm (-1)^{n'- 2 d_B}\Big)
~=~
\hbox{\large$\frac{1}{4}$}
(1 \pm (-1)^{n})(1 \pm (-1)^{n'}),
\end{eqnarray}
and $1 \leq n, n' \leq \infty$; $\sigma_{n\,n'}^{(+)}$ emerges in the
pseudo--scalar case.\footnote{Note that according to our above convention
on $n$ the local anomalous dimensions are defined in the {\sf unpolarized
case} at {\sf odd} integers $n$ and $n'$, whereas they are defined at
{\sf even} integers in the {\sf polarized case} in difference to other
conventions in the literature.} 
The
difference of the sign due to $(-1)^{d_B}$ is compensated by an additional
factor $-(-1)^{d_B}$ which results from the different powers of $w_2$ in
Eqs.~(\ref{localqq}\,--\,\ref{localGG}). The eigenvalues of these 
triangular
matrices are given by their diagonal elements $\gamma^{AB}_{n\,n}$. In a
way similar to Eq.~(\ref{APw}) the splitting functions in the case
of forward scattering are obtained for the singlet case \cite{BGR1,Diss}
\begin{eqnarray}
\label{APwAB}
P^{AB}(z)
=
2 \int^1_0 dw \, {\widetilde O}^{AB}(w,z)\,
\int^{1-w}_0 dw_1 \,
{\widetilde K}^{AB}_0(w_1, w),
\end{eqnarray}
(no summation) with
\begin{eqnarray}
\label{x10}
{\widetilde O}^{AB}(w,z)
&=&
\left( \begin{array}{cc}
     \delta(w-z)  &  \partial_w\delta(w-z)\\
     \theta(w-z)/z & \delta(w-z)/z
\end{array} \right).
\end{eqnarray}
With respect to the cross terms an additional differentiation or
integration has been introduced to remove the factors $n$ or $1/n$,
followed by partial integration or changing the order of integration,
respectively. The consideration of the pseudo--scalar case follows the
same line of reasoning.
\subsection{Anomalous dimensions of twist--2 operators at
${\cal O}(\alpha_s)$}
After this general exposition we turn to the explicit calculation for the
above operators (\ref{ONS}\,--\,\ref{O5G}) in ${\cal O}(\alpha_s)$. We 
will
consider first the {\em unsymmetrized} singlet anomalous dimensions
${\widehat K}^{AB}_0$ and $\Delta {\widehat K}^{AB}_0$ for the 
unpolarized
and the polarized case, respectively, in the $\alpha$--representation.
Here the parameters $(\alpha_1, \alpha_2)$ are nothing but the
usual Feynman parameters appearing in the calculation of the corresponding
Feynman diagrams. As far as relations are concerned which are valid both
for the unpolarized and polarized case we will, for brevity, only give
the results for the unpolarized case in the following. The non--singlet
anomalous dimensions obey in leading order
\begin{eqnarray}
{\widehat K}^{\rm NS}_0 = \Delta {\widehat K}^{\rm NS}_0 =
{\widehat K}^{qq}_0 = \Delta {\widehat K}^{qq}_0~.
\end{eqnarray}
For brevity we use the convention that
the common factors resulting from the order of perturbation theory, as
$\alpha_s/(2\pi)$, and from the $\Theta$--structure of the different
integration measures  will be separated off from the
anomalous dimension kernels. Then without introducing new symbols
for the reduced kernels ${\widehat K}^{qq}_0(\AAA,\AB)$ we obtain
\begin{eqnarray}
\label{factorA}
\hspace{-.5cm}
 K^{AB}_{(1)}(\AAA,\AB, \kappa_-)
&=&
\frac{\alpha_s(\mu^2)}{2\pi}(\kb - \ka)^{d_B - d_A}
\Big(
\Theta(\alpha_1) \Theta(\alpha_2)\Theta(1-\alpha_1-\alpha_2)
{\widehat K}^{AB}_0(\AAA,\AB)
\\
\hspace{-.5cm}
&+& (-1)^{d_B}
\Theta(1-\alpha_1)\Theta(1-\alpha_2)\Theta(\alpha_1+\alpha_2-1)
{\widehat K}^{AB}_0(1-\AAA,1-\AB)
\Big).
\nonumber
\end{eqnarray}
For the unsymmetrized kernels ${\widehat K}^{AB}_0(\AAA,\AB)$
in leading order we get:
\\
\\
1) Unpolarized anomalous dimensions
\begin{eqnarray}
\hspace{-1cm}
\label{eqK1}
{\widehat K}^{qq}_0(\AAA,\AB)
&=&
C_F
\Big\{ 1 - \delta(\AAA) - \delta(\AB)
+ \delta(\AAA) \left [ \frac{1}{\AB}\right]_+
+ \delta(\AB)  \left [ \frac{1}{\AAA}\right]_+
+ \hbox{\large $\frac{3}{2}$}
  \delta(\AAA)\delta(\AB) \Big\},
\\
\hspace{-1cm}
\label{eqK3}
{\widehat K}^{qG}_0(\AAA,\AB)
&=&
- \; 2 N_f T_R
\left \{ 1 - \AAA - \AB + 4 \AAA\AB
\right \}, 
\\
\hspace{-1cm}
\label{eqK2}
{\widehat K}^{Gq}_0(\AAA,\AB)
&=&
- \; C_F
\left \{  \delta(\AAA) \delta(\AB) + 2 \right \},
\\
\hspace{-1cm}
\label{eqK4}
{\widehat K}^{GG}_0(\AAA,\AB)
&=&
C_A \bigg\{ 4 ( 1 - \AAA -\AB) + 12 \AAA \AB
+ \delta(\AAA) \bigg(
 \left [\frac{1}{\AB}  \right ]_+ - 2 + \AB  \bigg)
\\
& & \;\quad
+ \,\delta(\AB)  \bigg(
 \left [\frac{1}{\AAA} \right ]_+ - 2 + \AAA \bigg)
  \bigg\} 
+ \; \hbox{\large $\frac{1}{2}$} \beta_0 \;
 \delta(\AAA)\delta(\AB),
\nonumber
\end{eqnarray}
where $C_F = (N_c^2-1)/2N_c \equiv 4/3,\; T_R = 1/2,\; C_A = N_c
\equiv 3$, and the $\beta$--function in leading order,
$\beta_0 = (11 C_A - 4 T_R N_f)/3$.~\footnote{Note that the emergence
of the term containing
$\beta_0$ is due to self--energy graphs only and does
not induce scale breaking effects at this order,~\cite{BN}.} The
$\big[\phantom{O}\big]_+$--prescription is defined as
%
\begin{equation}
\label{+pres}
\int_0^1 dx
\left [f(x,y)\right]_+ \varphi(x)
=
\int_0^1 dx f(x,y)\left[\varphi(x) - \varphi(y) \right],
\nonumber
\end{equation}
if the singularity of $f$ is of the type $\sim 1/(x - y)$.
\\
\\
2) Polarized anomalous dimensions
\begin{eqnarray}
\label{eqDK1}
\Delta {\widehat K}^{qq}_0(\AAA,\AB)
&=&
{\widehat K}^{qq}_0(\AAA,\AB),
\\
\Delta {\widehat K}^{qG}_0(\AAA,\AB)
&=&
- 2 N_f T_R
\left \{ 1 - \AAA - \AB \right \},
\\
\Delta {\widehat K}^{Gq}_0(\AAA,\AB)
&=&
-  C_F
\left \{\delta(\AAA) \delta(\AB) - 2 \right \},
\\
\label{eqDK4}
\Delta {\widehat K}^{GG}_0(\AAA,\AB)
&=&
{\widehat K}^{GG}_0(\AAA,\AB) - 12 C_A  \AAA \AB.
\end{eqnarray}
Here $(-1)^{d_B}$ has to be changed into $-(-1)^{d_B}$ in
Eq.~(\ref{factorA}).

The anomalous dimensions for the polarized
case have been derived for the first time in our previous paper
\cite{BGR1} and were later confirmed in Ref.~\cite{BARA}. Those for the
unpolarized case have been found several years before in
Refs.~\cite{BGR,BB} already. In Ref.~\cite{BGR} also the
necessity of the symmetrization Eq.~(\ref{symsymW}) has been pointed out.
The kernels ${\widehat K}^{AB}$ and $\Delta {\widehat K}^{AB}$ determine
the respective evolutions of the operators $O^{\rm NS}_{(5)},
O^{q}_{(5)}$, and $O^{G}_{(5)}$ according to Eq.~(\ref{RGa}) in
${\cal O}(\alpha_s)$.

As we have seen above the $w$--representation reveals important
properties of the kernels. It has been proven as a general
representation in~\cite{LEIP}. The connection between the $w$--variables
and the  $\alpha$--variables are given by Eqs.~(\ref{w}). Using these
relations we get the following expressions for the anomalous dimensions:
\begin{eqnarray}
\label{factorW}
\hspace{-.5cm}
 K^{AB}_{(1)}(w_1,w_2, \kappa_-)
&=&
\frac{\alpha_s(\mu^2)}{4\pi} (\kappa_-)^{d_B - d_A}
\Big(
\Theta(1-w_2-w_1)\Theta(1-w_2+w_1)\Theta(w_2)
{\widetilde K}^{AB}_0(w_1,w_2)
\nonumber
\\
\hspace{-.5cm}
&+& (-1)^{d_B}
\Theta(1+w_2-w_1)\Theta(1+w_2+w_1)\Theta(-w_2)
{\widetilde K}^{AB}_0(w_1,-w_2)
\Big).
\end{eqnarray}
For the unsymmetrized anomalous dimension kernels in leading
order we obtain:
\\
\\
1)  Unpolarized anomalous dimensions
\begin{eqnarray}
\label{anow}
{\widetilde K}^{qq}_0(w_1,w_2)
&=&
C_F\bigg\{1 - 2\delta(1-w_2+w_1)
\bigg(1- \frac{2}{(1-w_2-w_1)_+}\bigg)
\\
& & \quad
- 2\delta(1-w_2-w_1)\bigg(1- \frac{2}{(1-w_2+w_1)_+}\bigg)
\bigg\}
+ 3 C_F \delta(1-w_2)\delta(w_1),
\nonumber
\\
\label{anowqG}
{\widetilde K}^{qG}_0(w_1,w_2)
&=&
 - 2 N_f T_R \{w_2 +(1-w_2)^2 - w_1^2\} ,
\\
\label{anowGq}
{\widetilde K}^{Gq}_0(w_1,w_2)
&=&
- 2 C_F \{\delta(1-w_2) \delta(w_1) +1\},
\\
\label{anowGG}
{\widetilde K}^{GG}_0(w_1,w_2)
&=& C_A\bigg\{ 4 w_2 +3\Big((1-w_2)^2 -w_1^2\Big)
-2\delta(1-w_2+w_1)\bigg(1+w_2-\frac{2}{(1-w_2-w_1)_+}\bigg)
\nonumber\\
& & \quad
-2\delta(1-w_2-w_1)\bigg(1+w_2-\frac{2}{(1-w_2+w_1)_+}\bigg)
\bigg\}
+ \beta_0 \delta(1-w_2)\delta(w_1).
\end{eqnarray}
\\
2) Polarized anomalous dimensions
\begin{eqnarray}
\label{panow}
\Delta {\widetilde K}^{qq}_0(w_1,w_2)
&=& {\widetilde K}^{qq}_0(w_1, w_2),
\\
\label{panowqG}
\Delta {\widetilde K}^{qG}_0(w_1,w_2)
&=&
- 2 N_f T_R \{w_2\},
\\
\label{panowGq}
\Delta {\widetilde K}^{Gq}_0(w_1,w_2)
&=&
- 2 C_F
\{\delta(1-w_2) \delta(w_1) -1\},
\\
\label{panowGG}
\Delta {\widetilde K}^{GG}_0(w_1,w_2)
&=&  {\widetilde K}^{GG}_0(w_1, w_2)
     + 3 C_A \Big(w_1^2 - (1-w_2)^2 \Big).
\end{eqnarray}
Again the factor $(-1)^{d_B}$ in Eq.~(\ref{factorW}) has to be changed
into $-(-1)^{d_B}$.

Using Eqs.~(\ref{localqq}\,--\,\ref{localGG},
\ref{anow}\,--\,\ref{anowGG}),
as well as Eqs.~(\ref{eqDK1}\,--\ref{eqDK4},\,
\ref{panow}\,--\,\ref{panowGG}) in the pseudo--scalar 
case,
we now derive the anomalous dimensions which are obtained in the
{\em local LCE}  directly from the non--local evolution
kernels.
Thereby the well--known relations
\begin{eqnarray}
\int^1_0 dw_2 \frac{w_2^n}{(w_2 - 1)_+} &=& \sum_{\ell=0}^{n-1}
\frac{1}{\ell + 1} = \psi(n+1) + \gamma_E\\
\int^1_0 dw_2 w_2^a (1-w_2)^b
&=&
\frac{a!b!}{(a+b+1)!} = B(a+1,b+1),
\end{eqnarray}
are used. Here
$\psi(x) = d \log(\Gamma(x))/dx$, $\gamma_E$ denotes the
Euler--Mascheroni constant, and $B(a,b)$ Euler's Beta-function.
Again a common factor
\begin{eqnarray}
\label{factorL}
\frac{\alpha_s}{2\pi} \times 2 \sigma^{(-)}_{n \,n'}
\qquad {\rm and} \qquad
\frac{\alpha_s}{2\pi} \times 2 \sigma^{(+)}_{n \,n'}
\end{eqnarray}
in the unpolarized and polarized case
will be suppressed, respectively.
\\
\\
1) Unpolarized anomalous dimensions:~\footnote{In comparing these
expressions with the values given in the literature \cite{AP1,AP2} the
above mentioned shift of $(n, n')$ into $(n-1, n'-1)$ sometimes has to be
taken into account.}
\begin{eqnarray}
\hspace{-1.5cm}
\label{x111}
\gamma^{qq}_{nn'}
&=&
C_F\,\bigg\{
\bigg[ \frac{1}{2} - \frac{1}{(n+1)(n+2)}
       + 2 \sum^{n+1}_{j=2} \frac{1}{j} \bigg] \delta_{nn'}
     - \bigg[\frac{1}{(n+1)(n+2)}
       + \frac{2}{n-n'}\frac{n'+1}{n+1}\bigg]
\theta_{nn'}\bigg\}
\\
\hspace{-1.5cm}
\gamma^{qG}_{nn'}
&=&
- N_f T \,
\frac{1}{(n+1)(n+2)(n+3)}\Big[(n^2+3n+4)-(n-n')(n+1)\Big],
\\
\hspace{-1.5cm}
\gamma^{Gq}_{nn'}
&=&
- C_F
\frac{1}{n(n+1)(n+2)}
\Big[(n^2+3n+4)\delta_{nn'}+2\theta_{nn'}\Big],
\\
\label{x112}
\hspace{-1.5cm}
\gamma^{GG}_{nn'}
&=&
C_A\bigg\{
\bigg[\frac{1}{6} - \frac{2}{n(n+1)} - \frac{2}{(n+2)(n+3)}
     + 2 \sum^{n+1}_{j=2}\frac{1}{j} + \frac{2N_f T}{3 C_A}
\bigg]\delta_{nn'}
\\
& &~~+~~\bigg[ 2\bigg(\frac{2n+1}{n(n+1)}
     - \frac{1}{n-n'}\bigg)
     - (n-n'+2)\bigg(\frac{1}{n(n+1)}
     + \frac{1}{(n+2)(n+3)}\bigg)\bigg] \theta_{nn'}\bigg\},
\nonumber
\end{eqnarray}
with the following notation
\begin{eqnarray}
\sigma^{(\pm)}_{n\,n'}
&=&
\hbox{\large $\frac{1}{4}$}
(1 \pm (-1)^n)(1 \pm (-1)^{n'})
\nonumber\\
\label{x12}
\theta_{n\,n'}
&=&
\left\{
\begin{array}{ll}
     1    &    {\rm for}\;\; n'< n,     \\
     0    &    {\rm otherwise}~.
\end{array}\right.
\nonumber
\end{eqnarray}
From this it becomes obvious that the symmetrization (\ref{sym_KA}) and
(\ref{sym_KW}) of the anomalous dimension kernels only results into the
property of the local anomalous dimension matrices being either odd--odd
or even--even for the unpolarized or polarized case, respectively.
\\
\\
2) Polarized local anomalous dimensions:
\begin{eqnarray}
\label{x13}
\Delta\gamma^{qq}_{nn'}
&=&
\gamma^{qq}_{nn'},
\\
\Delta\gamma^{qG}_{nn'}
&=&
- N_f T \,
\frac{n'}{(n+1)(n+2)},
\\
\Delta\gamma^{Gq}_{nn'}
&=&
\frac{1}{(n+1)(n+2)}
\Big[(n+3)\delta_{nn'}- \frac{2}{n} \theta_{nn'}\Big],
\\
\label{x131}
\Delta\gamma^{GG}_{nn'}
&=&
C_A\bigg\{
\bigg[\frac{1}{6} - \frac{4}{(n+1)(n+2)}
     + 2 \sum^{n+1}_{j=2}\frac{1}{j} + \frac{2N_f T}{3 C_A}
\bigg]\delta_{nn'}
\\
& &~~+~~\bigg[ 2\bigg(\frac{2n+1}{n(n+1)}
     - \frac{1}{n-n'}\bigg)
     - (n-n'+2)\frac{2}{(n+1)(n+2)}
     \bigg] \theta_{nn'}\bigg\}.
\nonumber
\end{eqnarray}
As already noted in Section~\ref{sec6.1} the well--known forward
anomalous dimensions are obtained as the diagonal elements of
Eqs.~(\ref{x111}\,--\,\ref{x112},\,\ref{x13}\,--\,\ref{x131}).

The explicit computation of the anomalous dimensions is straightforward.
In lowest order we have to determine the one--particle irreducible
one--loop Feynman diagrams containing the considered operators as the
first vertex. The new Feynman rules and an example of a calculation in
the covariant gauge in the unpolarized case has been presented in
\cite{BGR}. Here we have performed the calculation in axial gauge which
leads to essential simplifications in this order of perturbation theory.
A sample calculation is given in Appendix~\ref{sec-B}.
\section{Evolution equations and evolution kernels}
\renewcommand{\theequation}{\thesection.\arabic{equation}}
\setcounter{equation}{0}
\label{sec-7}
The distribution amplitudes and partition functions defined in Section~4 
have to be determined experimentally at a given factorization scale
$\mu_0^2$. Their evolution in $\mu^2$, however, can be described within
perturbative QCD and is ruled by renormalization group equations. In this
Section we discuss generic evolution equations which directly follow from
the renormalization group equation (\ref{RGsing}).
\subsection{ General properties of twist--2 evolution equations}
Let us first discuss general properties of the evolution equations and
their kernels at any order of perturbation theory. For simplicity we
restrict the discussion to the {\em non--singlet}~(NS) case since the
arguments in the singlet case are quite analogous.

The evolution equations of the partition functions in question are
consequences of the renormalization group equations of the complete
twist--2 operators, i.e.~including the trace terms which are proportional
to $\xx_\sigma$. These operators are  summed up local operators
of definite twist. Despite the fact that the vector operator
$O_{\sigma}^{\rm twist 2}$ and the scalar operator $O^{\rm twist 2} =
\xx^\sigma O_{\sigma}^{\rm twist 2}(\kappa_1\xx,\kappa_2\xx)$ fullfil the
{\em same} renormalization group equation~(\ref{RGvec}), as has been shown
above, the evolution equations of their {\em matrix elements} are
different.

For general values of the arguments $\ka, \kb$ the operator matrix
elements read
\begin{eqnarray}
\label{kdecNS}
\langle p_2|O^{\rm twist2}(\kappa_1 x,  \kappa_2 x)|p_1\rangle
 =  i
\widetilde f_J(\kappa_- xp_+,\kappa_- x p_-,
\kappa^2_- x^2, p_1p_2, \mu^2) e^{i\kappa_+ xp_-}
{\cal Q}^J(x;p_2, p_1)
\end{eqnarray}
and
\begin{eqnarray}
\label{MAINX}
\lefteqn{
\hspace{-1cm}
\langle p_2|O^{\mu,\,\rm twist 2}
(\kappa_1 x, \kappa_2 x)|p_1\rangle
= i
{\cal Q}^{J\mu}(p_2, p_1) \widetilde{F}_J
(\kappa_- xp_+,\kappa_- x p_-, \kappa_+ x p_-,
 \kappa^2_- x^2, p_1p_2, \mu^2)
}\\
&+& i {\cal Q}^J(x;p_2, p_1) i
\left\{
   \kappa_- p_-^\mu \partial_{\kappa_- x p_-}
 + \kappa_+ p_-^\mu \partial_{\kappa_+ x p_-}
 + \kappa_- p_+^\mu \partial_{\kappa_+ x p_-}
 + 2 x^{\mu} \kappa^2_- \partial_{\kappa^2_- x^2}
\right\} \nonumber\\ & &
\hspace{6cm} \times
\widetilde{F}_J,
(\kappa_- xp_+,\kappa_- x p_-, \kappa^2_- x^2, p_1p_2, \mu^2)
\nonumber
\end{eqnarray}
which is nothing but the Eq.~(\ref{MAIN}) in the coordinate space.

Whereas for the representation of the Compton amplitude the operators and
their matrix elements can be taken at arbitrary values of $x$,
cf.~Section~5, for the renormalization properties considered in Section~6,
it is essential that the operators are defined on the light--ray
$\tilde x$, ${\tilde x}^2 =0$. Furthermore, in the subsequent
considerations the operators have to be taken on the light--ray
at general values of $\kappa_i$ .

The matrix elements of the scalar operators $O^{NS}$ obey the
renormalization group equation
\begin{eqnarray}
\label{evox1}
\mu^2 \frac{d}{d \mu^2}
\langle p_2| O^{\rm NS}(\ka\xx,\kb\xx)|p_1\rangle
&=&
\int d^2{\underline\kappa'} \,
\gamma^{NS} (\ka, \kb; \kap, \kbp)
\langle p_2|O^{\rm NS}(\kap\xx,\kbp\xx)|p_1\rangle
\end{eqnarray}
and correspondingly for $O^{NS}_5$ with
$\gamma^{NS}_5 \equiv \Delta \gamma^{NS}$. Having in mind the kinematic
decomposition of the matrix elements, Eqs.~(\ref{kdec1},\,\ref{kdecNS})
we get renormalization group equations for the partition functions
in coordinate space, $\tilde{f}_J$, separately:
\begin{eqnarray}
\label{evog1}
\mu^2 \frac{d}{d \mu^2}
{\tilde f}_J(\kappa_-\xx p_+,\kappa_-\xx p_-)
e^{i\kappa_+ \tilde x p_- }
=
\int d^2{\underline\kappa'} \;
\gamma^{NS} (\ka, \kb; \kap, \kbp)\,
{\tilde f}_J({\kappa'}_-\xx p_+,{\kappa'}_-\xx p_-)
e^{i{\kappa'}_+ \xx p_- }~.
\end{eqnarray}

In the same way we obtain the renormalization group equations for the
matrix elements of the twist--2 vector operator (\ref{MAINX}) and the
corresponding partition functions $\widetilde{F}_J$. After the
kinematic decomposition
the operator relation Eq.~({3.8}) implies
\begin{eqnarray}
\label{evoG1}
\mu^2 \frac{d}{d \mu^2} \widetilde{F}_J(\kappa_-\xx p_+,
\kappa_-\xx p_-, \kappa_+\xx p_-)
& = &
\int d^2{\underline\kappa'} \,
\gamma^{NS} (\ka, \kb; \kap, \kbp)\,
\widetilde{F}_J({\kappa'}_-\xx p_+,{\kappa'}_-\xx p_-,
{\kappa}'_+\xx p_-)~.
\nonumber\\
\end{eqnarray}
By differentiation w.r.t. $\xx p_-$ and $\xx p_+$ the
equations
\begin{eqnarray}
\label{evoG2}
\hspace{-.7cm}
\mu^2 \frac{d}{d \mu^2}
\left[
\kappa_-' \partial_{\kappa_-' \tilde{x} p_-} +
\kappa_+' \partial_{\kappa_+' \tilde{x} p_-}
\right] \widetilde{F}_J
\!\!& = &\!\!\!
\int d^2{\underline\kappa'} \;
\gamma^{NS} (\ka, \kb; \kap, \kbp)\,
\left[
\kappa_- \partial_{\kappa_- \tilde{x} p_-} +
\kappa_+ \partial_{\kappa_+ \tilde{x} p_-}
\right] \widetilde{F}_J,
\\
\hspace{-.7cm}
\label{evoG3}
\mu^2 \frac{d}{d \mu^2}
\left[
\kappa_- \partial_{\kappa_- \tilde{x} p_+} \right]
\widetilde{F}_J
\!\!& = &\!\!\!
\int d^2{\underline\kappa'} \;
\gamma^{NS} (\ka, \kb; \kap, \kbp)\,
\left[
\kappa_-' \partial_{\kappa_-' \tilde{x} p_+} \right]
\widetilde{F}_J,
\end{eqnarray}
are obtained with the same arguments as in Eq.~(\ref{evoG1}).
The renormalization group equations are valid in the
limit $x \rightarrow \tilde{x}$, i.e. $x^2 = 0$. Therefore a further
relation
\begin{eqnarray}
\label{evoG4}
\lefteqn{\hspace{-1cm}
\mu^2 \frac{d}{d \mu^2}
\left[ \kappa_-^2 \partial_{\kappa_-^2 x^2}
\widetilde{F}(\kappa_- {x} p_+, \kappa_- {x} p_-, \kappa_+
{x} p_-, \kappa_-^2 {x}^2)\right]_{x \rightarrow \xx}}
\\
&=&
\int d^2{\underline\kappa'} \;
\breve{\gamma}^{NS} (\ka, \kb; \kap, \kbp)\,
\left[ {\kappa'}_-^2 \partial_{{\kappa'}_-^2 x^2}
\widetilde{F}({\kappa'}_- {x} p_+, {\kappa'}_-
{x} p_-, {\kappa'}_+
{x} p_-, {\kappa'}_-^2 {x}^2)\right]_{x \rightarrow \xx}
\nonumber
\end{eqnarray}
holds. Here we will not discuss this special evolution equation in detail.
We remark however that the corresponding anomalous dimension belongs
to the operator
\bea
O'(\ka \xx, \kb \xx)
=
 \xx^* \frac{\partial}{\partial x}
\hbox{\Large $\frac{i}{2}$}
[\overline\psi(\ka x)(x\gamma) U(\ka,\kb) \psi(\kb x)
-
\overline\psi(\kb x)(x\gamma) U(\ka,\kb) \psi(\ka x)]
\big|_{x \rightarrow \xx}
\eea
and has to be calculated independently. Here $\xx^*$
denotes a second independent light--like vector which obeys
$\xx^* \xx = 1$.

Note that there are two evolution equations, (\ref{evog1}) and
(\ref{evoG1}), for different partition functions which are
non--trivially related by
\begin{eqnarray}
\widetilde{F}_J
(\kappa_- \xx p_+, \kappa_- \xx p_-,\kappa_+ \xx p_-)
\label{eqA1}
&=&
\int_0^1 d\lambda \,
\tilde{f}_J
(\lambda\kappa_- \xx p_+, \lambda\kappa_- \xx p_-)
 e^{i \lambda\kappa_+ \xx p_-}~.
\end{eqnarray}
We finally show that this  relation
is compatible
with applying
the renormalization group operator. The evolution equation
for $\widetilde{F}_J$ reads
\begin{eqnarray}
\label{x14}
\mu^2 \frac{d}{d \mu^2} \int_0^1 d\lambda
\tilde{f}_J(\lambda\kappa_- \xx p_+, \lambda\kappa_- \xx p_-)
    e^{i\lambda\kappa_+ \xx p_- }
=
\int \! d^2{\underline\kappa'}
\int_0^1 \! d\lambda
\gamma^{\rm NS}(\lambda \kappa, \kappa')
\tilde{f}_J({\kappa'}_- \xx p_+,{\kappa'}_- \xx p_-)
e^{i{\kappa'}_+ \xx p_- }
\nonumber
\end{eqnarray}
in the representation of Eq.~(\ref{eqA1}).
Changing the integration variables
$\underline\kappa' = \lambda \hat{\underline\kappa}$ and applying the
scaling relation Eq.~(\ref{scale})  for the anomalous dimensions
\begin{equation}
\gamma(\lambda{\underline\kappa},
\lambda \hat{\underline\kappa}) =
\lambda^{-2}\gamma({\underline\kappa},\hat{\underline\kappa})~,
\end{equation}
we obtain
\begin{eqnarray}
\label{x15}
\mu^2 \frac{d}{d \mu^2}
    \!\int_0^1 \!\!d\lambda \,
    f_J(\lambda\kappa_- \xx p_+, \lambda\kappa_- \xx p_-,
      \lambda\kappa_+ \xx p_-  ) = 
\!\int\! d^2{\underline {\hat\kappa}} \,
\gamma^{\rm NS}({\underline\kappa}, \hat{\underline\kappa})
\!\int_0^1 \!\!d\lambda \,
 f_J(\lambda \hat{\kappa}_- \xx p_+,
   \lambda \hat{\kappa}_- \xx p_-)
e^{i\lambda\hat{\kappa}_+ \xx p_-},
\nonumber
\end{eqnarray}
which is the evolution equation (\ref{evoG1}). This shows that
Eq.~(\ref{evoG1}) and Eq.~(\ref{evog1}) are equivalent.
The evolution equations~({\ref{evog1}\,--\,\ref{evoG3}) yield already
a complete description in the non--singlet case. However, commonly
the evolution of parton densities or distribution amplitudes is described
in terms of functions which depend on the related momentum fractions.
This description is obtained after a Fourier transformation of the
above functions to the functions $f(z_+,z_-)$ and $F(z_+,z_-)$ which have
been introduced in Section~4 before. In the following we will consider
the evolution equations of these quantities.
\subsection{Two-Variable Equations}
In this paragraph we derive the evolution equations for the double
distribution amplitude in the {\em singlet} case. Let us start our
consideration with the double--variable distributions
$f^{q(2)}_J(z_+, z_-)$ and
$f^{G(2)}_J(z_+, z_-)$. In accordance with
Eqs.~(\ref{kdec1}) and  (\ref{kdecg2}) we use the kinematic
decompositions:
\begin{eqnarray}
\label{kdecq2}
\hspace{-1cm}
\kappa_-
\langle p_2|O^{q,\,\rm twist2}(-\kappa_- \xx,
  \kappa_- \xx |p_1\rangle
&=&
(\xx p_+)^{-1}
\tilde f^{q(2)}_J (\kappa_- \xx p_+,\kappa_- \xx p_-, p_1p_2, \mu^2)
{\cal Q}^J(\xx;p_2, p_1),
\\
\label{kdecG2}
\hspace{-1cm}
\kappa_-^2
\langle p_2|O^{G,\,\rm twist2}(- \kappa_- \xx, \kappa_- \xx)|p_1\rangle
&=&
(\xx p_+)^{-1}
\tilde f^{G(2)}_J(\kappa_- \xx p_+,\kappa_- \xx p_-,
p_1p_2, \mu^2)\,
{\cal Q}^J(\xx;p_2, p_1).
\end{eqnarray}
Combining both cases we use the notation ${\tilde f}^A$, where $A=q,G$,
thereby suppressing the index (2). We are interested in the evolution
equations for the physically relevant Fourier transforms
\begin{eqnarray}
f_J^A(z_+,z_-)
=
\int_{-\infty}^\infty 
\frac{d\kappa_- \xx p_-}{2\pi}
\int_{-\infty}^\infty 
\frac{d\kappa_- \xx p_+}{2\pi}
{\tilde f}^A_J(\kappa_- \xx p_+,\kappa_- \xx p_-)
e^{i \kappa_- \xx (p_+z_+ + p_- z_-)},
\end{eqnarray}
where variables being irrelevant for the present considerations
have been omitted.
In order to get the evolution equations for the singlet case
in $z$--space we have to perform a Fourier transform  of
the equations analogous to (\ref{evog1}).
We take matrix elements of the renormalization group equations
(\ref{RGw}) with the corresponding evolution kernels
\begin{eqnarray}
K^{AB}(w_1, w_2, \kappa_-) 
= \kappa_-^{d_B - d_A} {\widetilde K}^{AB}(w_1, w_2)
\end{eqnarray}
and
perform the Fourier transform to get the evolution
equations for $f_J^A$. Inserting the definition (\ref{x1})
into the right hand side we obtain
\begin{eqnarray}
\lefteqn{\hspace{-1cm}
\mu^2 \frac{d}{d \mu^2}f_J^{A(2)}(z_+,z_-)
=
 \int_{-\infty}^\infty  \frac{d\kappa_- \xx p_+}{2\pi}
 \int_{-\infty}^\infty  \frac{d\kappa_- \xx p_-}{2\pi}
e^{i({\kappa'}_+ - {\kappa_+})\xx p_-}
e^{i \kappa_- \xx (p_+z_+ + p_- z_-)}
}\\
\quad &\times&
\int Dw \, K^{AB}(w_1,w_2,\kappa_-)\kappa_-^{d_A}
\int Dz'\, (\kappa'_-)^{-d_B}
e^{-i\kappa_-'\xx (p_+ z'_+ +  p_- z'_-)}
 f_J^{B(2)}({z'_+},{z'_-})~.
\nonumber
\end{eqnarray}
Carrying out the integration over $d(\kappa_- \xx p_{\pm})$ we get
\begin{eqnarray}
\label{evo22}
\mu^2 \frac{d}{d \mu^2}f_J^{A(2)}(z_+,z_-)
&=&
\!\int\!\! Dw {\widetilde K}^{AB}(w_1,w_2) w_2^{-d_B}
\nonumber \\ & & \times
\!\!\int \!\!d{z'_-}d{z'_+}
\delta(z_-\! -\! w_2 {z'_-}\! \! + w_1)
\delta(z_+\! -\! w_2 {z'_+})
 f_J^{B(2)}({z'_+},{z'_-}).
\end{eqnarray}
This finally leads to
\begin{eqnarray}
\mu^2 \frac{d}{d \mu^2}f_J^{A(2)}(z_+,z_-)
&=&
\int Dz' \Gamma^{AB}(z_+,z_-,z'_+,z'_-)
f_J^{B(2)}(z'_+,z'_-), \\
\Gamma^{AB}(z_+,z_-,z'_+,z'_-)
&=& \frac{1}{|z'_+|}
{\widetilde K}^{AB}
\bigg( w_1= -z_- +\frac{z_+}{z'_+} z'_-, w_2= \frac{z_+}{z'_+}
\bigg)\left(\frac{z'_+}{z_+}\right)^{d_B}.
\end{eqnarray}
%
Now we consider the kinematic decomposition for the first
parametrization:
\begin{eqnarray}
\label{kdecq1}
\hspace{-1cm}
\langle p_2|O^{q,\,\rm twist2}(-\kappa_- \xx,  \kappa_- \xx)|p_1\rangle
&=&
\phantom{(-\xx p_+)} i
\tilde f_J^{q(1)}(\kappa_- \xx p_+, \kappa_- \xx p_-, p_1p_2, \mu^2)
\,{\cal Q}^J(\xx;p_2, p_1)
\\
\hspace{-1cm}
\label{kdecG1}
\langle p_2|O^{G,\,\rm twist2} (- \kappa_- \xx, \kappa_- \xx)|p_1\rangle
&=&
i(i\xx p_+)
\tilde f_J^{G(1)}(\kappa_- \xx p_+,\kappa_- \xx p_-, p_1p_2, \mu^2)\,
{\cal Q}^J(\xx;p_2, p_1).
\end{eqnarray}
The derivation of the evolution equations for the distribution amplitudes
Eqs.~(\ref{kdecq1},\ref{kdecG1}) follows the same line.
The only difference is that the
integration with respect to $\kappa_- \xx p_+ $ does not lead directly
to the $\delta$-function and one obtains
\begin{eqnarray}
\mu^2 \frac{d}{d \mu^2}f_J^A(z_+,z_-)
=
\int Dz'\, \Gamma^{AB(1)}(z_+, z_-; z'_+, z'_-)
 f_J^B({z'_+},{z'_-}),
\end{eqnarray}
with
\begin{eqnarray}
\Gamma^{AB(1)}(z_+,z_-; z'_+, z'_-)
=
\int Dw  {\widetilde O}^{AB}( z_+ - w_2 z'_+ )
{\widetilde K}^{AB}\bigg(\!w_1,\! w_2 \bigg)
\delta(z_- - w_2 z'_- + w_1),
\end{eqnarray}
(no summation), where $\widetilde{O}^{AB}$ is given by
\bea
{\widetilde O}^{AB}(z)
&=&  \frac{1}{2 \pi }
\int\! \frac{d\kappa_+\xx p_+}{(i \kappa_- \xx p_+)^{d_B -d_A}}
e^{i\kappa_- \xx p_+ z}
 =
\left\{
\begin{array}{ccc}
\delta(z)
&{\rm for}&~d_B - d_A =~0
\\
\partial_{z} \delta(z)
&{\rm for}&~d_B - d_A =~1
\\
\hbox{\large$\frac{1}{2}$}
\varepsilon(z)
&{\rm for}&~d_B - d_A =-1.
\end{array}\right.
\eea
Here $\varepsilon(z)$ denotes the sign--function, cf.~Eqs.~(\ref{eqdist},
\ref{Theta}).
\subsection{One-Variable Equations}
The equations given above cover the most general case. These equations
depend on two partition variables $z_+$ and $z_-$. The question arises
whether it is possible to express them in terms of single--variable
distributions imposing a one--parameter constraint. Having (physical)
scale invariance in mind it is reasonable to introduce a kinematic
condition by
\begin{equation}
\tau = \frac{\xx p_-}{\xx p_+}~,
\end{equation}
under which the evolution equations are derived. This approach has been
outlined in~\cite{LEIP,BGR1}. It is possible to define the
matrix elements {\it formally} by
\begin{eqnarray}
\left.
\frac{
\langle p_2|O^{q}(-\kappa_- \xx,
\kappa_- \xx)|p_1\rangle}
{(i\xx p_+)}
\right|_{\xx p_- = \tau \xx p_+}
&=&
\int_{-\infty}^{+\infty} dt
e^{-i\kappa_- \xx p_+ t} \Phi_{q}(t,\tau),\\
\left.
\frac{
\langle p_2|O^{G}(-\kappa_- \xx,
\kappa_- \xx)|p_1\rangle}
{(i\xx p_+)^2}
\right|_{\xx p_- = \tau \xx p_+}
&=&
\int_{-\infty}^{+\infty} dt
e^{-i\kappa_- \xx p_+ t}~t \Phi_{G}(t,\tau)~.
\end{eqnarray}
Here $\Phi(t,\tau)$ denotes, in the above sense,
a partition function which
in the limit $\tau \rightarrow -1 $ tends to the standard
Brodsky--Lepage wave function.
Note that the parameter $\tau$ is related to $x$--space. This parameter
may change into the purely kinematic variable $\eta = qp_-/qp_+$ in some
applications after a Fourier transform. In Eq.~(\ref{x3})
\begin{equation}
\label{tt}
\int Dz
\delta(z_+ - t + \eta z_-)
 f_J^{A(1)}(z_+ =t-\eta z_- ,z_-) = 
{\hat f}_J^{A(1)}(t,\eta)t^{d_A -1} ,
\end{equation}
$\eta$ occurs naturally as a variable in momentum space.
The reverse formula reads
\begin{eqnarray}
\label{eqREV}
f_J^{A(1)}(z_+,z_-) = \frac{1}{(2 \pi)^2}
\int_{-\infty}^{+\infty} dt
\int_{-\infty}^{+\infty} d\eta\;\frac{|{z}_-|}{\eta^2}
  \exp\bigg\{i\frac{z_- (z_+ - t)}{\eta}\bigg\}t^{d_A - 1}
{\hat f}_J^{A(1)}(t,\eta).
\end{eqnarray}
Eq.~(\ref{eqREV})
has been derived by assuming, that the
restrictions of the support are completely included into the definition
of the functions $ {\hat f}_J^{A(1)}(t, \eta)$. This can be achieved
using the representation (\ref{eqdist}) for the Heaviside--functions
and performing the limits $\lim_{\varepsilon \rightarrow 0^+}$
later on, cf.~\cite{VLA}.
One may equally well consider the set of one--variable functions
${\hat f}_J^{A(1)}(t,\eta)$ instead of the two--variable functions
${f}_J^{A(1)}(z_+,z_-)$.
Although the Compton amplitude depends on the functions
${F}_J^{A(1)}(z_-,z_+)$ our evolution equations for simplicity have been
formulated with the help of the functions ${f}_J^{A(1)}(z_-,z_+)$.
However, both are related by
\begin{eqnarray}
\label{x16}
F_J^{A(1)}(z_+,z_-) =
\int_0^1 \frac{d\lambda }{\lambda^2}
f_J^{A(1)}\left(\frac{z_+}{\lambda},\frac{z_-}{\lambda}\right)
\Theta (\lambda - |z_+|) \Theta(\lambda - | z_- |). \nonumber
\end{eqnarray}
The relation for $f_J^{A(1)}$ corresponding to Eq.~(\ref{tt}) reads
\begin{eqnarray}
\widehat
{F}_J^{A(1)}(t,\eta) t^{d_A -1} = \int Dz
 {F}_J^{A(1)}(z_+, z_-) \delta(z_+ - t + \eta z_-),
\end{eqnarray}
which yields
\begin{eqnarray}
\widehat
{F}_J^{A(1)}(t,\eta) = \int_0^1\frac{d\lambda}{\lambda }
\widehat
{f}_J^{A(1)}\left(\frac{t}{\lambda}, \eta\right)
\left(\frac{t}{\lambda}\right)^{d_A - 1}~.
\end{eqnarray}
We finally obtain the
one--parameter evolution equations
for the functions $f^{NS}_J$ and  ${f}_J^{A(1)}$:
\begin{eqnarray}
\label{evo3t}
\mu^2 \frac{d}{d \mu^2} \widehat{f}^{\rm NS}_J(t,\tau)
&=&
\int_{-\infty}^{+\infty} d t'
V_{\rm ext}^{\rm NS}(t,t',\tau)\widehat{f}^{\rm NS}_J(t',\tau), \\
\label{evo3ta}
\mu^2 \frac{d}{d \mu^2}
\widehat{f}_J^{A(1)}(t,\tau) 
&=&
\int_{-\infty}^{+\infty} d t'
V^{AB}_{\rm ext}(t,t',\tau)
\widehat{f}_J^{B(1)}(t',\tau), 
\end{eqnarray}
with $\widehat{f}_J^{\rm NS} = \widehat{f}_J^{q_i} 
- \widehat{f}_J^{\overline{q}_j}$.
The corresponding
{\em extended} kernels $V^{AB}_{\rm ext}(t,t',\tau)$ read
\begin{eqnarray}
\label{ker3t}
V_{\rm ext}^{AB}(t,t',\tau)
&=&
\int_0^1 d \AAA \int_0^{1-\AAA} d \AB
{\widehat K}^{AB}(\AAA,\AB)
\int_{- \infty}^{+ \infty} \frac{d(\kappa \xx p_+)}{2 \pi}
\nonumber
\\
&\times&
\left(i \kappa \xx p_+ \right)^{d_B - d_A}
\frac{{t'}^{d_B}}{t^{d_A}}
\exp\left\{i \kappa \xx p_+
\Big[t - (1 - \AAA -\AB)t' + \tau (\AAA -\AB)\Big]
\right\}.
\end{eqnarray}
They obey the scaling relation
\begin{eqnarray}
\label{x44}
V^{AB}_{\rm ext}(t,t',\tau) = 
V^{AB}_{\rm ext}(t,t',-\tau) =
\frac{1}{\tau} V_{\rm ext}^{AB}\bigg(\frac{t}{\tau},
\frac{t'}{\tau},1 \bigg).
\end{eqnarray}
For convenience we rewrite the general expressions for the evolution
kernels in the variables
\begin{eqnarray}
\label{xy}
x= \frac{1}{2}\bigg( 1 + \frac{t}{\tau}\bigg),~
\overline x= \frac{1}{2}\left ( 1 - \frac{t}{\tau}\right),~
y= \frac{1}{2}\bigg( 1 + \frac{t'}{\tau}\bigg),~
\overline y= \frac{1}{2}\left ( 1 - \frac{t'}{\tau}\right).
\end{eqnarray}
In one--loop approximation
they are given by 
\begin{eqnarray}
V^{AB}_{\rm ext}(t,t',\tau) = \frac{\alpha_s(\mu^2)}{2 \pi} 
V^{AB}_{0,{\rm ext}}(t,t',\tau)
\end{eqnarray}
where
\begin{eqnarray}
\label{kerGE1}
V^{qq}_{0,{\rm ext}}(t,t',\tau) &=&
\frac{1}{2\tau}
\left \{  V^{qq}(x,y)\rho(x,y) + V^{qq}(\bx,\by) \rho(\bx,\by)
+\hbox{$\large \frac{3}{2}$} C_{F} \delta(x-y)
\right \}
\\
V^{qG}_{0,{\rm ext}}(t,t',\tau) &=&\
\frac{1}{2\tau} \left(\frac{2y - 1}{2}\right)
\left \{  V^{qG}(x,y)\rho(x,y) - V^{qG}(\bx,\by) \rho(\bx,\by)
\right \} 
\\
V^{Gq}_{0,{\rm ext}}(t,t',\tau) &=&\frac{1}{2\tau}
 \left(\frac{2}{2x - 1} \right)
\left \{  V^{Gq}(x,y)\rho(x,y)
- V^{Gq}(\bx,\by) \rho(\bx,\by)
\right \}
\\
\label{kerGE4}
V^{GG}_{0,{\rm ext}}(t,t',\tau)
&=&
\frac{1}{2\tau}
\left(\frac{2y - 1}{2x - 1}  \right)
\left \{
V^{GG}(x,y)\rho(x,y) + V^{GG}(\bx,\by) \rho(\bx,\by) \right \}
\\
& &+ \frac{1}{2\tau} \frac{1}{2} \beta_0 \delta(x-y)
\nonumber\\
\nonumber\\
\label{kerDGE1}
\Delta V^{qq}_{0,{\rm ext}}(t,t',\tau) &=&  V^{qq}_{\rm
ext}(t,t',\tau)\\
\Delta V^{qG}_{0,{\rm ext}}(t,t',\tau) &=&\frac{1}{2\tau}
\left(\frac{2y - 1}{2}\right)
\left \{\Delta  V^{qG}(x,y)\rho(x,y) - \Delta V^{qG}(\bx,\by)
\rho(\bx,\by)
\right \} 
\\
\Delta V^{Gq}_{0,{\rm ext}}(t,t',\tau) &=&\frac{1}{2\tau}
\left (\frac{2}{2x - 1} \right )
\left \{\Delta  V^{Gq}(x,y)\rho(x,y)
- \Delta V^{Gq}(\bx,\by)
\rho(\bx,\by)
\right \}
\\
\label{kerDGE4}
\Delta
V^{GG}_{0,{\rm ext}}(t,t',\tau) &=&
 \frac{1}{2\tau}
\left(\frac{2y - 1}{2x - 1}  \right)
\left \{
\Delta V^{GG}(x,y)\rho(x,y) + \Delta V^{GG}(\bx,\by)
\rho(\bx,\by) \right \}
 \\
& &+ \frac{1}{2\tau} \frac{1}{2} \beta_0 \delta(x-y)
\nonumber
\end{eqnarray}
with the $\Theta$--structure
\begin{equation}
\rho(x,y) =
\Theta\left(1-\frac{x}{y}\right)
\Theta\left(\frac{x}{y}\right)~{\rm sign}(y)~,
\end{equation}
The non--singlet kernels are given by
$V^{\rm NS}_{0,\rm ext} = V^{qq}_{0,\rm ext}$.
and
$\Delta V^{\rm NS}_{0,\rm ext} = \Delta V^{qq}_{0,\rm ext}$, respectively.
The different partial expressions $V^{AB}$ and $\Delta V^{AB}$
read
\begin{eqnarray}
V^{qq}(x,y) &=& C_F
\bigg[ \frac{x}{y} - \frac{1}{y} + \frac{1}{(y-x)_+}
 \bigg],\\
V^{qG}(x,y) &=& - 2 N_f T_R \frac{x}{y} \bigg[ 4(1 - x) +
\frac{1 - 2x}{y} \bigg], \\
V^{Gq}(x,y) &=&  C_F \bigg [1-  \frac{x^2}{y} \bigg], \\
V^{GG}(x,y) &=&  C_A
  \bigg [2\frac{x^2}{y}\bigg(3-2x + \frac{1-x}{y}\bigg)
+\frac{1}{(y-x)_+}
   - \frac{y+x}{y^2}\bigg],\\
\nonumber\\
\Delta V^{qq}(x,y) &=& V^{qq}(x,y),\\
\Delta V^{qG}(x,y) &=& - 2 N_f T_R \frac{x}{y^2}, \\
\Delta V^{Gq}(x,y) &=& C_F \bigg [\frac{x^2}{y} \bigg ], \\
\Delta V^{GG}(x,y) &=& C_A
 \bigg [2\frac{x^2}{y^2}  +\frac{1}{(y-x)_+}
   - \frac{y+x}{y^2}\bigg ].
\end{eqnarray}
Details on the $[~]_+$--prescription, which acts to the right, are
given in Appendix~\ref{sec-D}.
Note that the kernels given in Eqs.~(\ref{kerGE1}\,--\,\ref{kerDGE4}) 
apply
to the {\it full} range of variables, i.e.~they represent the kernels
completely. The function $V^{qq}_{\rm ext}(t,t',\tau)$ was already
derived in Refs.~\cite{ROBA,LEIP}.

The determination of these kernels is straightforward observing, however,
that these quantities are distribution--valued. In the calculation one
may set $\tau = 1 $ and later apply the scaling relation,
Eq.~(\ref{x4}),
taking into account the general $\Theta$--structure in the
$(t,~t')$--plane derived in  \cite{LEIP}.  Alternatively, the calculation
can be done in the different kinematic regions separately. A
sample--calculation for the kernel $V^{Gq}$ is given in
Appendix~\ref{sec-C}. All other known representations in the literature
consist explicitly of two or three parts being calculated separately.

Recently, the  kernels (\ref{kerGE1}\,--\,\ref{kerGE4}) 
were also calculated
in Ref.~\cite{RAD}, using a different notation, cf. \cite{R2} for related
work.
For $\tau = 1$,
corresponding to $\zeta = 1$ in \cite{RAD}, they agree with the results
obtained above.
A further independent representation has been given recently in
Ref.~\cite{XJ}. The parameter $\tau$ introduced above equals to $- \xi/2$
in the notation of \cite{XJ}.
\section{Special cases}
\renewcommand{\theequation}{\thesection.\arabic{equation}}
\setcounter{equation}{0}
\label{sec-8}

\vspace{1mm}
\noindent
The evolution kernels given above cover a series of limiting cases which 
were studied in the literature before. These are characterized by special
kinematic constraints for the processes, as in the case of forward
scattering $\langle p_2| \rightarrow \langle p_1| \equiv \langle p|$ or 
the transition from the vacuum state $\langle 0|$ to a hadron state
$\langle p|$ being related to the hadron wave functions.
\subsection{The Brodsky--Lepage limit}
For $\tau = \pm 1$ the equations~(\ref{kerGE1}\,--\,\ref{kerDGE4})
transform into the limit $\langle p_2| \rightarrow \langle p|,
\langle p_1| \rightarrow \langle 0|$, which is known as the
Brodsky--Lepage~\cite{BL} and Efremov--Radyushkin~\cite{ER} 
case.~\footnote{Several independent calculations of the evolution kernels
for the meson wave functions were performed in Refs.~\cite{BLX}.} This 
limit may be performed {\it formally} leaving $p_1 \rightarrow 0$, which
leads to correct results, cf.~\cite{ROBA}. The corresponding evolution
equations are obtained using as variables $x$ and $y$, Eqs.~(\ref{xy},
\ref{evo3t}, \ref{evo3ta}). As an example we consider the simplest case.
For $0 < x,y  < 1, |\tau| = 1$ one obtains
\begin{eqnarray}
\hspace{-.4cm}
V^{qq}(t,t')
&=&
\hbox{\large $\frac{1}{2}$} C_F \left\{
\Theta(y-x)
\left[\frac{x}{y} -\frac{1}{y} +\frac{1}{(y-x)_+}\right]\right.
+\Theta(x-y)
\left.\left[\frac{1-x}{1-y}-\frac{1}{1-y}+\frac{1}{(x-y)_+}\right]
\right. \nonumber\\ & & \left.
+ \hbox{\large $\frac{3}{2}$}     \delta(x-y)
\right\}.
\end{eqnarray}
\subsection{The `near--forward' representation}
This representation has been introduced by X.~Ji and has also been given 
in Ref.~\cite{RAD}. It contains the forward case as limiting case. 
A correct application of the evolution equation for near forward matrix
elements needs the representation for the region $t < \tau$ and $t>\tau$
separately. They are obtained from our general kernels as special cases.
As an example we show here that our general structure for $V_{ext}^{qq}$
covers also the case $t>\tau,\;t'>\tau$. Note that
\begin{equation}
{\rm sign}~\overline y = -{\rm sign}~y~~{\rm and}~~\Theta
(1-{\overline x}/{\overline y}) = \Theta (y-x)~.
\end{equation}
One therefore obtains
\begin{eqnarray}
V^{qq}_{0,\rm ext}(t,t',\tau)
&=&  
\hbox{\large $\frac{1}{2}$}
C_F\Theta(y-x)
\bigg\{\frac{x}{y} \bigg[1+\frac{1}{(y-x)_+}\bigg]
  -\frac{1-x}{1-y} \bigg[1+\frac{1}{(x-y)_+}\bigg]
\bigg\} 
+ \hbox{\large $\frac{3}{2}$} C_F \delta(x-y)
\nonumber \\
&=&
\hbox{\large $\frac{1}{2}$}
C_F\Theta(y-x)\frac{1}{(y-x)_+}
\bigg[1 + \frac{x\overline x}{y \overline y}\bigg]
+ \hbox{\large $\frac{3}{4}$} C_F \delta(x-y)
\end{eqnarray}
for $\tau = 1$.
This representation was also given in Ref.~\cite{RAD}. 

Starting from the 
anomalous dimension we reproduce the results of~\cite{XJ} and 
{\em generalize} them even to the region with $t' \neq 1$. We 
obtain~\footnote{In ref.~\cite{XJ} the variable $t$ was denoted
by $x$, having a different meaning in our notation. Therefore we use 
$t$ instead in Eqs.~(\ref{JI1}\,--\,\ref{JI4}). Here
our variable $\tau$ has to be identified with the variable $-\xi/2$
from \cite{XJ}.}:
\begin{eqnarray}
\label{JI1}
K^{qq}(t,t',\xi) &=&
C_F \frac{t^2 + {t'}^2 - \xi^2/2}{({t'}^2 - \xi^2/4)(t' - t)_+}
+ \frac{3}{2} \delta(t' - t),
\\
K^{qG}(t,t',\xi) &=&
 T_R N_f \frac{t^2 + (t' - t)^2 - \xi^2/4}{({t'}^2 -
\xi^2/4)^2}t',
\\
K^{Gq}(t,t',\xi) &=&
 C_F \frac{{t'}^2 + (t' - t)^2 - \xi^2/4}{t({t'}^2 -
\xi^2/4)},\\
K^{GG}(t,t',\xi) &=&  2 C_A \frac{t'}{t}
\frac{1}{({t'}^2 -\xi^2/4)^2}
 \bigg[\frac{({t'}^2 -\xi^2/4)^2}{(t'-t)_+}
      + t'({t'}^2 + \xi^2/4)  \\
& & - t(3{t'}^2 -\xi^2/4) - (t'+t)(t'-t)^2 \bigg]
      + \frac{\beta_0}{2} \delta(t' - t),
\nonumber\\
\nonumber\\
\Delta K^{qq}(t,t',\xi) &=&
     K^{qq}(t,t',\xi), \\
\Delta K^{qG}(t,t',\xi) &=&
     T_R N_f \frac{t^2 - (t' - t)^2 - \xi^2/4}{({t'}^2 -
\xi^2/4)^2}\;t', \\
\Delta K^{Gq}(x,\xi) &=&
C_F \frac{t' - (t' - t)^2 - \xi^2/4}{t({t'}^2 - \xi^2/4)}, \\
\label{JI4}
\Delta K^{GG}(x,\xi) &=&
2 C_A \frac{t'}{t}
\frac{1}{({t'}^2 -\xi^2/4)^2} \bigg[\frac{({t'}^2 -\xi^2/4)^2}
{(t'-t)_+}  +       t'({t'}^2 + \xi^2/4) .
 \\
& &
 - t(3{t'}^2 -\xi^2/4) -2t'(t'-t)^2 \bigg]
      + \frac{\beta_0}{2} \delta(t' - t)~.
\nonumber
\end{eqnarray}
To get a complete representation of the kernels the independently 
calculated parts for $t< -\xi/2$ and $|t|< \xi/2 $ have to be added,
which can be directly derived from the general expressions given above.
\subsection{The Altarelli--Parisi limit}
Let us consider the case of forward scattering $p_2 = p_1 \equiv p$. The 
corresponding evolution kernels can be obtained after some calculation 
from Eqs.~(\ref{kerGE1}\,--\,\ref{kerDGE4}) in the limit
$\tau \rightarrow 0,~t' > 0$,
\begin{eqnarray}
\lim_{\tau \rightarrow 0}  V^{AB}_{0,ext}(t,t',\tau) =
\frac{1}{t'} P^{AB}\left(z\right) \Theta(z) \Theta(1-z)
\quad {\rm with}\quad
z = \frac{t}{t'}~,
\end{eqnarray}
cf.~Appendix~\ref{sec-C}. Alternatively, the splitting functions for
deeply inelastic forward scattering can be obtained by  direct
integration of the kernels Eqs.~(\ref{eqK1}\,--\,\ref{eqDK4}), which is
performed in the following. 
The splitting 
functions are obtained by 
\begin{eqnarray}
\label{APSP}
\widehat{P}^{AB}(z)
&=&
P^{AB}(z) \theta(z) \theta(1 - z)
\\
P^{AB}(z)
&=&
\int_{- \infty}^{+ \infty}
du \,\theta(1 - u) \theta(u)\,(1 - u)
\widehat{O}^{AB}(u,z)
\int_0^1 d\xi
\widehat{K}^{AB}\big((1 - u)\xi, (1 - u)(1 - \xi)\big),
\nonumber
\end{eqnarray}
where
\begin{equation}
 \widehat{O}^{AB}(u,z) =
\bigg( \begin{array}{ll} ~~\delta(z - u)~~~&
\partial_z  \delta(z - u)~~~\\
- \theta(z - u)/z  &\delta(z-u)/z \end{array}
\bigg)~.
\end{equation}
The $\Theta$--structure resulting from the measure $D\alpha$
is universal. In deriving Eq.~(\ref{APSP}) it is useful to apply the 
relations
\begin{equation}
\label{eqdist}
\theta(x) = \lim_{\varepsilon \rightarrow 0^+}
\frac{1}{2\pi}
\int_{- \infty}^{+ \infty} d\xi \frac{e^{ix\xi}}{i\xi +
\varepsilon},
~~~~~\delta^{(k)}(x) =
\frac{1}{2\pi}
\int_{- \infty}^{+ \infty} d\xi~(i\xi)^{k} e^{ix\xi},
\end{equation}
which are valid for tempered distributions~\cite{VLA}.
$P^{AB}(z) \; (\Delta P^{AB}(z))$ are the well--known splitting 
functions~\cite{AP1,AP2} for unpolarized and polarized deep--inelastic 
forward scattering in ${\cal O}(\alpha_s)$:
\begin{eqnarray}
P^{qq}(z) &=&
C_F \bigg(\frac{1 + z^2}{1 - z} \bigg)_+,
\\
P^{qG}(z) &=& 2 N_f T_R \bigg[z^2 + (1 - z)^2 \bigg],\\
P^{Gq}(z) &=& C_F \frac{1 + (1 - z)^2}{z},\\
P^{GG}(z) &=& 2 C_A \bigg[ \frac{1}{z} + \frac{1}{(1 - z)}_+ -2
+ z(1-z)  \bigg] + \frac{\beta_0}{2} \delta(1 - z),\\
\nonumber\\
\Delta P^{qq}(z) &=& P^{qq}(z),\\
\Delta P^{qG}(z) &=& 2 N_f T_R \bigg[z^2 - (1 - z)^2 \bigg]
,\\
\Delta P^{Gq}(z) &=& C_F \frac{1 - (1 - z)^2}{z},\\
\Delta P^{GG}(z) &=& 2 C_A \bigg[ 1 - 2 z + \frac{1}{(1 - z)}_+
\bigg]
+ \frac{\beta_0}{2} \delta(1 - z).
\end{eqnarray}
The parton densities of the quarks $q(z)$ and
antiquarks $\overline{q}(z)$
are given in Eqs.~(\ref{eqPAR1},\,\ref{eqPAR2}) above, with $z$ denoting 
the momentum fraction carried by the partons. $G(z)$ denotes the gluon 
distribution. Their evolution equations read in leading order for the 
unpolarized case
\begin{eqnarray}
\label{evoAP}
\mu^2 \frac{d}{d \mu^2} q^{\rm NS}(z,\mu^2)
&=&
\frac{\alpha_s(\mu^2)}{2 \pi}
\int_{z}^{1} \frac{d z'}{z'}
\widehat{P}^{\rm NS}
\bigg(\frac{z}{z'} \bigg) q^{\rm NS}(z',\mu^2),
\\
\mu^2 \frac{d}{d \mu^2} \bigg(
\begin{array}{c}
\Sigma(z,\mu^2)\\
G(z,\mu^2) \end{array} \bigg)
&=&
\frac{\alpha_s(\mu^2)}{2 \pi}
\int_{z}^{1}
\frac{d z'}{z'}
\widehat{\Pvec}
\left (\frac{z}{z'} \right)
\bigg(\begin{array}{c}
\Sigma(z',\mu^2)\\ G(z',\mu^2)
\end{array} \bigg)~.
\end{eqnarray}
Here $q^{\rm NS}(z,\mu^2)$ denotes a non--singlet combination
of quark densities, as e.g. $q(z,\mu^2) - \overline{q}(z,\mu^2)$,
$\Sigma(z,\mu^2) = \sum_{j=1}^{N_f} \left[q_j(z,\mu^2) + \overline{q}_j
(z,\mu^2)\right]$ is the flavor--singlet combination of the quark 
densities, and $\widehat{\Pvec}$ is the matrix of the
leading order singlet splitting functions.
The respective evolution equations in the polarized case are obtained
by replacing $q,\, \overline{q}$  and $G$ by $\Delta q,\,
\Delta \overline{q}$, and $\Delta G$
and the splitting
functions
$\widehat{P}^{AB}(z)$ by $\Delta \widehat{P}^{AB}(z)$. 
\section{Solutions}
\renewcommand{\theequation}{\thesection.\arabic{equation}}
\setcounter{equation}{0}
\label{sec-9}

\vspace{1mm}
\noindent
Here we extend  the method used in Ref.~\cite{RADL} for the non--singlet 
case to the singlet case. The idea follows the original solution of
the Brodsky--Lepage equation given in \cite{ER}. We start from 
Eq.~(\ref{evo22}) and form the moments of the distribution functions
\begin{eqnarray}
 f_n^{A(2)}(z_-) = \int_{-1 +|z_-|}^{1-|z_-|}
 dz_+ z_+^n f^{A(2)}(z_+,z_-)~.
\end{eqnarray}
leading to
\begin{eqnarray}
\label{gne}
\mu^2 \frac{d}{d \mu^2} 
f_n^{A(2)}(z_-) =
\frac{\alpha_s(\mu^2)}{2 \pi}
\YINT dz'_- {\Gamma}_n^{AB}(z_-,z'_-)
f_n^{B(2)}(z_-'),
\end{eqnarray}
with
\begin{eqnarray}
\label{gan}
 {\Gamma}_n^{AB}(z_-,z'_-) &=& \int Dw \delta(z_- - w_2 z_-' + w_1)
w_2^{n-d_B} \widetilde{K}^{AB}_0(w_1,w_2) \nonumber\\
&=&
\hbox{\large $\frac{1}{2}$} \int_0^1 dw_2 \Theta(1+z_- - w_2(1+z_-'))
\Theta(1-z_- - w_2(1-z_-'))
\nonumber\\ & & \times \Theta(1-z_-') \Theta(1+z_-')
\widetilde{K}^{AB}_0(-z_-+w_2 z_-',w_2) w_2^{n-d_B}~.
\end{eqnarray}
In this way a first diagonalization is obtained. The kernels 
$\Gamma^{AB}_n(z_-,z_-')$ read in the unpolarized case
\begin{eqnarray}
\label{ganqq}
\Gamma_n^{qq} (z,z')
&=& 
C_F\left\{\Theta(z-z')\left(\frac{1-z}{1-z'}\right)^n
\left (\frac{1}{n} + \frac{2}{(z-z')_+}\right)\right.
\nonumber \\
& & \left. ~~~ +\Theta(z'-z)\left(\frac{1+z}{1+z'}\right)^n
 \left(\frac{1}{n} + \frac{2}{(z'-z)_+}\right)\right\}
 + 3 \delta(z-z'),\\
\label{ganqG}
\Gamma_n^{qG} (z,z')
&=&
-2N_f T_R \left\{\Theta(z-z')
\left(\frac{1-z}{1-z'}\right)^n \frac{n^2 +2n(z-z') -(2zz'-1)}
{(n+1)n(n-1)} \right.
\nonumber \\
& & \left. ~~~ + \Theta(z'-z)\left(\frac{1+z}{1+z'}\right)^n
 \frac{n^2 +2n(z'-z) -(2zz'-1)}
{(n+1)n(n-1)}\right\},\\
\label{ganGq}
\Gamma_n^{Gq} (z,z')
&=&
-C_F\left\{\Theta(z-z')\left(\frac{1-z}{1-z'}\right)^n
 \frac{1}{n} 
+ \Theta(z'-z)\left(\frac{1+z}{1+z'}\right)^n
 \frac{1}{n} 
 +  \delta(z-z')\right\}, \\
\label{ganGG}
\Gamma_n^{GG} (z,z')& = & C_A \left\{\Theta(z-z')
\left(\frac{1-z}{1-z'}\right)^n
\left(\frac{3(1+z)(1-z')}{n-1}
+ \frac{6zz'}{n} + \frac{3(1+z')(1-z)}{n+1} \right. \right.
\nonumber\\
& & ~~~~~~~~~~~~~~~ \hspace*{2.5cm}
\left.
-\frac{2}{n} +\frac{2}{(z-z')_+} \right) 
\nonumber \\
& & ~~~ +
\Theta(z'-z) \left(\frac{1+z}{1+z'}\right)^n
\left(\frac{3(1-z)(1+z')}{n-1} +\frac{6zz'}{n} 
+ \frac{3(1-z')(1+z)}{n+1} \right. 
\nonumber\\ 
& & ~~~~~~~~~~~~~~~ \hspace{2.5cm}
\left. \left.
- \frac{2}{n} +\frac{2}{(z'-z)_+} \right)
\right\}
+ \beta_0 \delta(z-z').
\end{eqnarray}
Similarly one obtains for the kernels in the polarized case
\begin{eqnarray}
\label{pganqq}
\Delta\Gamma_n^{qq} (z,z') 
&=& 
C_F\left\{\Theta(z-z')
\left[\left(\frac{1-z}{1-z'}\right)^n
\left(\frac{1}{n} + \frac{2}{(z-z')_+}\right)\right]\right.
\nonumber \\
& &\left. ~~~
+\Theta(z'-z)\left[\left(\frac{1+z}{1+z'}\right)^n
 \left(\frac{1}{n} + \frac{2}{(z'-z)_+}\right)\right]
+ 3 \delta(z-z')\right\},\\
\label{pganqG}
\Delta\Gamma_n^{qG} (z,z')& = &-2N_f T_R \left\{\Theta(z-z')
\left(\frac{1-z}{1-z'}\right)^n \frac{1}{n}
+ \Theta(z'-z)\left(\frac{1+z}{1+z'}\right)^n
 \frac{1}{n} \right\},\\
\label{pganGq}
\Delta\Gamma_n^{Gq} (z,z')
&=&
C_F\left\{\Theta(z-z')\left(\frac{1-z}{1-z'}\right)^n
 \frac{1}{n}
+\Theta(z'-z)\left(\frac{1+z}{1+z'}\right)^n
 \frac{1}{n}  -  \delta(z-z')\right\}, \\
\label{pganGG}
\Delta\Gamma_n^{GG} (z,z')
&=& 
C_A \left\{\Theta(z-z')
\left(\frac{1-z}{1-z'}\right)^n
\left(\frac{4}{n} +\frac{2}{(z-z')_+}\right) \right.
\nonumber \\
& &  \left. ~~~+\Theta(z'-z)
\left(\frac{1+z}{1+z'}\right)^n
\left(\frac{4}{n} +\frac{2}{(z'-z)_+}\right)\right\}
 + \beta_0 \delta(z-z').
\end{eqnarray}
As a
next step we have to find a solution of Eq.~(\ref{gan}).
One finds that
the following symmetry--relations hold for kernels
$\Gamma_n^{GG} (z,z')$ and $\Delta\Gamma_n^{GG} (z,z')$~:
\begin{eqnarray}
\label{sym}
   (1-z'^2)^n\Gamma_n^{AB}(z,z')
   &=&\Gamma_n^{AB}(z',z)(1-z^2)^n ,\\
   (1-z'^2)^n\Delta\Gamma_n^{AB}(z,z')
   &=&\Delta\Gamma_n^{AB}(z',z)(1-z^2)^n~.
\end{eqnarray}
This is expected because of similar problems studied
previously in \cite{RADL,ER}.
The final diagonalization of Eq.~(\ref{gne}), referring to the
partition functions $f_n^{A(2)}(z_-)$
 for which the quarkonic and gluonic operators are dealt with equally,
can thus be performed using
Gegenbauer polynomials $C_m^p(z)$, cf.~\cite{RG},
\begin{eqnarray}
\label{exp}
  \Gamma_n^{AB}(z,z')&=& \sum \Gamma^{AB}_{n,m} (1-z^2)^n
  C^{n+1/2}_m (z) C^{n+1/2}_m (z')N_{n,m}^{-1} ,\\
\label{pexp}
  \Delta\Gamma_n^{AB}(z,z')&= &\sum \Delta \Gamma^{AB}_{n,m}
(1-z^2)^n
   C^{n+1/2}_m (z) C^{n+1/2}_m (z')N_{n,m}^{-1}~.
\end{eqnarray}
The coefficients $N_{n,m}$  are the normalization factors of the
Gegenbauer polynomials
\begin{equation}
\int_{-1}^{+1} dz (1-z^2)^n C_l^{n+1/2}(z) C_k^{n+1/2}(z) =
\delta_{lk} N_{n,k} = \delta_{lk} \left [
2^{2n+1}
\Gamma^2\left( n + \frac{1}{2}\right) \frac{(k+n+1/2) \cdot
k!}{2 \pi
\Gamma(2n+1+k)}\right]^{-1}~.
\end{equation}
$\Gamma^{AB}_{n,m}$ and $\Delta \Gamma^{AB}_{n,m}$
are the respective expansion coefficients which can be easily calculated
and are found to be the diagonal elements of the triangular matrices
discussed in Section~6.

Another method of solution
consists in forming also Mellin moments in
the variable $z_-$,
\begin{equation}
 \YINT dz_- z_-^k f^{A(2)}_n(z_-) = f^{A(2)}_{n,k}~.
\end{equation}
The evolution equations may be written as, cf. also \cite{BGR},
\begin{eqnarray}
\label{gnk}
\mu^2 \frac{d}{d \mu^2}f^{A(2)}_{n k}=
 \frac{\alpha_s(\mu^2)}{2 \pi}\sum_{l=0}^k
 {\Gamma}_{n, k l}^{AB}f^{A(2)}_{n l},
\end{eqnarray}
with the transformed kernels given by
\begin{eqnarray}
 {\Gamma}_{n, k l}^{AB}= \int
Dw_+~w_2^{n+l-d_B}w_1^{k-l}
\widetilde{K}^{AB}_0(w_1,w_2)\left( \begin{array}{c}
 k \\ l \end{array} \right )~.
\end{eqnarray}
These evolution equations are not diagonal with respect to the
indices
$(k,l)$.
The explicit expressions for
${\Gamma}_{n, k l}^{AB}$ and
$\Delta{\Gamma}_{n, k l}^{AB}$ are given in 
Eqs.~(\ref{x111}\,--\,\ref{x112},\,\ref{x13}\,--\,\ref{x131}).
For fixed $n$
they form triangular matrices. The eigenvalues are
the diagonal elements $k=l$
\begin{eqnarray}
 {\Gamma}_{n, k k}^{AB}= \int Dw_+~w_2^{n+k-d_B}
\widetilde{K}^{AB}_0(w_1,w_2)= \gamma^{AB}_{n+k-d_B}~.
\end{eqnarray}
The coefficients $\gamma^{AB}_{n+k-h_j}$
are the anomalous dimensions of the forward case with a shifted
Mellin index. A solution for the singlet evolution equations in the
case of the single--variable equations was given in \cite{RAD1}
recently.

\section{Conclusions}
\renewcommand{\theequation}{\thesection.\arabic{equation}}
\setcounter{equation}{0}
\label{sec-10}

\vspace{1mm}
\noindent
The scaling violations of the various observables, which occur in
space--like processes, can be traced back to the renormalization of
light--ray quark and gluon operators, which define these quantities. 
Therefore one may obtain a general description by calculating first the
anomalous dimensions, or evolution kernels, of these operators and
establish their connection to the different observables through the
respective integral mappings and, in some cases, subsequent limiting 
procedures, afterwards. In the present paper we were limiting the 
investigation to the case of twist--2 operators and massless QCD. 
They are found by a corresponding twist--decomposition starting with a 
genuine operator representation. The space--like evolution kernels are 
process--independent. To form the observables one has to calculate also 
the process--dependent coefficient functions, which are trivial in lowest
order. Furthermore, expectation values of the operators have to be formed,
allowing for the description of a wide variety of processes.

In the light--cone 
expansion both vector operators $O^{\sigma}$ and scalar
operators $\xx_{\sigma} O^{\sigma}$ appear. As was shown they are equivalent
under renormalization and it suffices therefore to study the evolution
of the simpler scalar operators. The respective relations for the vector
operators can be obtained by integral relations. This holds also for
the relation between their general non--forward expectation values
between the states $\langle p_2|$ and $|p_1\rangle$. This connection is
an important one, as it forms the basis for the integral relations
by Wandzura and Wilczek~\cite{WW} and one of the authors and 
Kochelev~\cite{BLK}. These relations were derived earlier in the local
operator product expansion and a subsequent analytic continuation from the
integer moments and a Mellin--transform. Here it is obtained directly
by the integral relation between the scalar and vector operators.

One may derive a series of all--order relations for the evolution kernels
of the light--ray operators. In the local representation they are given
by infinite triangular matrices $\gamma_{n_1,n_2}^{AB}$, the diagonal
elements of which are the anomalous dimensions in the forward scattering 
case. The anomalous dimensions are independent of the the total spin 
$n_1+n_2$.

The expectation values of the operators between the states $\langle p_2|$
and $|p_1\rangle$ are complex--valued distribution amplitudes 
through which
the scattering cross section may be represented, which are, however, no 
observables. Nonetheless they obey evolution equations which are implied 
by the renormalization group equations of the operators. Commonly these 
matrix elements are represented by a two--fold Fourier transform as 
distribution amplitudes which depend on the two momentum fractions $z_+$ 
and $z_-$ along the momenta $p_+$ and $p_-$. This representation is a 
generalization of the case of forward scattering in which only one 
momentum fraction emerges. 

The two--variable amplitudes may be related to single--variable amplitudes
by implying a kinematic constraint to the general case, i.e. demanding 
$\tau = \xx p_-/\xx p_+ = {\sf const.}$ In this representation the kernels
$V^{AB}_{\rm ext}$ depend on the variables $t$ and $t'$ besides the
parameter $\tau$. We obtained a unique representation for the whole 
kinematic range of $t$ and $t'$. One may easily describe the scaling 
violations in the case of vacuum--meson transitions~\cite{BL,ER} in the 
limit $\tau \rightarrow -1$. With the help of the functions
$V^{AB}_{\rm ext}(t,t',\tau)$ other recent representations~\cite{RAD,XJ} 
can be verified after some calculation. The evolution kernels in the limit
of forward scattering, which are the well--known splitting functions, can 
be obtained by an integral relation form the two--variable kernels at one
hand. They result also from the single--variable representation in the
limit $\tau \rightarrow 0$ for $t' > 0$ from the kernels 
$V^{AB}_{\rm ext}(t,t',\tau)$.

The solution of the two--variable non--singlet and singlet evolution
equations require a diagonalization of the evolution kernels with respect
to the variables $z_+, z'_+$ and $z_-, z'_-$, respectively. Here the
$w$--representation shows, that the diagonalization with respect to the
first variables can be performed by a Mellin transform, as in the forward
case. This is an all--order property. The diagonalization with respect
to the second set of variables through Gegenbauer polynomials relies
on a symmetry property of the kernels, which exists at leading order, but
is broken in higher orders. Alternatively, one may perform a second
Mellin transform. The latter representation is, however, only suited
for discrete representations over integer moments. An easy analytic
continuation is lacking in the latter case since the eigenvectors do not
obey an analytic relation which can be continued to complex numbers.

\vspace{1mm}
\noindent
{\bf Acknowledgement}\\
We would like to thank P.~S\"oding and U.~Gensch for their constant 
support of the project. For discussions we thank
A.~Tkabladze, W.L.~van Neerven, D.~M\"uller, A.V.~Radyushkin, 
I.I.~Balitsky, and
W.~Schweiger. We are very much indebted to T.~Braunschweig for discussions
on the relation between non--local and local anomalous dimensions, and to
M.~Lazar for various discussions on the twist decomposition of light--ray
operators.
B.G. would like to thank DESY Zeuthen and the Institute of Theoretical
Physics at Graz University for support through the Scientific Exchange 
Program. D.R. likes to thank DESY Zeuthen, the Karl--Franzens--University
of Graz, and the Graduate College `Quantum Field Theory' at NTZ, Leipzig 
University, for the kind hospitality extended to him. We would like to 
thank J. Vermaseren for providing us a new version of {\tt AXODRAW}. The 
work was supported in part by EU contract FMRX--CT98--0194(DG12--MIHT).

\newpage
\begin{appendix}
\section{Kinematic Relations}
\label{sec-A}
\renewcommand{\theequation}{\thesection.\arabic{equation}}
\setcounter{equation}{0}

\vspace{1mm}
\noindent
The Compton process is characterized by the invariants
\begin{eqnarray}
s &=& (p_1 + q_1)^2 = (p_2 + q_2)^2 
= M_1^2 - Q_1^2 + 2 p_1 q_1 = ~~~M_2^2 - Q_2^2 + 2 p_2 q_2
\nonumber\\
t &=& (p_2 - p_1)^2 = (q_2 - q_1)^2 
= M^2_1 + M_2^2 - 2 p_1 p_2 = -Q_1^2 -Q_2^2 - 2 q_1 q_2~.
\end{eqnarray}
Here $M_{1,2}$ are the masses of the initial-- and final state hadron.
$q_{1,2}^2 =  - Q^2_{1,2}$ are the virtualities of the incoming and
outgoing photon. The initial--state photon is assumed to be space--like
\begin{eqnarray}
Q_1^2 = - q_1^2 > 0~,
\end{eqnarray}
while the final state photon can either be light-- or time--like with
\begin{eqnarray}
q^2_2 < m_{\pi}^2,
\end{eqnarray}
to avoid $s$--channel mass thresholds in the subsequent description, as
they occur in reactions like $\gamma^* + p \rightarrow V + p'$, where
$p$ and $p'$ are the initial and final--state hadrons, respectively, and
$V$ denotes a meson at the photon--side. In the latter case the process
is no longer space--like and the light cone expansion {\it cannot} be
applied. On the other hand, one may produce mesons on the hadronic side
in a space--like process.

The inelasticity $\nu$ is given by
\begin{eqnarray}
\nu &=& p_+q_1 + \frac{1}{2}(M_1^2-M_2^2) =
p_+q_2 - \frac{1}{2}(M_1^2-M_2^2)~,
\end{eqnarray}
and the cms--energy $s$ reads
\begin{eqnarray}
s = 2 \nu + \frac{1}{2} \left[q_1^2 + q_2^2 + M_1^2 + M_2^2 -t \right]~.
\end{eqnarray}
The virtuality $Q^2 = - q^2$ is
\begin{eqnarray}
Q^2 = \frac{t}{4} + \frac{1}{2}\left(Q_1^2 + Q_2^2\right)~.
\end{eqnarray}
We define
\begin{eqnarray}
\hat{x} = \frac{Q^2_1}{\nu}~.
\end{eqnarray}
The variable $\xi$ is given by
\begin{eqnarray}
\xi = \frac{\hat{x}}{2} + \frac{t + 2 Q^2_2}{4 \nu} =
\hat{x} + \eta + \frac{t}{4 \nu}~,
\end{eqnarray}
showing the relation of the variables $\xi$ and $\eta$. For
$t/\nu \rightarrow 0$, as assumed in the generalized Bjorken region,
their difference is $\hat{x}$. In the case of forward scattering, $\eta =
t = 0$, $\xi$ and $\hat{x}$ turn into the Bjorken variable 
$x_B = Q^2_1/(2p.q_1)$. 
Also other kinematic domains were
studied for the Compton amplitude, as e.g. wide--angel scattering
$p_1.p_2 \rightarrow \infty$~\cite{KN}, a
quantity being fixed in the present 
treatment. 
\section{Sample calculation of the anomalous dimensions}
\label{sec-B}
\renewcommand{\theequation}{\thesection.\arabic{equation}}
\setcounter{equation}{0}

\vspace{1mm}
\noindent
In this Appendix a sample calculation for the non--local anomalous
dimensions is performed. As an example we will consider the derivation
of the anomalous dimension $\Delta K^{Gq}$. The calculation is performed
in the $\overline{\rm MS}$--scheme using dimensional regularization in
$N = 4 - 2\varepsilon$ dimensions. The anomalous dimensions are given by 
the pole parts (${\sc p.p.}$) 
\begin{eqnarray}
\hat{I} = \frac{1}{\varepsilon}~\hat{I}_{\sc P.P.} + O(\varepsilon^0)
\end{eqnarray}
of the respective integrals. The gluon operator, Eq.~(\ref{O5G}), reads 
in the momentum representation
\begin{eqnarray}
\label{G5}
O^G_{\rm 5}(\ka,\kb)
&=&
\hbox{\large$\frac{1}{2}$}
\xx^{\mu}  \xx^{\mu'}
\left [{F_{a \mu}}^{\nu}(\ka\xx)\FFA_{\mu'\nu}(\kb\xx)
- {F^a}_{\mu\nu}(\kb\xx)
{\FFA_{\mu'}}{\vspace*{-1.2mm}^{\nu}}(\ka\xx)\right ]
\nonumber \\
&= &
\int\frac{dp_1 dp_2}{4(2\pi)^{2N}}
(e^{i\kappa_1 \xx p_1 -i\kappa_2 \xx p_2}
- e^{i\kappa_2 \xx p_1 -i\kappa_1 \xx p_2})
\nonumber \\
& & \times
\xx^\mu \varepsilon_{\mu \nu \rho \sigma}\xx p_1 A_a^\nu(p_1)
\Big(p_2^\rho A^ \sigma_a(p_2)
-p_2^\sigma  A^\rho_a(p_2)\Big)~.
\end{eqnarray}
The one--loop correction to this operator is calculated. We work in the 
axial gauge, so that
\begin{eqnarray}
\tilde x^\mu F_{\mu \nu }(\kappa \tilde x) &\rightarrow&
\tilde x \partial_{\kappa \tilde x} A_\nu(\kappa \tilde x),\nonumber\\
\tilde x^\mu \tilde F_{\mu \nu }(\kappa \tilde x) &\rightarrow&
\tilde x^\mu \varepsilon_{\mu \nu \sigma \rho}
\partial^\sigma_{\kappa \tilde x} A^\rho(\kappa \tilde x)~.
\end{eqnarray}
We are interested in pole--parts only, which for $\Delta K^{Gq}$
correspond to the counter term of the quark operator. Actually not the
complete gluon operator but only the term
\begin{eqnarray}
\tilde x^\mu \tilde x^{\mu'} \tilde F_{\mu'}^{\;\nu}(\kappa_1
\tilde x) F_{\mu \nu}(\kappa_2\tilde x)
\end{eqnarray}
needs to be used in the calculation which corresponds to the
analytic expression:
\begin{eqnarray}
\label{ZE}
T\Big(\xx^{\mu}  \xx^{\mu'} {F}_{\mu\nu}(\kb\xx) 
\tilde {F}_{\mu'}^{\;\;\nu}(\ka\xx) S \Big) =
\int\frac{d^N p_1 d^N p_2}{(2\pi)^{2N}} \frac{1}{(2\pi)^N}  
\overline \psi(p_1) \widehat{I}(p_2,p_1,\kappa_2,\kappa_1)
\psi(p_2)~,
\end{eqnarray}
with
\begin{eqnarray}
\widehat{I}(p_2,p_1,\kappa_2,\kappa_1) =
e^{i\kappa_1 \xx p_1 -i\kappa_2 \xx p_2}
I_1(p_1,p_2,\kappa_1,\kappa_2) + e^{i\kappa_2 \xx p_1 -i\kappa_1 
\xx p_2} I_2(p_2,p_1,\kappa_2,\kappa_1)  
\end{eqnarray}
and
\begin{eqnarray}
\label{I}
I_1 &=&
\int d^N k (i g \gamma_{\mu}) i\frac{\hat k}{k^2}
 (i g \gamma_{\nu}) \frac{1}{i}
 \left(g^{\mu \sigma}-\frac{\tilde x^{\mu}q_1^{\sigma}
+ \tilde x^{\sigma} q_1^{\mu} }{\xx q_1 }\right)
\frac{1}{q_1^2} \frac{1}{i}
\left(g^{\nu\rho}-\frac{\tilde x^{\nu} q_2^{\rho}
+\tilde x^{\rho}q_2^{\nu} }{\xx q_2}\right)
\frac{1}{q_2^2}
\nonumber \\
 & & \times
\left(-i q_1^{\kappa}\tilde x^{\lambda} i (\tilde xq_2)
\varepsilon_{\kappa\lambda\rho\sigma}\right)
\,e^{i \kappa \xx k},
\end{eqnarray}
where $ q_1 = p_1 - k $ and $ q_2=p_2-k $. 
$I_2(p_2,p_1,\kappa_2, \kappa_1)$ is obtained by interchanging
$ \gamma^\nu \leftrightarrow \gamma^\mu $, $p_1 \leftrightarrow p_2 $ and
$\kappa_1 \leftrightarrow \kappa_2$. Here contrary to our convention 
adopted in the main part we use  $\kappa = \kappa_2 - \kappa_1 = 
2\kappa_-$, and the color factors are suppressed. The individual steps of
the calculation are performed by a {\tt FORM}-program~\cite{FORM}~:

\vspace{3mm}
\noindent
(I) Firstly, the $\gamma$--algebra is evaluated. The result is represented
by projections onto the matrices 
\begin{eqnarray}
{\bf 1},~~\gamma_5,~~\gamma_{\mu},~~\gamma_5\gamma_{\mu},
~~\sigma_{\mu\nu},~~\gamma_5 \sigma_{\mu\nu}~.
\end{eqnarray}
Here the only contribution is due to the axial vector part. The resulting
expressions have been classified with the help of the pole--parts 
$J_{lm}^{\rho_1 \ldots \rho_n}$ of basic integrals,
\begin{eqnarray}
\label{eqI1}
I_1  = \hbox{\Large -$\frac{i g^2}{4}$} &
\gamma_5 \gamma_\mu
&\Big\{2 J_{33}^{\mu \rho\sigma}\tilde x_{\rho}
 \tilde x_{\sigma}\nonumber \\
& & + J_{32}^{\rho \sigma}
\Big[\tilde x^{\mu}
(p_{1~\rho}\tilde x _{\sigma}+p_{2~\rho}\tilde x _{\sigma})
- \tilde x_{\rho}\tilde x_{\sigma}(p_1^{\mu}+p_2^{\mu})
- g^\mu_{\rho}\tilde x_{\sigma}\tilde x(p_1 + p_2)\Big]
\nonumber \\
& & - J_{31}^{\rho}
\Big[\tilde x^{\mu}(p_{1\,\rho} \tilde x p_2
+ p_{2\,\rho} \tilde x p_1)
 -\tilde x_{\rho}(p_1^{\mu} \tilde x p_2 +
p_2^{\mu} \tilde x p_1)\Big]  \\
& & -2 J_{21}^{\rho} \tilde x^{\mu} \tilde x_{\rho}
+ J_{20}\tilde x^{\mu}\tilde x(p_1+p_2) \Big\}. \nonumber
\end{eqnarray}
In the calculation of the pole--parts of the individual integrals we
apply the $\alpha$-representation of the Feynman denominators and 
perform the momentum integrals. The exponentials are expanded
into Taylor series as
\begin{eqnarray}
\exp(i\kappa k \tilde x) = \sum_{n=0}^{\infty}
\frac{(i\kappa k \tilde x)^n}{n!}~.
\end{eqnarray}
Because of $\tilde x^2 = 0$ only the first few terms of these
expansions contribute.

\vspace{3mm}
\noindent
(II) The  terms $J_{lm}^{\rho_1 \ldots \rho_n}$
in Eq.~(\ref{eqI1}) are
\begin{eqnarray}
\label{eqJ33}
J_{33}^{\lambda \mu \nu}
&=&
{\sc p.p.}\int d^N k \frac {k^\lambda k^\mu k^\nu}{q_1^2 q_2^2 k^2}
 e^{ik\tilde x\kappa}
\\
&=&
\frac{i\pi^2}{2\varepsilon}\int D_0\alpha e^{i\xx P(\alpha)\kappa}
\Big\{
\Big(g^{\lambda\mu}P^\nu    (\alpha)
    +g^{\mu\nu}    P^\lambda(\alpha)
    +g^{\nu\lambda}P^\mu    (\alpha)\Big) \nonumber \\
& & +i\kappa
\Big(
  \tilde x^\lambda P^\mu(\alpha) P^\nu(\alpha)
+ \tilde x^\mu P^\nu(\alpha) P^\lambda(\alpha)
+ \tilde x^\nu P^\lambda(\alpha) P^\mu(\alpha) \Big) \nonumber
\\
& &
-i\kappa \frac{D(\alpha)}{2}
\Big(
  g^{\lambda \mu} \tilde x^\nu
 +g^{\mu \nu} \tilde x^\lambda
 +g^{\nu \lambda} \tilde x^\mu \Big) \nonumber \\
 & &
-  ( i \kappa)^2 \frac{D(\alpha)}{2}
\Big(
  P^\lambda(\alpha) \tilde x^\mu \tilde x^\nu
 +P^\mu(\alpha) \tilde x^\nu \tilde x^\lambda
 +P^\nu(\alpha) \tilde x^\lambda \tilde x^\mu \Big)
+ c_1 \tilde x^\lambda \tilde x^\mu \tilde x^\nu \Big\},
\nonumber
\\
\label{eqJ32}
J_{32}^{\mu \nu}
&=&
{\sc p.p.}\int d^N k \frac {k^\mu k^\nu }{q_1^2 q_2^2 k^2}
 e^{ik\tilde x\kappa}
\\
&=&
\frac{i\pi^2}{2\varepsilon}\int D_0\alpha e^{iP(\alpha) \tilde x
 \kappa}
 \Big\{\Big(g^{\mu \nu}
 + i\kappa (P^\mu (\alpha)\tilde x^\nu+
   P^\nu(\alpha) \tilde x^\mu \Big)
 + c_2 \tilde x^\mu \tilde x^\nu \Big\},
\nonumber \\
J_{31}^\nu
&=&
{\sc p.p.}\int d^N k \frac {k_\nu }{q_1^2 q_2^2 k^2}
 e^{ik\tilde x\kappa}
= 
\frac{i\pi^2}{2\varepsilon}
i \tilde \kappa \tilde x^{\mu}
\int D_0\alpha e^{i \xx P(\alpha) \kappa},
\\
J_{21}^\nu
&=&
{\sc p.p.}\int d^N k \frac {k_\nu }{q_1^2 q_2^2 }
e^{ik\tilde x\kappa}  \\
&=&
\frac{i\pi^2}{\varepsilon}\int D_0\alpha
\delta(1-\alpha_1 - \alpha_2)
e^{i \xx P(\alpha)\kappa} \Big(P^\nu(\alpha)
- \frac{i}{2} \kappa \tilde x^{\nu}D(\alpha)\Big),
\nonumber
\\
J_{20}
&=&
{\sc p.p.}\int dk \frac {1 }{q_1^2 q_2^2 }
e^{ik\tilde x\kappa_-}= 
\frac{i\pi^2}{\varepsilon}\int D_0\alpha
\delta(1-\alpha_1 - \alpha_2)
 e^{i\xx P(\alpha)\kappa}~,
\end{eqnarray}
with the abbreviations
\begin{eqnarray}
P^\mu(\alpha)
&=&
p_1^\mu  \alpha_1 + p_2^\mu \alpha_2
\nonumber
\\
D(\alpha)
&=&
p_1^2 \alpha_1 (1-\alpha_1)+
p_2^2 \alpha_2 (1-\alpha_2)-
2p_1p_2 \alpha_1 \alpha_2~.
\nonumber
\end{eqnarray}
The constants $c_1$ and $c_2$ in Eqs.~(\ref{eqJ33},\ref{eqJ32})
remain unspecified here, since they do not
contribute to the final result. Note that in the present case the two 
possible
axial denominators $1/k \tilde x$ are canceled  by corresponding
factors in the numerator. For completeness we note that in integrals 
with axial denominators, like
\begin{eqnarray}
 \int dk \frac{k_\mu k_\nu
 e^{ik\tilde x\kappa}}{
(p_1-k)^2 (p_2-k)^2 k^2} \left\{\frac{1}{k \tilde x} \right\},
\end{eqnarray}
we proceed as follows. The integrand is first rewritten as
\begin{eqnarray}
 \frac{e^{ik\tilde x\kappa}}{k\tilde x} =
\frac{ e^{ik\tilde x\kappa} - 1}{k\tilde x} + \frac{1}{k\tilde x}~.
\end{eqnarray}
By applying the improved gauge prescription given by Leibbrandt and 
Mandelstam \cite{LEIB} the integrals decompose into a part being
calculated by Leibbrandt already, whereas the other part, important for
the operator renormalization, can be calculated without difficulties.
For diagrams containing more than one axial dominator similar 
decompositions hold, as e.g.
\begin{eqnarray}
e^{ik\tilde x\kappa}\frac{1}{(k\tilde x)^2} \frac{1}{\sf N} &=&
\frac{ e^{ik\tilde x\kappa} - 1 - ik \tilde x \kappa}
{(k\tilde x)^2  \sf N} + \frac{1}{(k\tilde x)^2}\frac{1}{\sf N}
+ i\kappa\frac{1}{k\tilde x}\frac{1}{\sf  N}  \\
 e^{ik\tilde x\kappa}\frac{1}{(k\tilde x)\left[(k-p) \tilde x\right]}
 \frac{1}{\sf N} &=&
e^{ik\tilde x\kappa}
\frac{1}{p\tilde x}\frac{1}{\sf N}
\bigg[\frac{1}{(k-p)\tilde x}-\frac{1}{k\tilde x}\bigg]
\nonumber \\
& =&\frac{1}{p\tilde x}\frac{1}{\sf  N}
\bigg[\frac{e^{ik\tilde x\kappa}-e^{ip\tilde x\kappa}   }
{(k-p)\tilde x }\bigg]
+e^{ip\tilde x\kappa} \frac{1}{(k-p)\tilde x}
\frac{1}{p\tilde x}\frac{1}{\sf  N} \nonumber \\
&-&\frac{1}{p\tilde x}\frac{1}{\sf  N}
\bigg[\frac{ e^{ik\tilde x\kappa} - 1}{k\tilde x}
+ \frac{1}{k\tilde x}\bigg].
\end{eqnarray}
Here {\sf N} denotes a standard Feynman denominator. As a result for the
pole--part of  $I_1$, Eq. (\ref{I}),  we obtain
\begin{eqnarray}
I_1(\kappa)
& = &
- {\sc p.p.}
{\gamma}_5 (\gamma \tilde x) \frac{\pi^2 g^2}{4}
\int D_0\alpha e^{i\kappa \xx P(\alpha)}
\Big\{\delta(1- \alpha_1 -\alpha_2)\Big[2P\xx (\alpha)
   - \xx (p_1 + p_2)\Big]\nonumber \\
& &  + \Big[i\kappa (\xx p_1)(\xx p_2) - 2 \xx P(\alpha)
     -i\kappa (\xx P(\alpha))^2\Big]\Big\}~.
\end{eqnarray}

\vspace{3mm}
\noindent
(III) Now some appropriate identities are applied to cast the expressions
into a form containing the operators again. In our case these rules can 
be derived by partial integration. We use the following relations:
\begin{eqnarray}
\int D_0\alpha e^{i\kappa \xx P(\alpha)}
{\left[
(i\kappa \xx P(\alpha))^2\right]
}
& = &
\int D_0\alpha e^{i\kappa \xx P(\alpha)}
[\delta(1-\alpha_1 -\alpha_2)
\big(i\kappa \tilde x P(\alpha) - 3  \big)  +6]
\\
\int D_0\alpha e^{i\kappa \xx P(\alpha)}
{i\kappa \xx P(\alpha)}
& = &
\int D_0\alpha e^{i\kappa \xx P(\alpha)}
[\delta(1-\alpha_1 -\alpha_2) - 2]
\\
\int D_0\alpha e^{i\kappa \tilde x P(\alpha)}
i\kappa \xx p_1 \, \xx p_2
& = &
\int D_0\alpha e^{i\kappa \xx P(\alpha)}
\left\{\delta(1-\alpha_1 -\alpha_2)\left[
\xx p_2
-\frac{1}{i\kappa}\delta(\alpha_1)\right]  \right.
\nonumber\\ & & \hspace*{5.7cm}  \left.
+ \frac{1}{i\kappa}\delta(\alpha_1) \delta(\alpha_2) \right\}
\end{eqnarray}
\vspace{-.6cm}
\begin{eqnarray}
\int D_0\alpha e^{i\kappa \tilde x P(\alpha)}
\delta(1-\alpha_1 -\alpha_2)\alpha_2
 (\tilde x p_1- \tilde x p_2)      = 
\int D_0\alpha e^{i\kappa \tilde x P(\alpha)}
 \frac{1}{i\kappa}
  \delta(1-\alpha_1 -\alpha_2)
 (1 - \delta(\alpha_1))~. \nonumber\\
\end{eqnarray}

\vspace{3mm}
\noindent
(IV) As the result of this calculation we obtain
\begin{eqnarray}
\label{Ip1p2}
I_1
&=& \frac{1}{\varepsilon}
\gamma_5 (\xx\gamma) \,
\frac{i\pi^2 g^2}{4}
\int D_0\alpha e^{i (\kappa_2 -\kappa_1) \xx (p_1\alpha_1
+p_2\alpha_2 )}
\frac{1}{\kappa_2 -\kappa_1}
\Big\{ \delta(\alpha_1)\delta(\alpha_2) - 2 \Big\}
+ {\rm finite~terms}~.\nonumber\\
\end{eqnarray}
Collecting all terms one gets
\begin{eqnarray}
\label{TS0}
\Big(T O^G_{\rm 5   }(\ka,\kb) S \Big)
&=&
\frac{1}{2}\int\frac{d^N p_1 d^N p_2}{(2\pi)^{3N}}
\overline \psi(p_1)
\Big\{~e^{i\kappa_1 \xx p_1 -i\kappa_2 \xx p_2}
[- I_1(p_1,p_2,\kappa_1,\kappa_2)
 + I_1(p_1,p_2,\kappa_2,\kappa_1)]
\nonumber \\
& &\hspace{2.5cm} +e^{i\kappa_2 \xx p_1 -i\kappa_1 \xx p_2}
[- I_2(p_2,p_1,\kappa_2,\kappa_1)
 + I_2(p_2,p_1,\kappa_1,\kappa_2)] \Big\}
 \psi(p_2)~.\nonumber\\
\end{eqnarray}
After inserting the explicit expressions and exploiting the symmetry
relations of Eq.~(\ref{Ip1p2}) one obtains finally
\begin{eqnarray}
\label{TSF}
{\sc  p.p.}\Big(T O^G_{\rm 5   }(\ka,\kb) S \Big)
&=& -
\int\frac{d^N p_1 d^N p_2}{(2\pi)^{8}} i\overline \psi(p_1)
\gamma_5 \left(\xx  \gamma\right)
 \,\psi(p_2) \\
& & \times \frac{\pi^2 g^2}{4(2\pi)^4} \int D_0\alpha
\frac{C_F}{\kappa_2 -\kappa_1}
\Big\{ \delta(\alpha_1)\delta(\alpha_2) - 2 \Big\}
\nonumber \\
& & \times \Big\{
e^{i\kappa_1 \xx p_1 -i\kappa_2 \xx p_2}
+e^{i\kappa_2 \xx p_1 -i\kappa_1 \xx p_2}\Big\} \nonumber\\
& & \times
\Big\{
e^{i (\kappa_2 -\kappa_1) \xx (p_1\alpha_1+ p_2\alpha_2 )}
+e^{-i (\kappa_2 -\kappa_1) \xx (p_1\alpha_1+ p_2\alpha_2 )}
 \Big\}~.
\nonumber
\end{eqnarray}
Here we have introduced the color factor $C_F$.
Now we introduce new variables ${\kappa'}_i$, use the possibility of 
changing $\alpha_1 \leftrightarrow \alpha_2 $ and obtain
\begin{eqnarray}
\label{TSF1}
{\sc  p.p.}\Big(T O^G_{\rm 5   }(\ka,\kb) S \Big)
&=& -
\frac{\pi^2 g^2}{4(2\pi)^4}
\int d{\kappa'}_1 d{\kappa'}_2
\int\frac{d^N p_1 d^N p_2}{(2\pi)^{8}} i\overline \psi(p_1)
\gamma_5 \left(\xx  \gamma\right)  \,\psi(p_2)  \\
&\times&
\Big\{
e^{i{\kappa'}_1 \xx p_1 -i{\kappa'}_2 \xx p_2}
+e^{i{\kappa'}_2 \xx p_1 -i{\kappa'}_1 \xx p_2}\Big\}
\int D_0\alpha
\frac{C_F}{\kappa_2 -\kappa_1}
\Big\{ \delta(\alpha_1)\delta(\alpha_2) - 2 \Big\}
\nonumber \\
&\times&\Big [
\delta({\kappa'}_1 - [\kappa_1 +\alpha_1(\kappa_1
- \kappa_2)])
\delta({\kappa'}_2 - [\kappa_2 -\alpha_2(\kappa_1
- \kappa_2)]) \nonumber \\
& & +
\delta({\kappa'}_1 - [\kappa_1 -\alpha_1(\kappa_1
- \kappa_2)])
\delta({\kappa'}_2 - [\kappa_2 +\alpha_2(\kappa_1
- \kappa_2)])
\Big ]. \nonumber
\end{eqnarray}
Performing the momentum integration we finally end up with
\begin{eqnarray}
\label{TSF2}
{\sc  p.p.}\Big(T O^G_{\rm 5   }(\ka,\kb) S \Big)
&=& -
\int d{\kappa'}_1 d{\kappa'}_2 \frac{i}{2}
[ \overline \psi({\kappa'}_1 \tilde x)
\gamma_5 (\xx  \gamma) \,\psi({\kappa'}_2 \tilde x)
+
\overline \psi({\kappa'}_2 \tilde x)
\gamma_5 \left(\xx \gamma\right) \,\psi({\kappa'}_1 \tilde x)]
\nonumber \\
& & \times
\frac{2 \pi^2 g^2 }{4(2\pi)^4}
\int D_0\alpha
\frac{1}{\kappa_2 -\kappa_1} \left[
C_F
\Big\{ \delta(\alpha_1)\delta(\alpha_2) - 2 \Big\} \right]
\nonumber \\
& & \times
\Big [
\delta({\kappa'}_1 - [\kappa_1 +\alpha_1(\kappa_1
- \kappa_2)])
\delta({\kappa'}_2 - [\kappa_2 -\alpha_2(\kappa_1
- \kappa_2)]) \nonumber \\
& & +
\delta({\kappa'}_1 - [\kappa_1 -\alpha_1(\kappa_1
- \kappa_2)])
\delta({\kappa'}_2 - [\kappa_2 +\alpha_2(\kappa_1
- \kappa_2)])
\Big ]~.
\end{eqnarray}
From this expression we can now read off the anomalous dimension,
\begin{eqnarray}
\Delta \widehat{K}^{Gq}_0(\alpha_1,\alpha_2)  = 
C_F \left[\delta(\alpha_1) \delta(\alpha_2) -2\right]~.
\end{eqnarray}
Finally we would like to add a remark concerning the Feynman rules.
It is important to note that for non--local operators the involved
field operators  appear with unrestricted  momentum variables, e.g. 
$A^\nu (p_1),$ so that we are not enforced to determine operator
vertices explicitly. We  can simply apply the standard Feynman rules if 
we have in mind that the non--local operators are in fact $x$--space 
operators so that for diagrams in momentum space there remain certain 
exponential functions as relic terms from the lacking 
$\tilde x $--integration of the Fourier representation.
\section{The $[~~]_+$--prescription}
\renewcommand{\theequation}{\thesection.\arabic{equation}}
\label{sec-D}
\setcounter{equation}{0}

\vspace{1mm}
\noindent
The anomalous dimensions naturally appear in a regularized form,
Eqs.~(\ref{eqK1}\,--\,\ref{eqK4}) and (\ref{eqDK1}\,--\,\ref{eqDK4}) with the
$[~~]_+$--prescription given by (\ref{+pres}). This prescription 
influences the subsequent relations, especially those of the evolution
kernels $V^{AA}_{\rm ext}(t, t',\tau)$. Let us consider the typical term
\begin{eqnarray}
V_{\rm sing}(t,t',\tau) &=&
\int D_0\alpha \left[\delta(\alpha_1) + \delta(\alpha_2)\right]
\bigg[\frac{1}{\alpha_1 +\alpha_2}\bigg]_+ 
\delta \Big(t - t'(1-\alpha_1 -\alpha_2) +\tau(\alpha_1 - \alpha_2)
\Big)
\end{eqnarray}
which appears in $V^{qq}_{\rm ext} = \Delta V^{qq}_{\rm ext},~
V^{GG}_{\rm ext}$ and  $\Delta V^{GG}_{\rm ext}$. We introduce the 
variables 
\begin{eqnarray}
\alpha_1 = \lambda \xi,~~~~~\alpha_2 = \lambda (1- \xi)~.
\end{eqnarray}
The corresponding measures are
\begin{eqnarray}
\int D_0 \alpha \left [ \delta(\alpha_1) + \delta(\alpha_2) \right]
= \int_0^1 d\lambda \int D\xi = \int_0^1 d\lambda \int_0^1 d\xi
\left[\delta(\xi) + \delta(1-\xi)\right]~.
\end{eqnarray}
One obtains
\begin{eqnarray}
V_{\rm sing}(t,t',\tau) &=&  \int D\xi \, \int_0^1 d\lambda \,
\bigg[\frac{1}{\lambda}\bigg]_
+ \,\delta \Big(t-t'(1-\lambda) + \tau \lambda (2\xi-1)\Big) \nonumber\\
&=& \int_0^1 d\lambda \left[\frac{1}{\lambda}\right]_+ \left[
\delta(T(\lambda)-\tau \lambda) + \delta(T(\lambda) + \tau 
\lambda)\right] \nonumber\\
&=& \int_0^1 \frac{d\lambda}{\lambda} \left[\delta(T(\lambda) - \tau
\lambda) + \delta(T(\lambda) + \tau \lambda) - 2 \delta(t-t')\right],
\end{eqnarray}
with $T(\lambda) = t - t'(1- \lambda)$. The latter integral may now be
rewritten in terms of the variable $x$ and $y$, Eq.~(\ref{xy}), yielding
\begin{eqnarray}
V_{\rm sing}(t,t',\tau) 
&=&  \frac{1}{2|\tau|} \frac{1}{\left[y-x\right]_+}
\left[\rho(x,y) - \rho(\overline{x},\overline{y})\right]\nonumber\\
&=&
\frac{1}{2|\tau|} \left[\frac{\rho(x,y)}{\left[y-x\right]_+} +
\frac{\rho(\overline{x},\overline{y})}
{\left[\overline{y}-\overline{x}\right]_+} \right],
\end{eqnarray}
where the $[~~]_+$--prescription acts to the right.
\section{Sample calculation of extended evolution kernels}
\label{sec-C}
\renewcommand{\theequation}{\thesection.\arabic{equation}}
\setcounter{equation}{0}

\vspace{1mm}
\noindent
This Appendix is devoted to show how the extended evolution kernels
are obtained. As an example we consider
$V_{\rm ext}^{Gq}(t,t',\tau) $.
We evaluate Eq.~(\ref{ker3t})
\begin{eqnarray}
\label{kerGqa}
V_{\rm ext}^{Gq}(t,t',\tau)
&=&
\int_0^1 d \AAA \int_0^{1-\AAA} d \AB
{\widehat K}^{Gq}_0(\AAA,\AB)
\int_{- \infty}^{+ \infty} \frac{d(\kappa\xx p_+)}{2 \pi}
\nonumber\\
&\times&
\left(i \kappa\xx p_+~t\right)^{-1}
\exp\left\{
i\kappa\xx p_+\left[t-(1-\AAA-\AB)t' + \tau (\AAA -\AB)\right]
\right\},
\end{eqnarray}
with
\begin{eqnarray}
\widehat{K}^{Gq}_0(\AAA,\AB) &=&  -\, C_F
\left \{  \delta(\AAA) \delta(\AB) + 2 \right \}.
\end{eqnarray}
Because of the scaling relation
\begin{eqnarray}
V^{AB}_{\rm ext}(t,t',\tau)
 =
\frac{1}{\tau} V_{ext}^{AB}\left(\frac{t}{\tau},
\frac{t'}{\tau},1 \right),
\end{eqnarray}
and reminding the general investigations in \cite{LEIP}, we perform the 
calculations for $\tau =1$ and $ 0<x,y <1$. The relation of the variables
$x$ and $y$ to $t,t'$ and $\tau$ is given in Eqs.~(\ref{xy}).
The internal integral over $\kappa\xx p_+$ yields
\begin{eqnarray}
\label{Theta}
\frac{1}{2 \pi} \int_{- \infty}^{+ \infty}
 \frac{d(\kappa\xx p_+)}
{i \kappa\xx p_+ }
\exp \{i \kappa\xx p_+ X\}
=
\frac{1}{2}\left[\Theta(X)- \Theta(-X)\right] = \frac{1}{2} 
\varepsilon(X)~.
\end{eqnarray}
The latter equation is distribution--valued in the sense of tempered
distributions~\cite{GELFAND,VLA}. The l.h.s of Eq.~(\ref{Theta}) is
the Fourier--transform of the distribution
\begin{eqnarray}
\frac{1}{2\pi i} {\sf P} \frac{1}{z}~.
\end{eqnarray}
This distribution can be represented by
\begin{eqnarray}
\frac{1}{2\pi i}
{\sf P} \frac{1}{z} = \lim_{\varepsilon \rightarrow 0^+}
\frac{1}{4\pi i}\left[
\frac{1}{z-i\varepsilon} + \frac{1}{z+i\varepsilon}\right]~.
\end{eqnarray}
From Eqs.~(\ref{eqdist}a) and (\ref{sochotz}) follows then 
Eq.~(\ref{Theta}).

In terms of the variables $x,y$ we  solve
\begin{eqnarray}
t V_{\rm ext}^{Gq}(t,t',\tau)
&=&
\int_0^1 d \AAA \int_0^{1-\AAA} d \AB
{\widehat K}^{Gq}_0(\AAA,\AB) 
\\
& &
\hbox{\large$\frac{1}{2}$}
\left\{\Theta\left[x- (1-\AAA -\AB)y - \AB\right]
-\Theta\Big(-[x- (1-\AAA -\AB)y - \AB]\Big)\right\}~,
\nonumber
\end{eqnarray}
where
\begin{eqnarray}
\int D_0\alpha \,
\Theta\left[x- (1-\AAA -\AB)y - \AB\right]
&=&
 \left (x-\frac{y}{2}\right)+ \Theta(y-x)\frac{(y-x)^2}{2 y} \nonumber \\
& & \hspace{1.7cm}
-\Theta(x-y)\frac{(\overline y -\overline x)^2}{2\overline y}
\\
\int D_0\alpha
\Theta\left[x- (1-\AAA -\AB)y - \AB\right]
\delta(\AAA) \delta(\AB)
&=&
 \Theta(x-y)~.
\end{eqnarray}
Furthermore we observe that
\begin{eqnarray}
\int D_0\alpha \Theta\Big(-[x- (1-\AAA -\AB)y - \AB]\Big)
& =&
\int D_0\alpha \Theta\Big(\overline x- (1-\AAA -\AB)\overline y
- \AB\Big)~.
\nonumber
\end{eqnarray}
Using these equations and taking into account the general structure of 
$V_{\rm ext}^{Gq}(x,y)$ we obtain finally
\begin{eqnarray}
\label{eqVGQ}
 (2x-1)V_{\rm ext}^{Gq}(x,y)
&=&  C_F\,
\frac{1}{\tau}
\left\{ 
\Theta\left(\frac{x}{y}\right)\Theta\left(1-\frac{x}{y}\right)
{\rm sign}(y)\left(1-\frac{x^2}{y}\right) \right.
\nonumber \\
& & ~~~~~~\left.
- \Theta\left(\frac{\overline x}{\overline y}\right)
\Theta\left(1-\frac{\overline x}{\overline y}\right)
{\rm sign}(\overline y)
\left(1-\frac{{\overline x}^2}{\overline y}\right) 
\right\}~.
\end{eqnarray}
Eq.~(\ref{eqVGQ}) also
allows to derive the splitting function in the case
of forward Compton scattering. It is obtained in the the limit
$\tau \rightarrow 0$  for $t' > 0$, with ${\rm sign}(y) =
-{\rm sign}(\overline{y}) = 1$. 
Here the variables $x,y$ are expressed in terms of the variables
$(t,t')$ and $\tau$, Eqs.~(\ref{xy}).
In this limit $x/y = \overline{x}/
\overline{y} \rightarrow z = t/t'$ holds.
Thus $V_{\rm ext}^{Gq}(t,t',\tau)$ is given by
\begin{eqnarray}
\left.
V_{\rm ext}^{Gq}(t,t',\tau)\right|_{t' > 0}
= \frac{1}{t'~z} C_F
\Theta(z) \Theta(1-z) \left\{2 + \frac{
2 t^2 - 4 tt'}{2{t'}^2}\right\} +O(\tau)
\end{eqnarray}
for small values of $\tau$, from which
\begin{eqnarray}
 \lim_{\tau \rightarrow 0} \left.
 V_{\rm ext}^{Gq}(t,t',\tau)\right|_{t' > 0}
 =
\frac{1}{t'} C_F\,
\bigg(\frac{1+(1-z)^2}{z}\bigg) \Theta(z) \Theta(1-z)
\end{eqnarray}
results.
\end{appendix}
\newpage


\begin{thebibliography}{999}
%
\bibitem{OPEL}
K.G.~Wilson, Phys. Rev. {\bf 179} (1969) 1699;\\
R.A.~Brandt and G.~Preparata, Fortschr. Phys. {\bf 18} (1970)
249;\\
W.~Zimmermann, {\sf Lect. on Elementary Particle Physics and
Quantum
Field Theory}, Brandeis Summer Inst., Vol.~1,
(MIT Press, Cambridge, 1970),~p. 395;\\
Y.~Frishman, Ann. Phys. {\bf 66} (1971) 373.
%
\bibitem{AS}
S.A. Anikin, M.C. Polivanov, and O.I. Zavialov, Fortschr. Physik {\bf 27}
(1977) 459;\\
S.A. Anikin and O.I. Zavialov, Ann. Phys. (NY) {\bf 116} (1978)
135.
%
\bibitem{ZAV}
O.I.~Zavialov, {\sf Renormalized Feynman Diagrams},
(Nauka, Moscow, 1979), in Russian;  \\
{\sf Renormalized Quantum Field Theory} (Kluwer Academic Press,
Dordrecht, 1990).
%
\bibitem{BNL}
T. Ohrndorf, Nucl. Phys. {\bf 198} (1982) 26;\\
I.I.~Balitsky Phys. Lett. {\bf 124B} (1983) 230;\\
I.I.~Balitsky and V.M.~Braun,
{\sf Proc. XXV LNPI Winter School on Physics}, Leningrad (1990) p. 105.
%
\bibitem{LEIP}
D.~M\"uller, D. Robaschik, B.~Geyer, F.~Dittes, and J.~Ho\v{r}ej\v{s}i,
Fortschr. Phys.  {\bf 42} (1994) 2.
%
\bibitem{SLAC}
M.~Bordag and D.~Robaschik, Nucl. Phys. {\bf B169} (1980) 445;\\
M.~Bordag, B.~Geyer, J.~Ho\v{r}ej\v{s}i and D.~Robaschik,
Zs. Phys. {\bf C26} (1985) 591;\\
B.~Geyer, D.~M\"uller, and D.~Robaschik, preprint SLAC-Pub-6495 (1994).
%
\bibitem{EARLY}
R. Gatto and G. Preparata, Nucl. Phys. {\bf B47} (1972) 313;\\
G. Altarelli and G. Preparata, Phys. Lett. {\bf 39B} (1972) 371.
%
\bibitem{DRDR}
A. De R\'ujula  and E. De Rafael, Ann. Phys. (NY) {\bf 78} (1973) 132.
%
\bibitem{WMR}
E. Wieczorek, V.A. Matveev, and D. Robaschik, Teor. Mat. Fiz. {\bf 19}
(1974) 14.
%
\bibitem{RADL}
A.V.~Radyushkin, Phys. Lett. {\bf B380} (1996) 417.
%
\bibitem{RAD}
A.V.~Radyushkin,
Phys. Lett. {\bf B385} (1996) 333; Phys. Rev. {\bf D56} (1997) 5524.
%
\bibitem{BGR2}
J. Bl\"umlein, B. Geyer, and D. Robaschik, in~: Proc. of the Int Workshop
{\sf Deep Inelastic Scattering off Polarized Targets : Theory Meets 
Experiment}, DESY Zeuthen, September 1997, Eds. J. Bl\"umlein et al., 
(DESY, Hamburg, 1997),~p.~196; DESY 97--209 (1997) and 
{\tt hep-ph/9711405}.
%
\bibitem{BGR1}
J.~Bl\"umlein, B.~Geyer, and D.~Robaschik, Phys. Lett. {\bf B406} (1997) 
161 and Erratum.
%
\bibitem{JJ}
X.~Ji, Nucl. Phys. {\bf B402} (1993) 217;\\
R.~Jaffe and X.~Ji, Nucl. Phys. {\bf B375} (1992) 527.
%
\bibitem{GROSS}
D.J.~Gross and S.B.~Treiman, Phys. Rev. {\bf D4} (1971) 1059.
%
\bibitem{LAZAR}
M.~Lazar, {\sf Konstruktion und Anwendung irreduzibler nichtlokaler 
Lichtkegeloperatoren in der Quantenchromodynamik}, Diplomarbeit, Leipzig 
1998.
%
\bibitem{GLR}
B. Geyer, M.~Lazar, and D. Robaschik, {\tt hep-th/9901090}.
%
\bibitem{BB}
I.I.~Balitsky and V.M.~Braun, Nucl. Phys. {\bf B311} (1988/89) 541.
%
\bibitem{WW}
S. Wandzura and F. Wilczek, Phys. Lett. {\bf B72} (1977) 95.
%
\bibitem{BLK}
J. Bl\"umlein and N. Kochelev, Phys. Lett. {\bf B381} (1996) 296; 
Nucl. Phys. {\bf B498} (1997) 285.
%
\bibitem{CG}
C.G. Callan and D.J. Gross, Phys. Rev. Lett. {\bf 22} (1969) 156.
%
\bibitem{COM1}
R. Tarrach, Nuovo Cim. {\bf 28A} (1975) 409.
%
\bibitem{COM2}
D. Drechsel, G. Kn\"ochlein, A.Y. Korchin, A. Metz, and S. Scherer,
Phys. Rev. {\bf C57} (1998) 941.
%
\bibitem{BATU}
W.A. Bardeen and Wu--Ki Tung, Phys. Rev. {\bf 173} (1968) 1423.
%
\bibitem{GELFAND}
I.M.~Gel'fand and G.E.~Schilow, {\sf Verallgemeinerte Funktionen 
(Distributionen), Vol. I}, (Dt. Verlag d. Wissenschaften, Berlin, 1960);\\
Yu.A.~Brychkov and A.P.~Prudnikov, {\sf Integraltransforms of 
Generalized Functions} (Nauka, Moscow, 1977), in Russian.
%
\bibitem{BT}
J. Bl\"umlein and A. Tkabladze, DESY 98--181, {\tt hep-ph/9812478},
Nucl. Phys. {\bf B} in print.
%
\bibitem{BV}
J. Bl\"umlein and A. Vogt, Phys. Rev. {\bf D58} (1998) 014020.
%
\bibitem{BGR}
T.~Braunschweig, B.~Geyer, and D.~Robaschik, Ann. Phys. (Leipzig) {\bf 44}
(1987) 407.
%
\bibitem{Diss}
T.~Braunschweig, {\sf Einige Anwendungen der nichtlokalen 
Lichtkegelentwicklung auf die nicht--vorw\"arts Streuung in der QCD 
(Singlett--Fall)}, PhD Thesis, Leipzig, 1984.
%
\bibitem{BN}
J. Bl\"umlein, V. Ravindran, W.L. van Neerven, and A. Vogt, DESY 98--036,
{\tt hep-ph/9806368}, in~: {\sf Proc. of the 6th Int Workshop on Deep 
Inelastic Scattering and QCD}, Brussels, April 1998, eds.~Gh. Coremans
and R. Rosen, (World Scientific, Singapore, 1998) 211.
%
\bibitem{BARA}
I.I. Balitsky and A.V. Radyushkin, Phys. Lett. {\bf B413} (1997) 114.
%
\bibitem{AP1}
D.~Gross and F.~Wilczek, Phys. Rev. {\bf D8} (1973) 3633; {\bf D9} (1974)
980;\\
H.~Georgi and D.~Politzer, Phys. Rev. {\bf D9} (1974) 416;\\
G.~Altarelli and  G.~Parisi, Nucl. Phys. {\bf B126} (1977) 298.
%
\bibitem{AP2}
H.~Ito, Prog. Theor. Phys. {\bf 54} (1975) 555;\\
K.~Sasaki, Progr. Theor . Phys. {\bf 54} (1975)  1816;\\
M.A.~Ahmed and G.G.~Ross, Phys. Lett. {\bf B56} (1975) 385; Nucl. Phys.
{\bf B111} (1976) 441.
%
\bibitem{VLA}
V.S.~Vladimirov, {\sf Gleichungen der mathematischen Physik}, (Dt. Verlag
d. Wissenschaften, Berlin, 1972).
%
\bibitem{ROBA}
F.-M. Dittes, D.~M\"uller, D.~Robaschik, B.~Geyer, and J.~Ho\v{r}ej\v{s}i, 
Phys. Lett. {\bf B209} (1988) 325.
%
\bibitem{R2}
L. Mankiewicz, G. Piller, and T. Weigl, Eur. J. Phys. {\bf C5} (1998) 
119;\\
L.~Frankfurt, A.~Freund, V.~Guzey, and M.~Strikman, Phys. Lett. {\bf B418}
(1998) 345;\\ Erratum~: {\bf B429} (1998) 414.
%
\bibitem{XJ}
X.~Ji, Phys. Rev. Lett. {\bf 78} (1997) 610, Phys. Rev. {\bf D55} (1997) 
7114.
%
\bibitem{BL}
G.P.~Lepage and S.J.~Brodsky, Phys. Lett. {\bf B87} (1979)359; Phys. Rev.
{\bf D22} (1980) 2157;\\
S.J. Brodsky, Y. Frishman, G.P. Lepage, and C. Sachrajda, Phys. Lett.
{\bf B91} (1980) 239.
%
\bibitem{ER}
A.V.~Efremov and A.V.~Radyushkin, Phys. Lett. {\bf 94B} (1980) 245;
Teor. Mat. Fiz. {\bf 42} (1980) 97.
%
\bibitem{BLX}
M.K.~Chase, Nucl. Phys. {\bf B174} (1980) 109;\\
T.~Ohrndorf, Nucl. Phys. {\bf B186} (1981) 153;\\
M.A.~Shifman and M.I.~Vysotsky, Nucl. Phys. {\bf B186} (1981) 475;\\
M.V.~Terentjev,  Yad. Fiz. {\bf 33} (1981) 1692;\\
V.N.~Baier and A.B.~Grozin, Sov. J. Nucl. Phys. {\bf 35} (1982) 899;
Nucl. Phys. {\bf B192} (1981) 476.
\
\bibitem{RG}
I.M. Ryshik and I.S. Gradstein, {\sf Tables of Series, Products, and
Integrals}, (Dt. Verlag d. Wiss., Berlin, 1957).
%
\bibitem{RAD1}
A.V. Radyushkin, {\tt hep-ph/9805342}.
%
\bibitem{KN}
A.S. Kronfeld and B. Nizic, Phys. Rev. {\bf D44} (1991) 3445.
%
%
\bibitem{FORM}
J. Vermaseren, {\tt FORM 2.3}, (CAN, Amsterdam, 1996);\\
{\tt AXODRAW}, Comput. Phys. Commun. {\bf 83}  (1994) 45.
%
%
\bibitem{LEIB}
S.~Mandelstam, Nucl. Phys. {\bf B201} (1983) 149;\\
G.~Leibbrandt, Phys. Rev. {\bf D49} (1984) 1699; Rev. Mod. Phys.
{\bf 59} (1987) 1067;\\
G.~Leibbrandt, {\sf Noncovariant Gauges}, (World Scientific, Singapore,
1994).
\end{thebibliography}
\end{document}